\documentclass{JHEP3}    
\usepackage{epsfig,amsfonts,amsmath}

\newcommand{\eref}[1]{(\ref{#1})}
\newcommand{\sref}[1]{\S\ref{#1}}

\newcommand{\fref}[1]{Figure~\ref{#1}}
\newcommand{\cref}[1]{Chapter~\ref{#1}}
\newcommand{\beq}{\begin{equation}}
\newcommand{\eeq}{\end{equation}}
\newcommand{\ba}{\begin{array}}
\newcommand{\ea}{\end{array}}
\newcommand{\bcenter}{\begin{center}}
\newcommand{\ecenter}{\end{center}}

\def\C{\mathbb{C}}
\def\IC{\mathbb{C}}

\def\IF{\mathbb{F}}

\def\IGa{\relax\hbox{${\rm I}\kern-.18em\Gamma$}}

\def\IN{\mathbb{N}}
\def\IP{\mathbb{P}}
\def\IR{\mathbb{R}}

\def\IZ{\mathbb{Z}}

\def\diff{{\rm diff}}

\def\smiley{\hbox{\large$\bigcirc$\hspace{-0.80em}\raise.2ex
\hbox{$\cdot\cdot$}\kern-.61em\lower.2ex\hbox{\scriptsize$\smile$}}\ }
\def\frowny{\hbox{\large$\bigcirc$\hspace{-0.80em}\raise.2ex
\hbox{$\cdot\cdot$}\kern-.635em\lower.2ex\hbox{\scriptsize$\frown$}}\ }

\newcommand{\mat}[1]{\left( \matrix{#1} \right)}

\makeatletter
\let\hangafter\@hangfrom
\makeatother

\def\cN{{\cal N}}
\def\cM{{\cal M}}
\def\cW{{\cal W}}
\renewcommand{\diff}[2]{\frac{\partial #1}{\partial #2}}
\renewcommand{\mat}[1]{\left( \begin{matrix} #1 \end{matrix} \right)}
\newcommand{\be}{\begin{equation}}
\newcommand{\ee}{\end{equation}}
\newcommand{\bea}{\begin{eqnarray}}
\newcommand{\eea}{\end{eqnarray}}
\newcommand{\bean}{\begin{eqnarray*}}
\newcommand{\eean}{\end{eqnarray*}}
\newcommand{\bc}{\begin{center}}
\newcommand{\ec}{\end{center}}

\preprint{HUTP-05/A0049\\
{\tt hep-th/0511287}}

\title{Dimer Models from Mirror Symmetry and Quivering Amoeb\ae}
\author{Bo Feng$^{1,2}$, Yang-Hui He$^{3,4}$, Kristian~D.~Kennaway$^5$ and
  Cumrun Vafa$^6$\\
~\\
\begin{tabular}{rl}
$^1$ & Blackett Laboratory, Imperial College, London, SW7 2AZ, UK.\\
$^2$ & The Institute for Mathematical Sciences, Imperial College
  London, \\
  & 48 Princes Gardens, London SW7 2AZ, UK\\
$^3$ & Merton College, Merton St., Oxford, OX1 4JD, UK\\
$^4$ & Mathematical Institute, University of Oxford,
  24-29 St.\ Giles', Oxford, OX1 3LB, UK\\
$^5$ & Department of Physics, University of Toronto, 
  Toronto, ON M5S 1A7, CANADA.\\
$^6$ & Jefferson Physical Laboratory, Harvard University,
  Cambridge, MA 02138, USA
\end{tabular}
}

\abstract{Dimer models are 2-dimensional combinatorial systems that
  have been shown to encode the gauge groups, matter content and
  tree-level superpotential of the world-volume quiver gauge theories
  obtained by placing D3-branes at the tip of a singular toric
  Calabi-Yau cone.  In particular the dimer graph is dual to the
  quiver graph.  However, the string theoretic explanation of this was
  unclear.  In this paper we use mirror symmetry to shed light on
  this: the dimer models live on a $T^2$ subspace of 
  the $T^3$ fiber that is involved in mirror symmetry and is wrapped by
  D6-branes. These D6-branes are mirror
  to the D3-branes at the singular point, and geometrically encode the
  same quiver theory on their world-volume.}


\begin{document}

%=========================================================================
\section{Introduction}
%=========================================================================

Non-compact Calabi-Yau threefolds corresponding to toric geometries
have been a source of many interesting insights for string theory,
including geometric engineering, local mirror symmetry, topological
strings and large $N$ dualities.  One important question in this
context is how one describes for type IIB superstrings the gauge
theory living on a D3-brane placed at a singular point of a toric
threefold.

Some special cases of this correspondence have been understood for a
while \cite{Douglas:1996sw, Douglas:1997de, Hori:2000ck,
Hanany:2001py, Wijnholt:2002qz, Cachazo:2001sg,Feng:2001xr,Feng:2000mi}.  
More recently it was
shown that dimer models on $T^2$ encode the full data of the quiver
gauge theory \cite{Hanany:2005ve,Franco:2005rj}.  This simple picture
led to a number of new insights
\cite{Franco:2005zu,Benvenuti:2005cz,Benvenuti:2005ja,Franco:2005sm,Butti:2005vn}.
However, a direct explanation of the relevance of the dimer models as
systems deriving from string theory was unclear.  This link
between dimer models and the quiver gauge theory from toric geometries
must have a more direct explanation.  This is because it has been well
known that dimer models are naturally associated to toric geometries
(see \cite{Okounkov:2003sp} for a discussion of this point).

The problem we will address in this paper is to understand the
relation of the dimer models to string theory and their relation to
previous geometrical constructions of quivers \cite{Hanany:2001py}.
Following \cite{Hori:2000ck}, we use mirror symmetry to relate the
D3-brane to a system of intersecting D6-branes in the mirror geometry.
These are mirror to the wrapped branes on the exceptional cycle of
the singular geometry, 
which carry the gauge groups of the quiver theory, and whose pattern
of intersection encode the bifundamental matter of the quiver via the
massless strings localized at their intersection points.  We propose
that the geometry of the mirror D6-branes, which by mirror symmetry
should correspond to $T^3$, are further divided up to a number of
intersecting $S^3$'s.  Moreover, this collection of $S^3$'s admit a
projection to a $T^2\subset T^3$, with fiber being an $S^1$.  This
projection gives a certain tiling of $T^2$ where the boundary of faces
are identified with loci where the fiber $S^1$ vanishes.  This tiling
of the $T^2$ is the dimer model!

We explain how the geometry of the mirror D6-branes induces the
``rules'' of the dimer models on $T^2$, in particular why the dimer
models encode the gauge groups, matter content and tree-level
superpotential of the quiver theory coming from the D6-brane
intersections.  Specifically, matter arises from the intersection loci
of the D6 branes wrapping $S^3$'s and superpotential terms arise when
a collection of D6-branes meet at a point.  This will explain the
origin of the dimer model and the way it encodes the quiver gauge
theory.
  
%%%%%%%%%%%
%% THE SET-UP
\subsection{Setup and Organization}
\label{s:setup}
We study the physics living on the world-volume of D3-branes probing
transversely a toric Calabi-Yau threefold $\cM$. The theory is a gauge
theory whose matter content is summarised in a {\bf quiver diagram}.
One applies local mirror symmetry and translates this setup to
D6-branes wrapping Lagrangian $T^3$ in the mirror Calabi-Yau threefold
$\cW$. We will chiefly work in the mirror manifold $\cW$. The
intersection of the 3-cycles in $\cW$ gives the spectrum corresponding
to the quiver.  This is, by now, a standard construction
\cite{Hori:2000ck,Hanany:2001py,Cachazo:2001sg,Feng:2002kk}; however, 
we will propose a
more detailed description which we will summarise in
\sref{sec:proposal}.

%First, in order to familiarise the reader with the geometry, we begin
%with a review of the toric Calabi-Yau threefold $\cM$, how local
%mirror symmetry can be applied to make $\cW$ a double fibration and
%how precisely the D6-branes wrap 3-cycles in $\cW$.

%Bearing this setup in mind,
The organization of the paper is as follows.
In \sref{s:toric} we review
some relevant aspects of mirror symmetry as applied to toric
threefolds.  In \sref{sec:proposal}
we discuss the general structure of our
proposal.  In \sref{s:dimer} we discuss some facts about dimer models and local
toric threefolds.  We then analyse how one lifts the dimer model from
$T^2$ to a graph on a Riemann surface in \sref{sec:untwisting}; 
this constitutes a central idea of our proposal.
In \sref{s:amoeba} we discuss projections of the local
mirror geometry on the base of the mirror fibration (the so-called
`Amoebae') as well as the $T^2\subset T^3$ of the mirror fiber (the
so-called `Algae').  Finally, in \sref{sec:alga}
we show discuss how dimer models arise from Algae and also how our
proposal is concretely realized through examples.
We conclude with prospects in \sref{s:conc} and in 
Appendix \ref{app:counting} we give some mathematical results on Newton
polytopes and critical points.

%%%%%%%%%************
%*** TORIC DIAGRAMS, (p,q)-WEBS
%%%%%%%%%%%%%%%%%%%%%%%%%%%%%%%%%%%%%%%%%%%%%%%%%%%%%%%%%%%%%%%%%%
\section{Toric Geometry and Mirror Symmetry}\label{s:toric}
%%%%%%%%%%%%%%%%%%%%%%%%%%%%%%%%%%%%%%%%%%%%%%%%%%%%%%%%%%%%%%%%%%
In this section we discuss aspects of mirror symmetry in the context
of local toric 
threefolds following the derivation in \cite{Hori:2000kt} and its application
to the study of mirror of the D-branes discussed in \cite{Hori:2000ck}.
For a review of toric geometry and mirror symmetry see
e.g. \cite{claymirror}.

Local toric threefolds $\cM$ are specified by a convex integer sublattice
$Q\subset \IZ^2$ (the so-called {\bf toric diagram}). Our D3-branes
probe the tip of this singular $\cM$. 
It was shown in \cite{Hori:2000kt} (see also  \cite{Hori:2000ck}) that
the mirror geometry corresponding to it 
is given by a local threefold $\cW$ specified by
\bea
W &=& P(z,w) := \sum_{(p,q)\in Q}
 c_{(p,q)} z^p w^q
\label{WPzw} \\
W &=& u v 
\label{Wuv}
\eea
where $w,z$ are $\IC^*$ variables and $u,v$ are $\IC$ variables.
The coefficients $c_{(p,q)}$ are complex numbers and are mirror
to the K\"ahler moduli of the local toric geometry.  Three of the
$c_{(p,q)}$'s 
can be set to 1 by rescaling of the variables.   The mirror manifold
$\cW$ is therefore a 
double fibration over the $W$-plane.  Even though $W$ can be eliminated
to give the local geometry $P(z,w)-uv=0$, it is convenient for us
to view it as a double fibration over the $W$ plane.
We exemplify this in the diagram below, where we draw the toric
diagram of the Hirzebruch surface $\IF_0 \simeq \IP^1 \times \IP^1$;
the Calabi-Yau threefold $\cM$ is the affine cone over this surface:
\begin{equation}\label{F0toric}
\begin{array}{cc}
\epsfxsize = 4cm \epsfbox{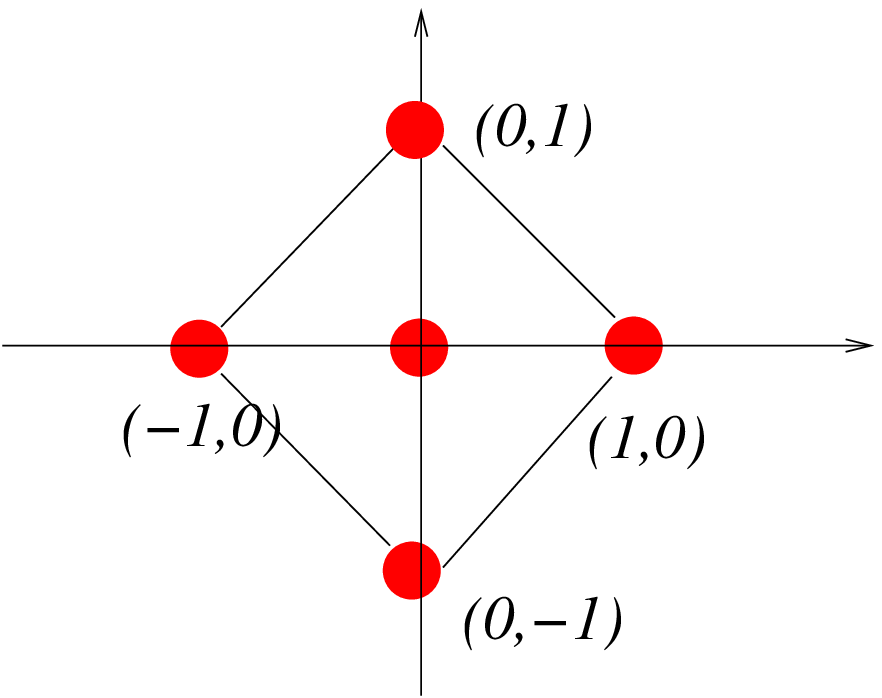}
& P_W(z,w) = c_{(0,0)} + c_{(0,1)} w +  c_{(0,-1)} \frac1w +  c_{(1,0)} z +
c_{(-1,0)} \frac1z
\end{array}
\end{equation}   
Out of the five $c$'s three can be set to 1 and the other two are
mirror to K\"ahler moduli of the two $\IP^1$'s. Another terminology to
which we shall later refer is that we call the interior lattice point(s)
in the toric diagram (such as the point $(0,0)$
in \eref{F0toric}) the {\em internal point(s)} and those on the boundary, the
{\it external points.} The number of internal points will turn out to
be very important to us.

The $T^3$ symmetry, denoted by three phases
$(\alpha, \beta, \gamma )$ which is used in the mirror symmetry derivation
(in accord with the conjecture \cite{Strominger:1996it}) act
on the geometry as
\bea
z &\rightarrow& \alpha z \nonumber \\
w &\rightarrow& \beta w \nonumber \\
u/|u| &\rightarrow& \gamma u/|u| 
\eea
(the action on $v$ is specified by requiring that $P(z,w)-uv=0$ remain
valid).  Since $u\in \IC$, the circle action on $u$ degenerates
when $u=0$.  In particular the $T^3$ mirror fiber
can be viewed as fibered over the $(\alpha, \beta)\in T^2$ with an $S^1$
fiber which degenerates at $u=0$, i.e.~at loci where $P(z,w)=0$.

%%--------------------------
\subsection{The Curve $P(z,w) = W$ and D6-branes}
\label{s:branes}
We are now ready to describe the mirror D6-branes and its relation to
the $T^3$ geometry described above, 
following \cite{Hori:2000ck}.
The most important equation, which encodes the essential content of
the toric geometry,
is the curve $P(z,w) = W$, the study of which shall be our primary concern.
The definition \eref{WPzw} dictates that $P(z,w)$ is the {\bf
Newton polynomial} associated with the toric diagram; conversely,
the diagram is the {\bf Newton polytope}\footnote{Strictly, the
Newton polytope is
  the convex hull of the lattice points of exponents. Here, since our
  original toric data is itself a convex polytope, the hull over $\IZ$
  is simply the toric points themselves.}
for the polynomial $P(z,w)$. These concepts are widely used in
combinatorial geometry. The coefficients $c_i \in \IC^*$, which parametrize
the complex structure deformations of $\cW$, correspond to
K\"ahler deformations of the mirror $\cM$. For our example in
\eref{F0toric}, the Newton polynomial for the associated toric diagram
of $\IF_0$ is indicated.

Now, this curve $P(z,w)-W=0$ is a genus $g$ (punctured) {\bf Riemann
surface}, which we will denote as $\Sigma_W$, fibred over each point
$W$. The genus is prescribed by the
simple relation \cite{mikhalkin-2000-2,Khovanskii} 
\beq\label{g=int} g
= \mbox{number of internal points in the toric diagram}.  
\eeq
Of particular importance to us is the fibre above the origin,
$P(z,w)=0$, which we call $\Sigma$.

Whenever we are at a critical point of $P(z,w)$, given by
\beq\label{critical}
(z_*, w_*) \mbox{ s.t. } \left. \left(
\diff{}{z}P(z,w) = \diff{}{w}P(z,w) = 0
\right)\right|_{(z_*,w_*)},
\eeq
or, in the $W$-plane, over the point $W_* = P(z_*,w_*)$, a cycle
in
$\Sigma_W$ degenerates and pinches off. On the other hand, the point
$W=0$ is, as mentioned above, special in that here the $S^1$-fibre
governed by $u,v$ pinches. We may therefore join, by a
straight-line,
$W=0$ to each of the critical points $W_*$ with a fiber
which is an $S^1\times S^1$; at the $W=0$ one $S^1$ shrinks and
at $W=W_*$ an $S^1\subset\Sigma$ degenerates.  The total space over
this interval thus has the topology of an $S^3$.  We
illustrate this structure in \fref{f:W}. These $S^3$'s form a
basis $\{S_i\}$ for $H^3(\cW; \IZ)$ and thus the $T^3$ class 
can be expanded therein as
\beq\label{t3}
[T^3] = \sum_i a_i S_i \ .
\eeq
\FIGURE[h]{\centerline{\epsfig{file=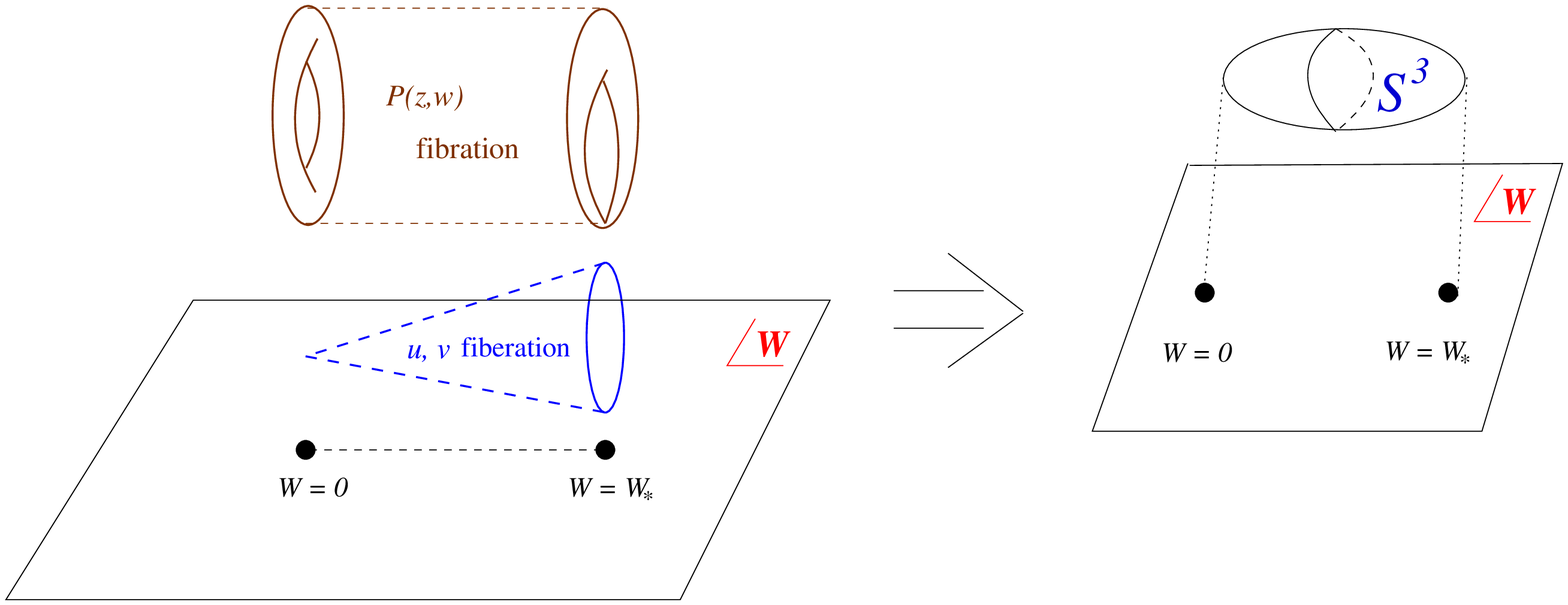,width=10cm}}
\caption{The geometry of the mirror of the toric threefold $\cM$. It is
  here shown explicitly as a double fibration over the $W$-plane: one
  being a circle $W = u v$ degenerating at $W=0$ and another being a
  fibration of a Riemann surface $\Sigma$ defined by
  $W = P(z,w)$ degenerating at critical points
  $W=W_*$. Together the two fibrations constitute 3-spheres over lines
  joining $0$ and $W_*$.}
\label{f:W}}
\subsubsection{D6-branes Wrapping the $T^3$-Class}
The D6-branes wrap the $S^3$-classes in \eref{t3} and therefore
will intersect one another in the fibre above the
origin where all the $S^3$ meet. This is shown in \fref{fig:gluing}. 
The graph $\Gamma$ where the
intersection takes place is of great significance and will be
explained in detail in \sref{sec:untwisting}.
\FIGURE[h]{\centerline{\epsfig{file=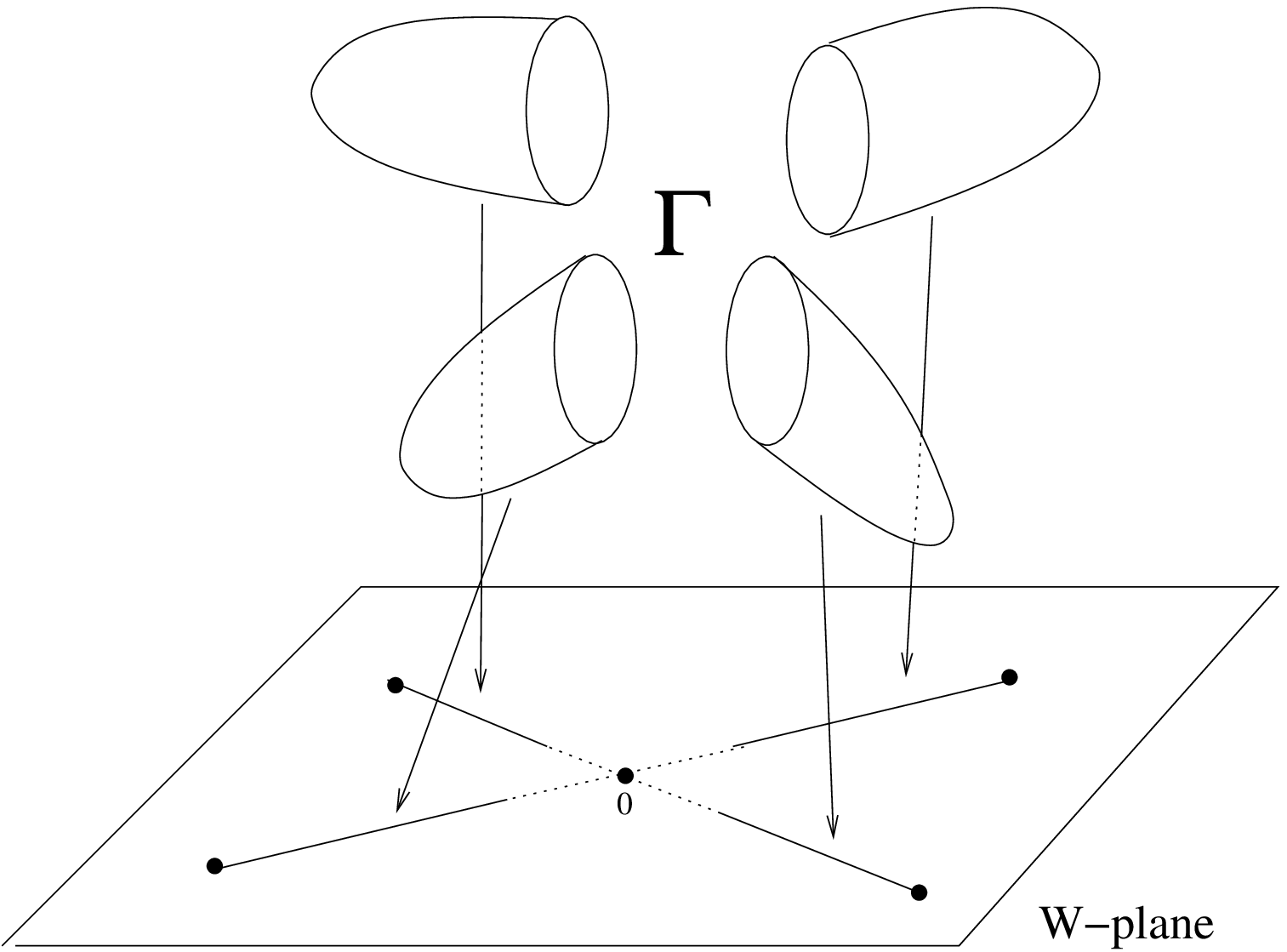,width=7cm}}
\caption{Above each straight-line path from the critical point of $W$
  to $W=0$ the vanishing cycle in the fibre $P(z,w)=W$ sweeps out a
  disc.  Above the origin the boundaries of these discs (hats)
  intersect the
  fibre $P(z,w)=0$ along some graph $\Gamma$.  Not shown is the other $S^1$
  fibre, which vanishes above the origin and is nontrivial elsewhere.
  The total space of each path is an $S^3$ in the mirror geometry.} 
\label{fig:gluing}}
Equation \eref{critical} tells us that the number of D6-branes,
which should be the number of gauge groups in the quiver theory,
should be equal to the number of critical points of $W=P(z,w)$, 
In fact, it was pointed out in \cite{Hori:2000ck} that
the matter content of the quiver theory is actually
captured by the soliton spectrum of a 2-dimensional $\cN = (2,2)$
Landau-Ginzburg (LG) theory with superpotential $W = P(z,w)$. 

We show in Appendix \ref{app:counting} that $W=P(z,w)$ has the correct
number of critical points to produce the basis of wrapped D6-branes as
long as the genus of the curve $P(z,w)=0$ is greater than 0 (i.e.~the
toric diagram contains 1 or more internal point).  This number of
critical points is equal to twice the area of the toric diagram, which
is in turn equal to the number of gauge groups in the quiver, as it
should.
This is equivalent to saying that
the original geometry $\cM$ contains a vanishing 4-cycle. If it
contains only vanishing 2-cycles (such as the conifold), then the
mirror curve has genus 0 and we seem not to obtain enough critical points
from $W=P(z,w)$ to describe the basis of D6-branes of the quiver
theory.

Thus it appears that the description of the mirror geometry as a
fibration over the $W$-plane is best suited to the case of vanishing
4-cycles, and there may exist another presentation better suited to
the case of vanishing 2-cycles.  This is related to the fact that
$W=P(z,w)$ is the superpotential of the massive LG model mirror to a
sigma model with target space given by the 4-cycle.  In the vanishing
2-cycle cases, it suggests that there should be a way to relate the
mirror geometry to the LG model mirror to the 2-cycle.
However, even in the case of vanishing 2-cycle, we are still able to
obtain the full data of the quiver theory\footnote{When the curve has genus 0
the resulting quiver is always {\it non-chiral}, an observation which
does not appear to have been made before.} from intersections of
1-cycles on the curve $P(z,w)=0$. Thus, the majority of our
results hold in generality.

%--
\subsection{Toric Diagrams and $(p,q)$-Webs}\label{s:pqweb}
Let us investigate the above picture from another well-known
perspective, namely that of $(p,q)$-webs.  The field theory
associated with the $(p,q)$-webs was studied in \cite{Aharony:1997bh},
where $(p,q)$ refers to a 5-brane, with $(1,0)$ being the D5-brane
and $(0,1)$ being the NS5-brane.  Suppressing 3 of the transverse
directions thus represents our branes as intersections of lines in
a two-dimensional co\"{o}rdinate system with axes $(p,q)$ so that
the charge of the
$(p,q)$ fivebrane is aligned to its slope.
These configurations are the so-named {\bf $(p,q)$-webs} of
fivebranes. Thus 5-branes are associated to edges in the web; one
too could associate D3-branes with vertices and D7-branes with
faces. In other words, the web, though generated from a
5-dimensional theory, essentially corresponds to D3, D5  and D7
branes wrapping 0, 2 and 4 cycles respectively in $\cM$, resulting
in an encoding of the original quiver gauge theory living on the
world-volume of the D3-brane probe. 

It was shown in \cite{Leung:1997tw} that this web of 5-branes
is S-dual to M-theory on a toric Calabi-Yau threefold.  Moreover,
\begin{quote}
{\em The $(p,q)$ web diagram is the graph dual of the toric
diagram of
  $\cM$.}
\end{quote}
This is really because there is a special Lagrangian $T^2 \times \IR$
fibration of $\cM$, where the $(p,q)$ cycle of $T^2$ vanishes along
the corresponding edge of the web.  The vanishing cycles turn into the
5-branes using S-dualities. In fact, the Riemann surface $\Sigma$
defined by $P(z,w)=0$ can be thought of as a thickening of the
$(p,q)$-web. We will make this statement clear in \sref{s:amoebapq}.
For now, let us illustrate the above discussion in \fref{f:F0pq}. For
our example of $F_0$ presented in \eref{F0toric}, we draw the
corresponding $(p,q)$-web in part (b) of \fref{f:F0pq} by graph
dualising the toric diagram which we redraw in part (a). In part (c)
we draw the Riemann surface $\Sigma$ defined by $P(z,w)=0$ and see
that it is a thickening of the $(p,q)$-web;
the punctures correspond to the legs while the handle corresponds to the
internal point. Far away along the $(p,q)$-directions $\Sigma$ looks
like cylinders.

\FIGURE[h]{\centerline{\epsfig{file=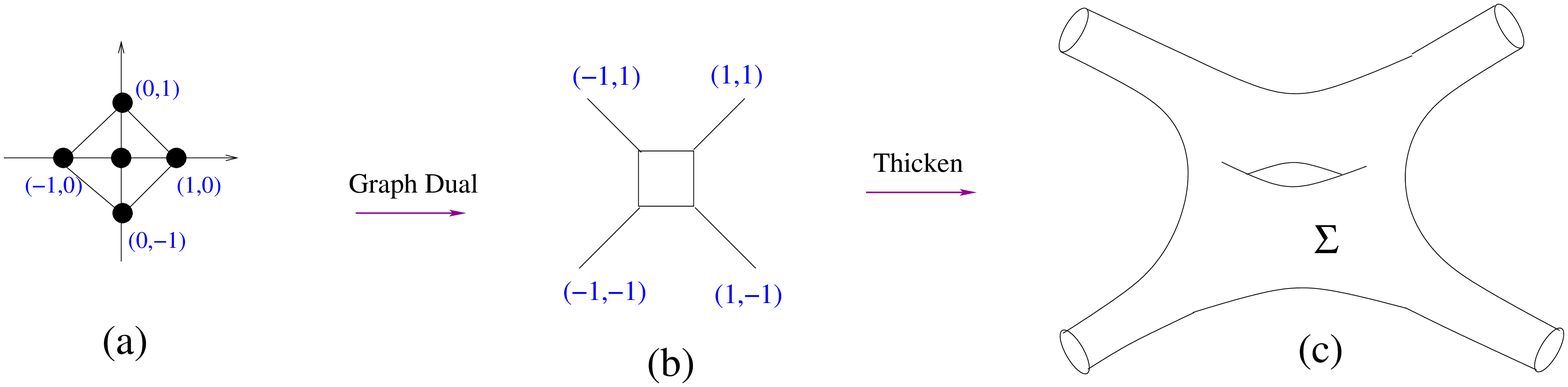,width=14cm}}
\caption{(a) The toric diagram of $F_0$; (b) the corresponding
  $(p,q)$-web as a graph dual, with the charges marked; (c) the
  Riemann surface $P(z,w)=0$ defined by the Newton polynomial of (a)
  is a thickening of (b); far away it looks like cylinders and in the
  interior it has holes depending on the number of internal points in
  (a).}
\label{f:F0pq}}

In the case of the toric diagram having one internal point
\cite{Hori:2000ck,Hanany:2001py}, the antisymmetric part of the
adjacency matrix of the quiver, i.e., the topological intersection
number, $S_i \circ S_j$, is given by a simple expression. In terms of
the $(p,q)$-charges of the legs, it is
\beq\label{pqdet}
A_{ij} = S_i \circ S_j = \det \mat{p_i & p_j \cr q_i & q_j } \ ,
\eeq
which we see to be explicitly antisymmetric due to the determinant.
This formula does not hold for more than one internal point.
However, the total number of fields in the quiver given by
\beq\label{totfield} 
N_f = 1/2 \sum_{i,j \in {\mbox{ legs}}} \left|
\det \mat{
p_i & q_i \\
p_j & q_j} \right| 
\eeq still holds for at least one phase of the
quiver theory. The reason for this is different though. As we
explain later, we will relate these $(p,q)$ charges to some
auxiliary curves (so-called zig-zag paths) instead of the
$S^3$-basis in \eref{t3}. Then, the formula (\ref{totfield}) counts
the total topological intersection number among these curves.  All
matter fields will be in 1-1 correspondence
with the intersection of these auxiliary curves.

%%%%%%%%%%%%%%%%%%%%%%%%%%%%%%%%%%%%%%%%%%%%%%%%%%%%%%%%%%%%%%%%%%%%%%%%%%%%
\section{Our General Proposal}
\label{sec:proposal}
%%%%%%%%%%%%%%%%%%%%%%%%%%%%%%%%%%%%%%%%%%%%%%%%%%%%%%%%%%%%%%%%%%%%%%%%%%%
Having reviewed the requisite knowledge and the
various inter-relations amongst them,  
we are now at a position to discuss our proposal in detail.
Our general strategy is very simple: we are interested in studying the gauge
theory on $N$ D3-branes filling the 3+1 dimensional
spacetime, and placed at the tip of a singular locus of a toric
Calabi-Yau threefold $\cM$.  Since
there are worldsheet instanton corrections to the geometry of toric threefolds
it is natural to use mirror symmetry where the classical geometry is reliable.
In this context we can study what gauge theory lives on the mirror of
the D3 brane. 

The mirror of the $N$ D3-branes will be $N$ D6-branes filling the spacetime and
wrapping the mirror of a point.  The mirror of a point will be a $T^3$
inside a Calabi-Yau $\cW$.  
For a generic point inside the Calabi-Yau, the mirror
of a point would be a smooth $T^3$ and thus the mirror geometry would be
the maximally supersymmetric $U(N)$ gauge theory in 6+1 dimensions,
reduced on a $T^3$ to 3+1 dimensions.  However, we are interested
in placing the D3-brane at a singular locus of the toric threefold.  Clearly
as we vary the point where the D3-brane is placed the fact that the mirror
is a $T^3$ does not change.  However, what may happen is that at special
points the $T^3$ degenerates -- similar to the degenerations familiar
for a $T^2$, where it can become pinched.  For example $T^2$ can get
pinched to a bunch of $S^2$'s joined back to back.

A similar thing is happening here for the $T^3$: placing $N$ D3-branes
at the singular point is the mirror image of $N$ D6-brane's wrapping a
singular $T^3$ made up of a collection of intersecting $S^3$'s.  The
reduction of $N$ D3-branes on $S^3$ gives rise to an ${\cal N}=1$,
$U(N)$ gauge theory in 4 dimensions.  So if we have the $T^3$ being
made up of $k$ component $S^3$'s we will have a theory with a gauge
group $G=U(N)^k$.

Furthermore, the $S^3$'s intersect one another. If they intersect over
an $S^1$ then we get a massless bifundamental hypermultiplet
$(N,\overline N )$ between the corresponding $U(N)$'s.  However, in the
cases discussed here we find that they intersect not over $S^1$'s but
over intervals. Since two intervals glued back to back make up an
$S^1$, we should be getting half of a hypermultiplet from an
interval, i.e., an ${\cal N}=1$ chiral $(N,\overline N )$ multiplet.
Furthermore, if a number of $S^3$'s intersect at a point, i.e., where
the intervals meet, then there would be world-sheet disc instantons
which get mapped to the intersection point of the $S^3$'s and can give
rise to a superpotential for the corresponding chiral multiplets.  
The result, is an ${\cal N}=1$ quiver gauge theory in 4 dimensions,
which is the familiar world-volume theory of the D3-brane probe
\cite{Douglas:1996sw}.

In order to flesh out the scenario presented in \sref{s:setup},
we will have to find a convenient
way of encoding the intersecting geometry of the $S^3$'s that make up
the $T^3$.  We recall that it is the $T^3$ which the D6-branes wrap
and the intersection of the $S^3$'s give the quiver matter content of
our gauge theory.
Our proposal will be to demonstrate intimate connections
between the mirror geometry, especially the Riemann surface $\Sigma$
and a certain corresponding dimer model the general properties of which
we review in the next section.

Specifically, we will show that in the fibre $\Sigma$ above $W=0$
(which we recall is given by the equation $P(z,w)=0$) 
there exists a
graph $\Gamma$ that admits non-trivial 1-cycles with the correct
properties to encode the gauge groups and matter content of the quiver
theory.  Thus, $\Gamma$ describes the intersection of the D6-branes
with the curve $\Sigma$ (cf.~\fref{fig:gluing}).

Next, we need to show how the intersecting D6-branes map to a $T^2
\subset T^3$.  This $T^2$ is embedded in the geometry according to
mirror symmetry, and we will exhibit it in two ways.
%----------------
\paragraph{Mapping $S^3$ to Discs on $T^2$}
Firstly, we describe a map between the intersection of the $S^3$'s
with the curve $\Sigma$, and a graph on $T^2$.  This $T^2$ is {\it
topologically} identified with the $T^2 \subset T^3$ of the D6-brane
world-volume.  We will find that the $S^3$'s map to polygons
(topologically, discs) that span the entire $T^2$, and join up along
the graph of a certain dimer model.  The remaining circle of the $T^3$
is fibred over the $T^2$ so that it vanishes precisely along this
graph, i.e., the intersection locus of $\Sigma$ with this $T^2$.

Moreover, we find that the discs intersect on intervals.  Thus the
intervals, i.e.~the edges of the graph obtained by the intersection of
$P(z,w)=0$ with the $T^2$, give rise to bifundamental chiral
multiplets.  Furthermore, the vertices of the dimer model, which
correspond to specific points on $T^2$ give rise to superpotential,
via open string disc instantons attached to the vertices.

One may read off the superpotential by going around the boundary of
the disc, which is a circle around the vertex of the graph and writing
the corresponding chiral multiplet associated with each edge the
circle intersects in an ordered fashion.  Furthermore, there is a sign
associated to the generated superpotential term.  It is easy to see
that it has to be there: vertices which are the boundary of a given
edge give rise to opposite signs for the superpotential contribution.

In order to argue for this, a heuristic reasoning is as follows:
Suppose we have an $S^1$ intersection region which can be {\it
artificially} divided to two intervals, from each of which we obtain a
chiral multiplet: $X:(N,{\overline N})$ and ${\tilde X}=({\overline
N}, N)$.  Then we will get superpotential terms from the two vertices
proportional to $Tr X {\tilde X}$ and $Tr {\tilde X}X$.  If they come
with the same phase these two terms will add and would correspond to a
massive hypermultiplet.  However, we know that the theory with an $S^1$
intersection must give rise to a massless hypermultiplet, therefore
these two terms should come with opposite phases.  So if we normalize
the field so one of them comes with a $+$ sign, the other one should
come with a $-$ sign so they would cancel.
%-------------------
\paragraph{Projections: Realising the $T^2$ Concretely}
Secondly,
we also find that in suitable cases the $T^2$ of the dimer model may
be obtained by projecting (the so-called alga projection
which we will discuss in \sref{sec:alga})
the $S^3$'s onto the $T^2 \subset T^3$ defined
by the phases of $(z,w)$ in our local model.  This gives a concrete
embedding of the $T^2$ into the geometry.

Since these D6-branes (faces of the dimer model) together span the
entire $T^2$ and the transverse $S^1$ that vanishes over the locus of
the dimer model graph within this $T^2$, we have identified the
singular $T^3$ that is mirror to the D3-brane at the
singular point of the toric Calabi-Yau threefold $\cM$.

%%%%%%%%%%%%%%%%%%%%%%%%%%%%%%%%%%%%%%%%%%%%%%%%%%%%%%%%%%%%%%%%%%%%%%%%%%%%%%%
\section{Dimer Models and Quiver Gauge Theories}\label{s:dimer}
%%%%%%%%%%%%%%%%%%%%%%%%%%%%%%%%%%%%%%%%%%%%%%%%%%%%%%%%%%%%%%%%%%%%%%%%%%%%%%%
We have reviewed the mirror geometry for toric threefolds as well as
$(p,q)$-webs,
the last requisite ingredient to our story are the dimer models.
In this section, 
we review some relevant properties concerning dimer
models \cite{Kenyon:2003uj,Kenyon:2002a} 
and then review their recently-uncovered
combinatorial relation to gauge theories
\cite{Hanany:2005ve,Franco:2005rj,Franco:2005sm}.

%%%%%%%%%%%%%%%%%%%%%
\subsection{Some Rudiments on Dimer Models}
%%%%%%%%%%%%%%%%%%%%%
Let us start with some basic terminologies.  A {\bf bipartite graph}
is a graph with the property that all vertices can be divided into
black or white, such that every black vertex is only adjacent (i.e.,
linked by an edge) to white vertices, and vice versa.  A {\bf perfect
matching} of a bipartite graph is a subset of edges such that every
vertex in the graph is an endpoint of precisely one such edge.  To the
chemist, the matching consists of white-black pairs ( ``dimers'')
linked by a single edge (bond). To the mathematical physicist, a {\bf
dimer model} is the statistical mechanics of such a system, viz., a
system of random perfect matchings of the bipartite graph with some
appropriately assigned weights for the edges. 

In principle, dimer
models can be discussed with arbitrary boundary conditions, but the
one in which we are particular interested are graphs on oriented
Riemann surfaces.  In fact, two such Riemann surfaces will be relevant
to us: one is the torus $T^2$ and the other is $\Sigma$ defined by
$P(z,w)=0$. We will discuss dimer models on $T^2$, and later,
isomorphic dimer models on $\Sigma$.

Many important properties of the dimer models are encoded in the {\bf
  Kasteleyn matrix} $K(z,w)$ \cite{Kasteleyn}.  
It is a weighted adjacency
matrix of the graph with (in our conventions) the rows indexed by
the white nodes, and the columns indexed by the black nodes 
constructed by the following rules:

\begin{enumerate}
\item Associate, to each edge, a number $e_i$ called the {\bf
  edge weight}. In previous literature on dimer models, the $e_i$ were
  taken to be real and subject to a constraint on the parity of the
  product of edges around the faces of the dimer model.  We shall see
  later than it is natural for us to take
  $e_i$ to be $\IC^*$-valued, and not to impose any sign constraint.

\item Now, for dimer models defined on $T^2$, because there are two
  nontrivial cycles $(0,1)$ and $(1,0)$ on the torus, 
we can introduce two complex
variables $z,w$ to count their nontrivial effects by assigning them
as weights on edges. The way of doing it is following. First, the colouring
of vertices in the graph induces an orientation to the edges, for
example, choose the orientation ``black'' to ``white''. 
Second, we can construct paths
$\gamma_w$, $\gamma_z$ in $T^2$ that wind once around the
$(0,1)$ and $(1,0)$ cycles of the torus, respectively.  Such paths
would cross the graph edges: for every
edge crossed by $\gamma_w$, multiply the edge weight by a factor of
$w$ or $1/w$ (similarly for $\gamma_z$, one multiplies by $z$ or
$1/z$) according to the relative
orientation of the edge crossed by $\gamma$.
\end{enumerate}

The adjacency matrix of the graph weighted by the above factors is the
Kasteleyn matrix $K(z,w)$ of the graph. The determinant of this matrix
$P(z,w) = \det K$ is called the {\em characteristic polynomial} of the
graph, and defines a {\bf spectral curve} via $P(z,w) = 0$.  The
astute reader may with foresight see why we have named this curve as
$P$; indeed later on it will be identified with the mirror curve in
\eref{WPzw}.  The fact that dimer models implicitly know about local
mirror symmetry may be taken as a hint that mirror symmetry should be
involved in their string theoretical realization.

To make these above concepts concrete, let us present an illustrative
example. This is actually the theory of the so-called phase one of the cone
over the zeroth Hirzebruch surface $F_0 \simeq \IP^1 \times \IP^1$
(cf.~\cite{Franco:2005rj,Feng:2001xr}) with the toric diagram and the
Newton polynomial given in \eref{F0toric}:
\begin{equation}\label{F0dimer}
\hspace{-2cm}
\begin{array}{|c|c|c|}\hline
\begin{array}{c} \epsfxsize = 11cm \epsfbox{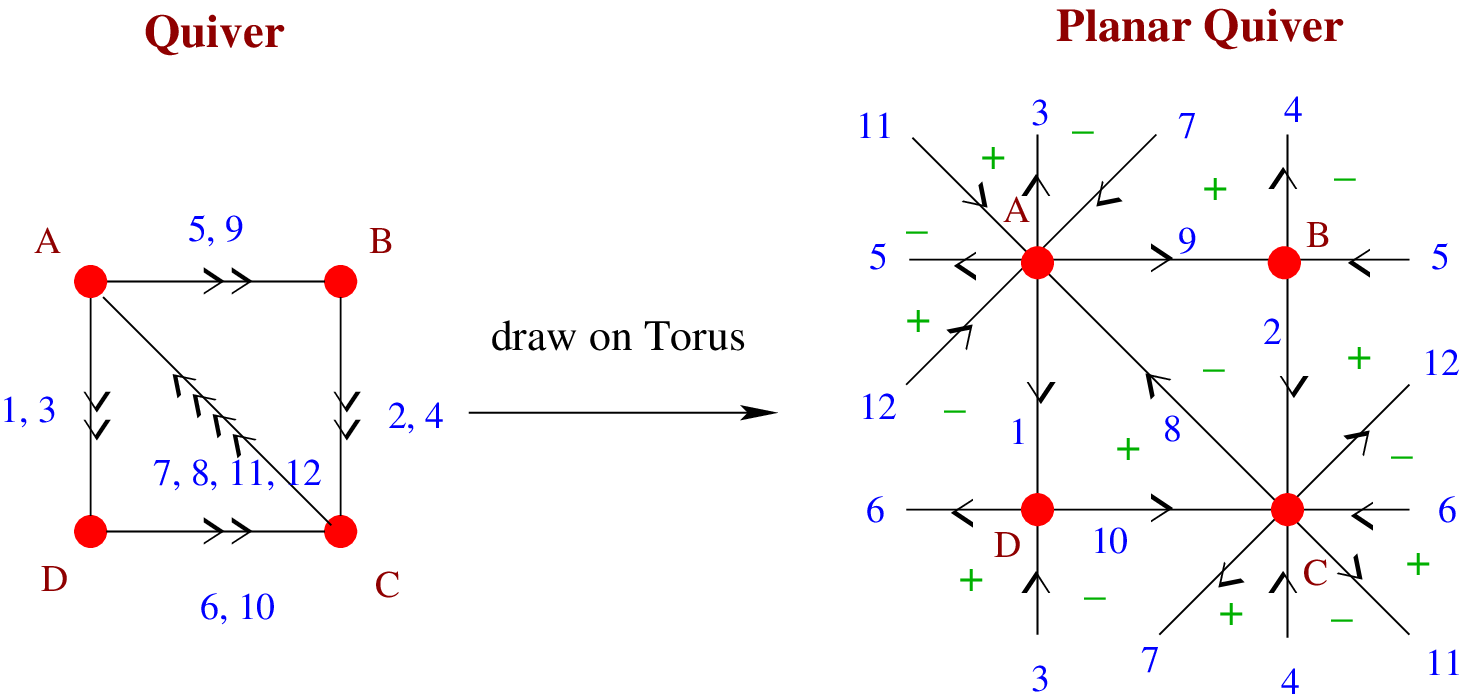} \end{array}
& \stackrel{\begin{array}{c}\mbox{Graph}\\ 
    \mbox{Dualise}\end{array}}{\Longrightarrow} & 
\begin{array}{c} \epsfxsize = 5cm \epsfbox{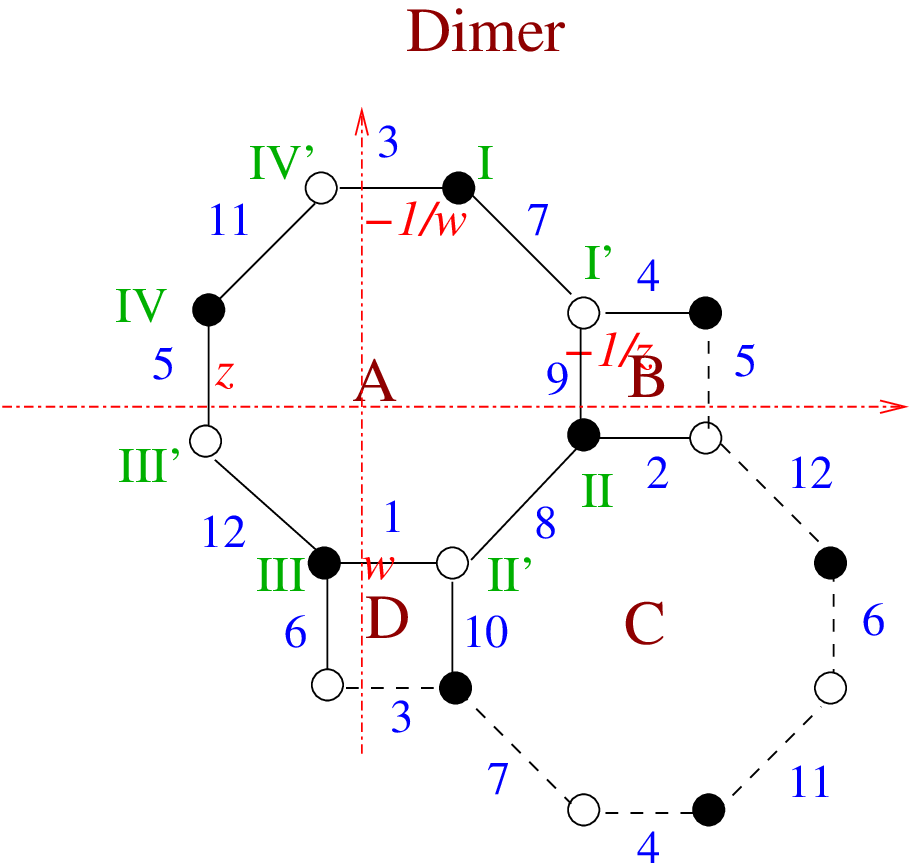} \end{array}
\\
\begin{array}{ccc}
W &=& X_1X_{10}X_8 - X_3X_{10}X_7 - X_2X_8X_9 - X_1X_6X_{12}\\
  && + X_3X_6X_{11} + X_4X_7X_9 + X_2X_{12}X_5 - X_4X_{11}X_5
\end{array}
& & 
\begin{array}{c}
K(z,w) = \begin{array}{c|cccc}
   & I'& II' & III' & IV' \\ \hline
 I & 1 & 1 & 0 & -1/w \\
 II& 1/z & 1 & 1& 0 \\
III& 0 & w & 1 & 0 \\
IV & 1 & 0 & -z & 1 \\
\end{array}
\\
\begin{array}{rcl}
P(z,w) &=& \det K(z,w) \\ 
&=& -1/w-w-1/z-z-5
\end{array}
\end{array}
\\
\hline\end{array}
\end{equation}

In fact, $P(z,w)$ can be calculated
by another way. Taking an arbitrary perfect matching $M_0$ on the graph
as a reference and another perfect matching $M$, 
we can get a set of closed curves in the graph using the
difference $M- M_0$. This in turn defines a {\bf height function}
on the faces of the graph: when a path (non-trivial cycle on $T^2$)
crosses
the curve, the height is increased or decreased by 1 according to
the orientation of the crossing. In terms of the height function,
the characteristic polynomial takes the following form:
\begin{equation}
\label{eq:det}
P(z,w) = z^{h_{x0}} w^{h_{y0}} \sum c_{h_x,h_y} (-1)^{h_x + h_y +
h_x h_y} z^{h_x} w^{h_y}
\end{equation}
where $c_{h_x,h_y}$ are integer coefficients that count the number of
paths on the graph with height change $(h_x,h_y)$ around the two
fundamental cycles of the torus.  It is precisely due to \eref{eq:det}
that the spectral curve is useful to the statistics of dimer models:
the function $P(z,w)$ is the generating function for the perfect
matchings $M$, counted according to the winding number of $M-M_0$.

The similarity between \eref{eq:det} and the Newton polynomial in
\eref{Wuv} was one of the initial inspirations that led to the
investigation of the correspondence between toric gauge theories and
dimer models. We point out, however, that the coefficients in
\eref{Wuv} are arbitrary complex parameters while those in
\eref{eq:det} are integers. We need general moduli in order to make
the dimer model explore the full Calabi-Yau geometry \cite{CYmoduli}.

Let us make some final remarks on the construction of $K(z,w)$ and
$P(z,w)$. There is a freedom in the choice of paths
$\gamma_z,\gamma_w$ as well as the reference matching $M_0$. This
freedom may result in different overall factors $z^{i_0} w^{j_0}$ in
front of $P(z,w)$, and an $SL(2,\IZ)$ transformation of the Newton
polytope. This should not trouble us since for the Newton polytope
(toric diagram) both transformations are induced by $SL(3,\IZ)$
transformations of the lattice $\IZ^3$, which does not effect the
underlying geometry.  Moreover, with the above choices of edge weights
in the construction of $K(z,w)$, the coefficients $c_{h_x,h_y}$ in
$P(z,w)$ are integers (as they were originally devised by
\cite{Kasteleyn} as a counting problem).  

However, as we shall
identify $P(z,w)$ with the definition fibration of a mirror geometry,
in order to account for the full moduli of the geometry\footnote{It is
  also necessary to impose the D-term constraints coming from the
  linear sigma model, which restrict the moduli space of the
  world-volume theory to the geometrical phases
  \cite{Douglas:1997de}.}, it is 
preferable to have the edge weights, and hence $c_{h_x,h_y}$, to be
arbitrary complex numbers, modulo the gauge degree of freedom
corresponding to multiplying the weights of all edges incident to a
given vertex by the same factor.

%%%%%%%%%%%%%%%%%%%%%%%
\subsection{Relation to Toric Gauge Theories}
%%%%%%%%%%%%%%%%%%%%%%%
Having reviewed the basics of dimer models, we continue recalling what
has already been spelled out in recent literature on the relationship
between dimer models and toric gauge theories
\cite{Hanany:2005ve,Franco:2005rj,Franco:2005sm}.

It is observed that for toric gauge theories, that is, the setup
discussed in \sref{s:toric} of placing D3-brane probes transverse to a
toric Calabi-Yau singularity, there are some
special properties. First, it is a quiver theory. This means that
every matter field carries only two gauge
indices, i.e., they are only charged under two gauge groups, as the
fundamental of one and the anti-fundamental of another. Therefore,
we can represent the matter content in terms of a finite graph called
the quiver diagram, where vertices denote gauge groups, and
edges the bi-fundamental matter fields \cite{Douglas:1996sw}.

Second, and this is special to toric singularities,
every matter field shows up two and only two times in the
superpotential, {\it one with plus sign and one with minus sign}. This
was referred to as the ``toric condition'' in \cite{Feng:2000mi}.
Furthermore, for {\it any} toric theory one may normalize the
coefficient of each term in the superpotential to 1 by a rescaling of
the fields.  This is associated to the fact that after blowing up the
toric singularity (i.e.~deforming the K\"ahler moduli away from the
singular point), there are no complex structure deformations of the
resulting non-singular Calabi-Yau manifold, and these complex
structure deformations would have shown up as coupling constants in
$W$.  It is only for non-toric singularities that one finds coupling
constants that cannot be normalized away (e.g. for the non-toric del
Pezzo cones, see \cite{Wijnholt:2002qz}).

Now, every monomial term in the
superpotential is gauge-invariant, which means that in the quiver
diagram it corresponds to a closed path, traced by the edges
(matter fields constituting this term) according to the
order of gauge group index contractions.
Combining this fact and the toric condition,
it is easy to see that if we think of the superpotential terms as
defining the boundary of {\it polygonal faces}, we can glue these
faces (superpotential terms) together along their common edges to
construct a tiling of an oriented Riemann 
surface without boundary, dubbed the ``periodic quiver'' in
\cite{Franco:2005rj,Franco:2005sm}. 
Furthermore, every face can be assigned a plus or a minus, according
to the sign of the term in the superpotential.

If we take the {\it planar dual graph of the periodic quiver}, that
  is, faces, edges, vertices are mapped to vertices, edges, faces
  respectively, we obtain a bipartite graph. In other words, {\em The
  planar graph dual of the periodic quiver is a dimer model on a
  Riemann surface.}

Finally, it can be shown \cite{Franco:2005rj}
that the superconformal conditions for
toric gauge theories, viz., the $R$-charge of every term in
superpotential must be 2 and the beta-function of every gauge
group must be 0, imply the relation
\beq\label{FVE}
F+V-E = 0 \ .
\eeq 
Here, $F,V,E$ are
the numbers of faces, vertices and edges, respectively, in the periodic
quiver diagram. Equivalently, they are the numbers of superpotential
terms, gauge groups and bifundamental matter fields, respectively, in
the gauge theory.
Of course, \eref{FVE} is the famous Euler relation,
$F+V-E = 2 - 2g$. Thus, $g=1$, and our periodic quiver actually lives
on a torus $T^2$.

Now let us use one simple example to demonstrate the above
procedure. The famous ${\cal N}=4$ $SU(N)$ theory has the
superpotential, in $\cN=1$ language, ${\rm Tr}(X_1 X_2 X_3-X_1 X_3
X_2)$. This is the theory of D3-branes probing the flat space
$\IC^3$. It is easy to construct the periodic quiver as shown in
\fref{f:C3dimer}, where after we identify edges $X_2$ and $X_3$ we get
a torus. Since there is only one black node and one white node, the
$K$-matrix is a one-by-one matrix, with 3 monomial terms in $z,w$;
this defines the toric data, as required.
\FIGURE[h]{\centerline{\epsfig{file=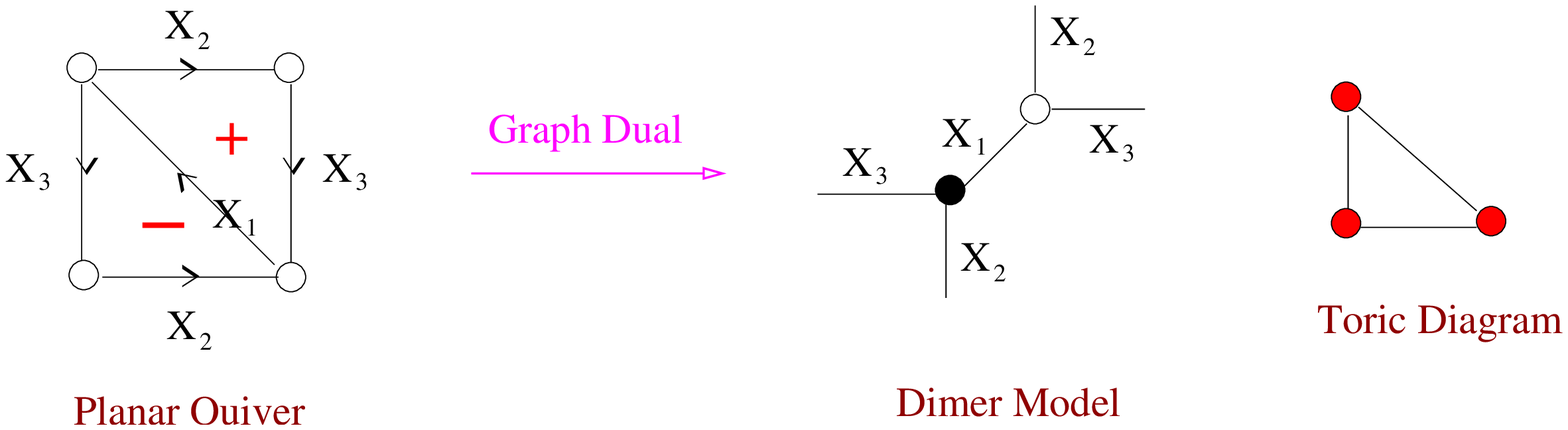,width=13cm}}
\caption{The planar periodic quiver for the theory on D3-branes on
  $\IC^3$, i.e., the famous, $\cN=4$ theory has one gauge group
  $SU(N)$ and superpotential ${\rm Tr}(X_1 X_2 X_3-X_1 X_3 X_2)$. Its
  graph dual is a dimer model with just one pair of nodes. The toric
  diagram consist of the lattice points $\{(0,0), (1,0), (0,1)\}$ and
  we have included it on the right. Indeed we can read out the toric
  diagram from the Kasteleyn matrix of the dimer model.  }
\label{f:C3dimer}}

In the dimer model, vertices correspond to superpotential terms.  By
multiplying every edge around a vertex by a common factor, the
normalization of each term in the superpotential can be fixed to 1 as
discussed above.  Since this operation corresponds to multiplying a
row or column of the Kasteleyn matrix by a common factor, it amounts
to rescaling the determinant $P(z,w)$, which is a gauge transformation
that does not affect the physics of the dimer model. 
%%
%-------------------
\subsubsection{Summary: Dimer Models and Quiver Theories}
Combining the above facts, we summarise the relation thus far known
between dimer models and toric quiver gauge theories:
\begin{description}
\item[(a)] Every toric gauge theory can be encoded into a
{\em periodic quiver diagram} living on $T^2$ where gauge groups are
represented by vertices, matter fields by edges, and
superpotential terms by faces;

\item[(b)] The planar dual of the periodic quiver diagram is a 
bipartite graph, where now gauge groups are represented by
faces, matter fields by edges, and superpotential terms with plus
(minus) sign, by white (black) vertices.

\item[(c)] The dimer model on this bipartite graph is equivalent to
  the linear sigma model description of the world-volume quiver theory
  on the D3-branes \cite{Douglas:1997de}.  In particular, the perfect
  matchings of the graph are in 1-1 correspondence with the fields of
  the linear sigma model \cite{Franco:2005rj}.
\end{description}
\subsubsection{Zig-zag Paths and the Double-line Notation}
There is one more crucial concept that will be useful in the
subsequent discussions. This is the so-called ``zig-zag'' paths in a
dimer model \cite{Kenyon:rhombic,Vegh:2005}.
We define a ``zig-zag'' path on a graph as follows: starting
with some edge on the graph, the zig-zag path starts to the left of
the edge and follows parallel to it, then turns to the right and
crosses the edge, before turning left so it is parallel to the edge
again.  Upon reaching a vertex the path turns right and follows
parallel to the next edge incident to the vertex, then crosses the
edge to the left side and continues parallel, and so on.  
We demonstrate this in \fref{f:doubleline}. We can trace the various
arrows and walk a zig-zag path along the edges of the dimer
graph\footnote{The reader may be more familiar with the definition of 
walking along an edge, such that
every time we reach a vertex, we choose to walk to the next
vertex which is alternatingly the rightmost and leftmost of the
present vertex. Our notation is an equivalent way of representing
this.
}.

Such a walk along the edges of the graph indeed traces out a zig-zag
pattern.  For an arbitrary bipartite graph on $T^2$, a zig-zag path
can intersect itself.  But for a dimer model describing a toric gauge
theory, it is shown \cite{Kenyon:2002a,Kenyon:rhombic,Vegh:2005} that
these zig-zag paths are simple closed curves in $T^2$ and never
intersect themselves. These dimer models belong to a special subclass
called ``isoradial embeddings'' \cite{Vegh:2005}.  

There is good way to
visualize the zig-zag paths using a double line notation (see
\fref{f:doubleline}).  In this notation, we see that, for our toric
dimer models, every edge has
two and only two zig-zag paths passing through it with opposite
directions. In fact, as we will show later, this double line notation
is not just a convenient tool, but has geometrical and physical
meaning.
\FIGURE[h]{\centerline{\epsfig{file=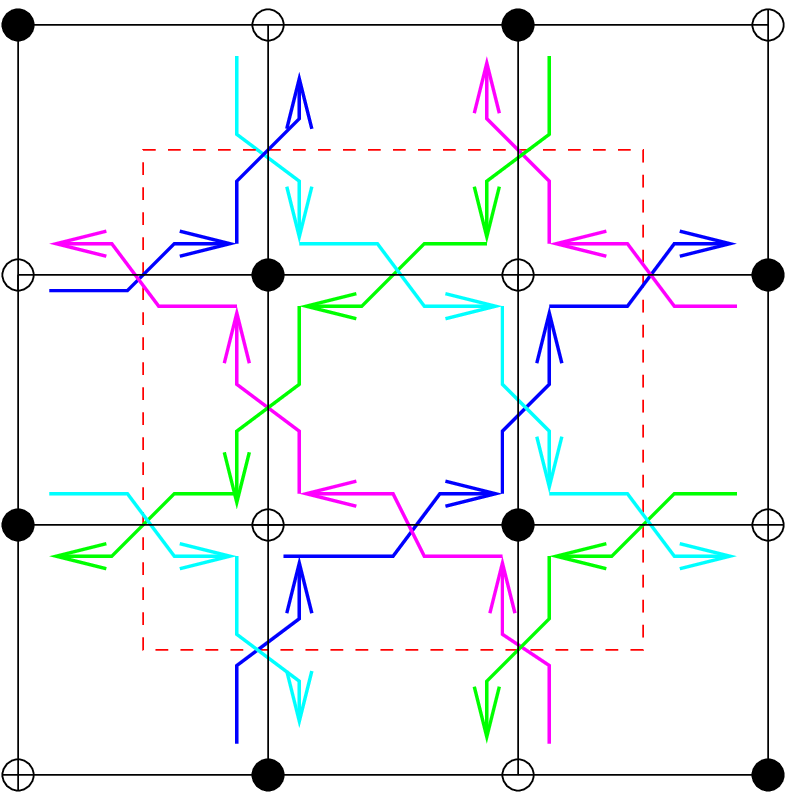,width=5cm}}
\caption{Given a dimer model, we can construct a zig-zag path of
  alternating leftmost and rightmost directions. A convenient way to
  represent it is the oriented double line notation where the zig-zag
  path crosses an edge at the mid-point. The region inside the dotted
  line represents a single fundamental domain of the $T^2$ on which
  the dimer model is defined.}
\label{f:doubleline}}
%

%%-------------------------------------------------------
\subsection{From Dimer Models to Planar Quivers via $(p,q)$-Cycles}
\label{s:interpolate}
%---------------------------
It was noted in equations \eref{pqdet} and
\eref{totfield} that the intersections of
certain $(p,q)$ cycles on $T^2$ (determined by the slope of external
lines in the 5-brane web) count the fields in the quiver.  This
observation will be given geometrical meaning in
\sref{sec:alga}.  Here we show that such abstract $(p,q)$ cycles allow
one to interpolate between the graph of the dimer model, and its dual
graph, the planar quiver.

\FIGURE[h]{\centerline{\epsfig{file=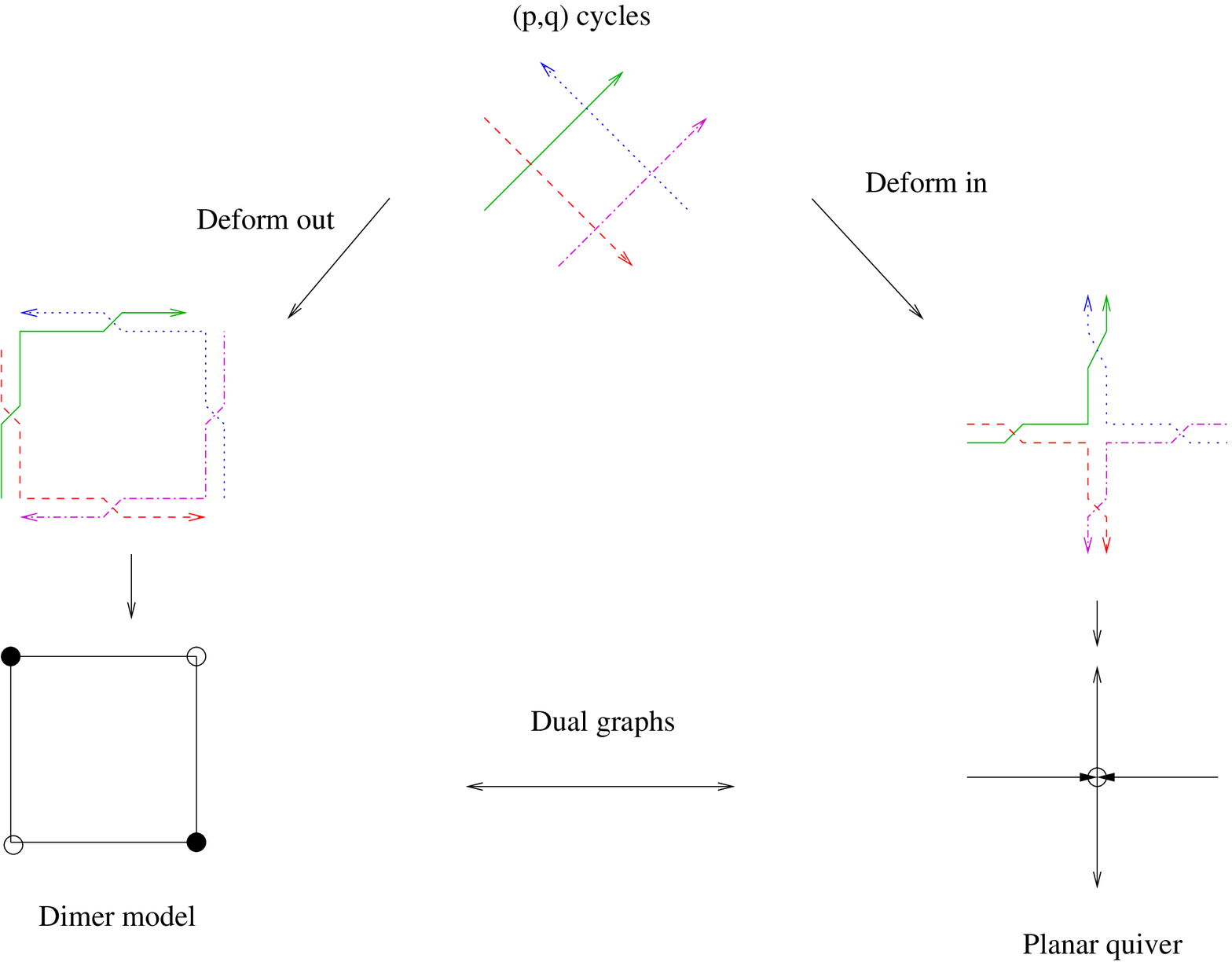,width=11cm}}
\caption{The $(p,q)$ winding cycles on $T^2$ interpolate between the
dimer model and its dual graph, the planar quiver. This crucial
observation will be a cornerstone to \sref{sec:untwisting}.}
\label{f:pq-deform}}
This is a crucial observation and we summarise it in
\fref{f:pq-deform}.  The dimer models are
obtained in the limit where the $(p,q)$ cycles form zig-zag paths on
an underlying graph.  In this limit the zig-zag paths form closed
loops around the vertices of the dimer model, with consistent
orientation that alternates at adjacent vertices.  By consistent we
mean that the zig-zag path traces a loop either clockwise or
counter-clockwise.
This orientation
induces the bipartite colouring of the vertices.  In the opposite
limit the vertices of the dimer model expand to become faces;
conversely the faces shrink to vertices, and the edges of the graph
become dual edges; thus, we obtain the planar quiver, which is the
graph dual.  The $(p,q)$
winding cycles allow us to interpolate between the two.

%%%%++++++++++++++++++++++++++++++++++++++++++++++++++++++
\section{Untwisting the Dimer Model from $T^2$ to $\Sigma$}
\label{sec:untwisting}
Our first task is to identify the dimer model which naturally arises
in our construction. We have a dimer model which is related to the
gauge theory by simply being the graph dual of the planar quiver. How
does this dimer graph relate to our mirror geometry?

In this section we discuss how to isomorphically map between a dimer
model on $T^2$ and a bipartite graph $\Gamma$ on the curve $\Sigma$
with the same adjacency matrix.  Furthermore, the mapped dimer graph
admits an alternative basis of 1-cycles whose intersection properties
reproduce the data of the quiver theory, which we may therefore
interpret as being part of the locus of the wrapped D6-branes.  In
fact the D6-branes are obtained by attaching certain discs with
boundary along these 1-cycles (together with another $S^1$ fibre that
vanishes along the boundary), as in \fref{fig:gluing}.  These
discs map to the faces of the dimer model on $T^2$ under the inverse
mapping.  Together, this will explain the physical relevance of the
dimer models to the quiver theories.

The procedure rests on the observation made in \sref{s:interpolate}
that the dimer model graph may be obtained by taking a limit of
intersecting $(p,q)$ winding cycles on $T^2$, where the polygonal
regions enclosed by these cycles with clockwise or counterclockwise
orientations are retracted to produce the bi-coloured vertices of the
graph.  Conversely, given such a bipartite graph, we can ``un-glue''
it to produce such a set of intersecting $(p,q)$ cycles (the zig-zag
paths) by merely reversing this process.  Each edge of the dimer model
is produced from two such cycles, which cross each other along the
edge.

%%---------------------------------------------
\subsection{The Untwisting Procedure}
\label{s:untwist}

%%%%%%%%+===============
Given a consistent dimer graph (one whose Kasteleyn determinant
defines a convex polygon), the associated zig-zag paths are uniquely
determined.  In fact, consistency of the zig-zag paths may be used to
constrain the allowed dimer graphs (since not every doubly-periodic
bipartite graph produces a Newton polygon that is convex and therefore
defines a toric CY).  As illustrative examples, we draw in
\fref{fig:zigzag-f0} the zig-zag
paths associated to the two dimer models describing quiver theories
for the CY cone over $\IP^1 \times \IP^1$, which are related by
Seiberg duality \cite{Franco:2005rj,Feng:2001bn}. 
We will have more to say on Seiberg duality
later on.
\FIGURE[h]{\epsfig{file=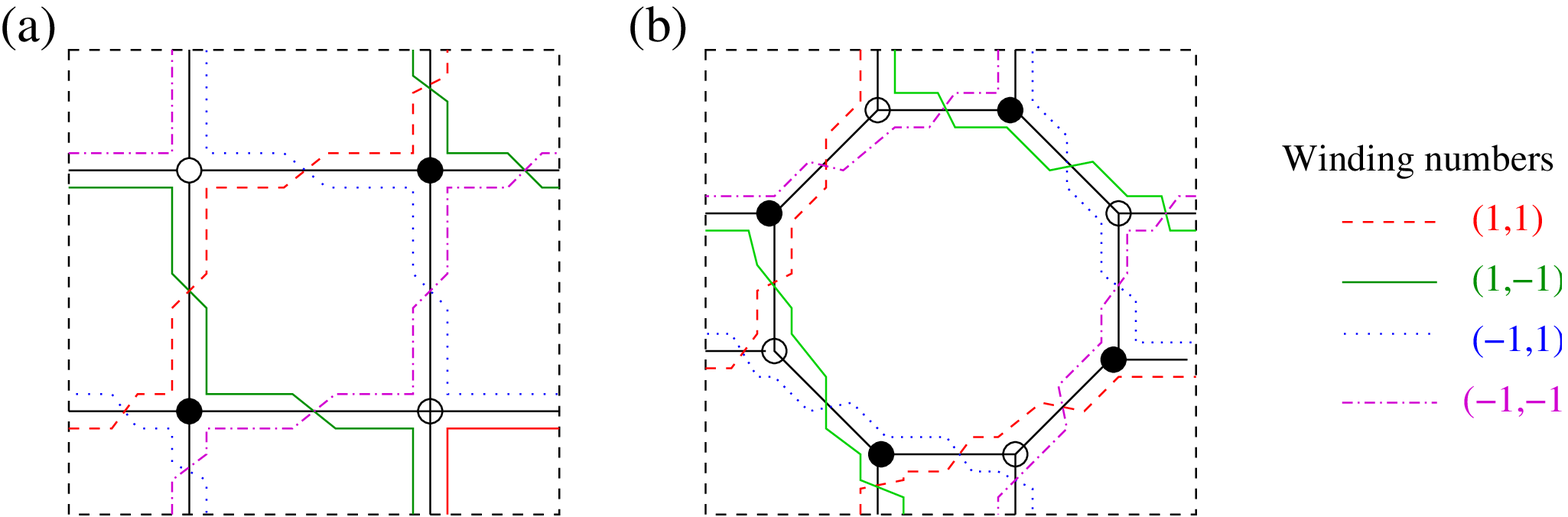,width=15cm}
\caption{The dimer models for the two Seiberg-dual phases of $F_0$ are
  shown in black in (a) and (b) respectively.  
The zig-zag paths are drawn in colour according to
  their winding number (direction of paths not indicated explicitly).}
\label{fig:zigzag-f0}}

To map the dimer model from $T^2$ to $\Sigma$, recall that each
zig-zag path is a winding cycle with $(p,q)$ winding on the $T^2$ of
the dimer model.  Furthermore, recall from \sref{s:pqweb} 
(especially \fref{f:F0pq})
that each such $(p,q)$ winding cycle is
canonically associated to a contour encircling one of the punctures on
the curve $\Sigma$.  
The precise sense in which this is true will be
discussed in section \ref{s:amoebapq}, but for now we observe that the
$(p,q)$ external legs of the web are mapped to a $(p,q)$ cylindrical
region in the neighbourhood of a puncture on the curve.

This suggests that the $(p,q)$ zig-zag path should be associated to an
$S^1$ contour around the corresponding cylinder in $\Sigma$.  
This mapping is
realized by a certain ``untwisting'' operation shown in
\fref{fig:untwist}.  On the left, we show the edge of the dimer model
(which separates two faces), given by a crossing of two zigzag paths.
We untwist by flipping this crossing, the zigzag paths now bound a
segment of a closed polygon, whereas the boundary of the dimer model
faces (solid and dashed lines) now cross one another.  When we later
relate these crossing paths to D6-branes, we will obtain a massless chiral
multiplet from each crossing.
\FIGURE[h]{\epsfig{file=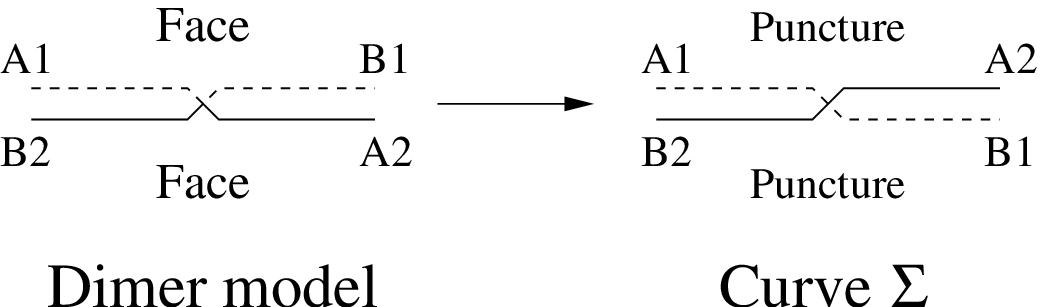,width=10cm}
\caption{The local untwisting operation that maps the dimer graph to a
  tiling of $\Sigma$. The crossing of two zigzag paths $A, B$ in the
  dimer model (left) can be flipped to the boundary of a closed
  polygon (right).}
\label{fig:untwist}}

This is a local operation performed at each crossing of zig-zag paths.
It cannot be done as a planar operation on $T^2$.  Indeed, as we will
show, it converts a tiling of $T^2$ to a new tiling of the punctured
genus $g$ Riemann surface $\Sigma$.

As we successively untwist each edge in the dimer model, the $(p,q)$ winding
paths become the boundary of {\bf closed polygons} in the new tiling.
These polygons are identified with the cylinders of $\Sigma$, so they
have a puncture at a point in the interior, and the $(p,q)$ winding
paths have mapped to contours encircling the punctures, as
anticipated.  This procedure amounts to rewriting each zig-zag path on
$T^2$ as the boundary of a new polygon in the plane.  Doing this for
each zig-zag path, we obtain a collection of such polygons, and when
glued together according to the gluing of zig-zag paths on the dimer
model, we obtain a tiling of a new Riemann surface.  We now show that
this new Riemann surface is topologically equivalent to $\Sigma$.

Let us demonstrate this in detail, still adhering to the $F_0$ example
above. First, let us re-draw, in \fref{fig:zigzag-f0-2}, 
the zig-zag paths of the two phases (a) and (b) as defined in 
\fref{fig:zigzag-f0}; we have now carefully labeled each piece of the
zigzag paths.
\FIGURE[h]{\epsfig{file=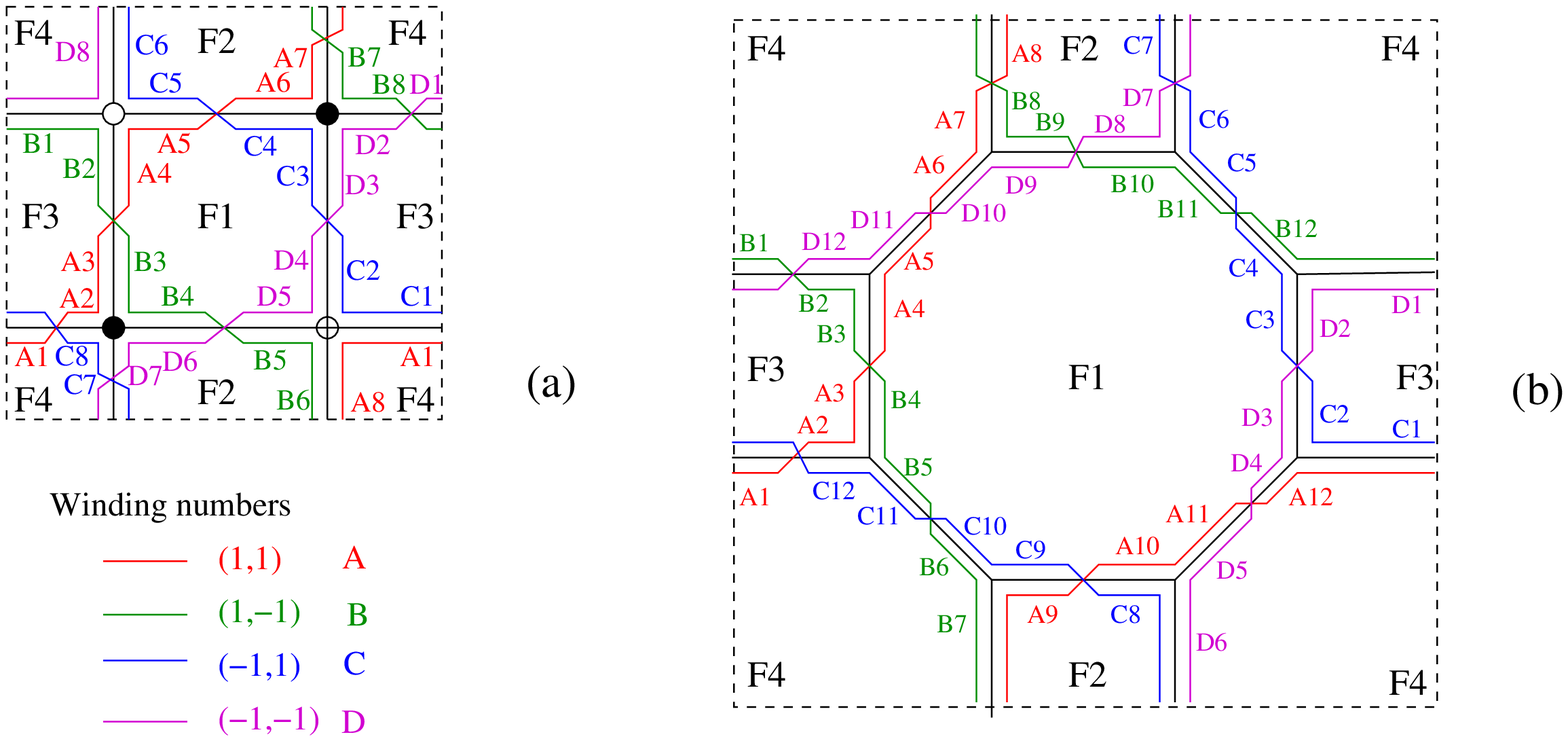,width=15cm}
\caption{Labeling the path segments on the zig-zag paths.  Each
  segment of the zig-zag paths is glued to another to form the dimer
  graph in black.} 
\label{fig:zigzag-f0-2}} 
Now, we can perform the untwisting procedure. The result is given in
\fref{fig:zigzag-f0-4}. Let us follow, for example, the segments $A7$
to $A8$, and the juxtaposed $B6$ to $B7$ in part (a) of
\fref{fig:zigzag-f0-2}. We see that $A7$ and $B7$ are glued together
while $B6$ and $A8$ are so glued. Therefore, in part (a) of
\fref{fig:zigzag-f0-4}, we see that along the boundary of regions $A$
and $B$, we have the pairs $7-7$ and $8-6$ being adjacent. So too,
are, for example, $C2-D4$ and $C3-D3$, etc. In this picture, the
zigzag paths are simply the boundaries of the polygonal regions
$A,B,C,D$, which correspond to the punctures in $\Sigma$.
\FIGURE[h]{\epsfig{file=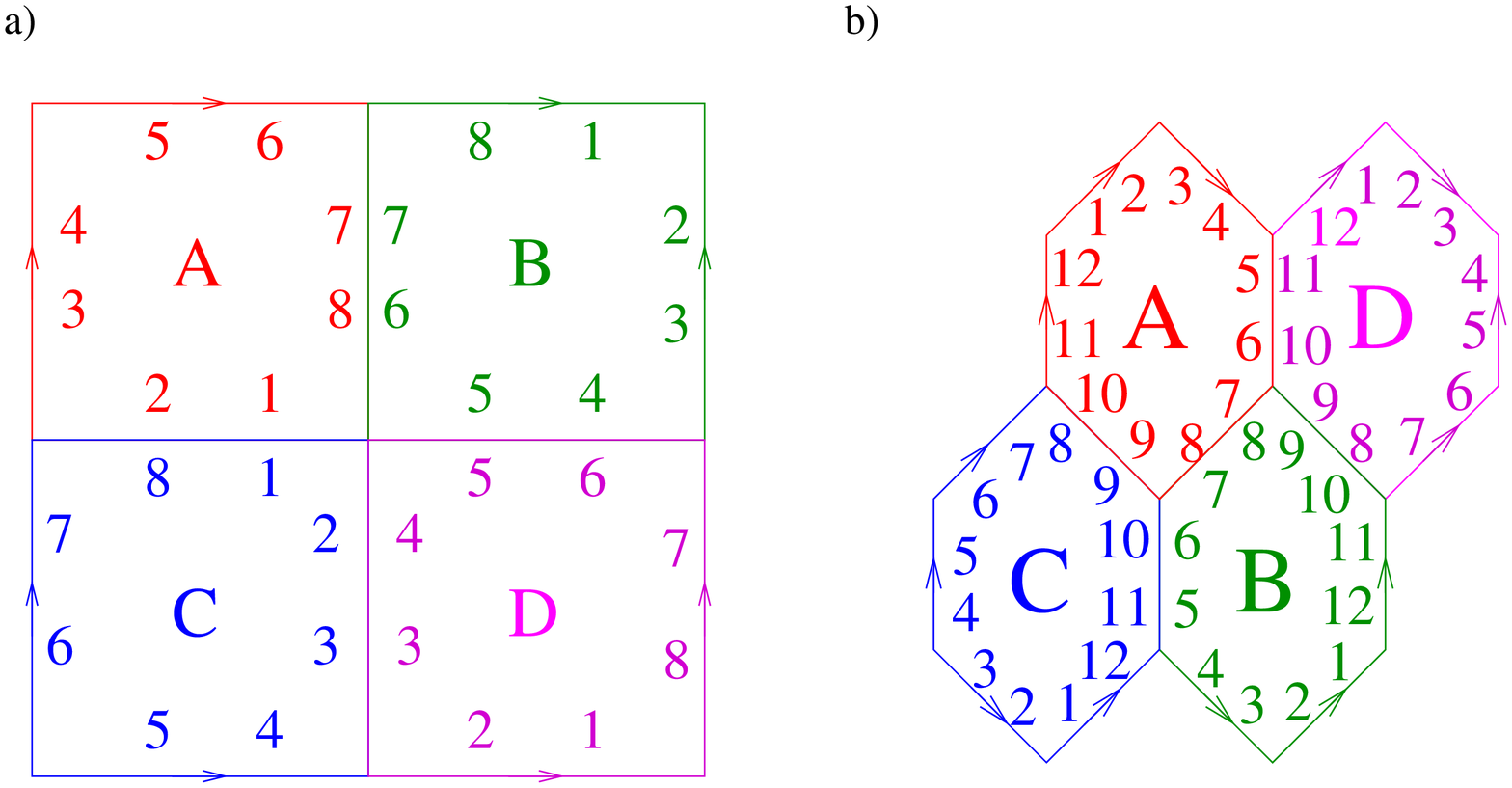,width=10cm}
\caption{Redrawing the zig-zag paths as bounding polygons in the
  plane.  Since this is a tiling of a genus 1 curve, opposite sides in
  the tiling are to be identified. From the two dimer models of $F_0$
  shown in \fref{fig:zigzag-f0-2}, we obtain two tilings of a
  genus 1 curve with a puncture in the interior of each of the four
  faces, which we identify with the curve $P(z,w)=0$.  The two
  Seiberg-dual dimer models correspond to two different choices of
  gluing.  Note that the quartic and trivalent vertices of the dimer
  model are preserved in this tiling of $\Sigma$ by construction, so
  strings may interact locally around such a vertex to produce
  superpotential terms.} 
\label{fig:zigzag-f0-4}}

Now let us discuss the genus of the surface which results from this
gluing.
Recall, for the dimer model on $T^2$ we had, from \eref{FVE},
\begin{equation}
V + N_g - N_f = 2 - 2 \times 1 = 0
\end{equation}
where $V$ are the vertices of the dimer model, $N_g$ the number of
faces (which correspond to gauge groups of the quiver theory), and
$N_f$ the number of edges of the dimer model (fields in the quiver
theory).

Each puncture on the curve $P(z,w)=0$ comes from a semi-infinite
external line in the $(p,q)$ web, equivalently to a line segment on
the boundary of the toric diagram.  The number $N_p$ of such segments
is equal to the number of lattice points on the boundary on the toric
diagram (what we called external points). The number of gauge
groups in the quiver theory is equal to twice the area of the toric
diagram, which by Pick's theorem \cite{Pick}
(see e.g.,\cite{Vegh:2005} in this context) 
is given by
\begin{equation}\label{pick}
N_g = 2 \mbox{Area} = 2 I + N_p - 2 
\end{equation}
where $I$ is the number of internal points.
Thus,
\begin{equation}
V + N_p - N_f = 2 - 2I \ .
\end{equation}
Therefore, the result of the untwisting is a Riemann surface of genus
$g=I$. Now, the vigilant reader would recall the fact from
\eref{g=int} that the genus of a curve is equal to the number of
internal points of its Newton polygon. Hence, we have reconstructed
the surface $\Sigma$, of genus $I$.
Indeed, the faces of this polygonal tiling are in 1-1 correspondence
with the punctures of $\Sigma$, as desired.

The above construction amounts to giving a double meaning to each
$(p,q)$ zig-zag path: {\it it is the $(p,q)$ winding cycle in the
$T^2$ of the dimer model; it is also the boundary of a face in the
  tiling of the curve $\Sigma$, associated to a punctured region.}

Indeed, there is another set of 1-cycles constructed from the
combination of various segments of zigzag paths. For example, in (a)
of \fref{fig:zigzag-f0-2}, we can take the closed loop formed by
$A4-A5-C4-C3-D4-D5-B4-B3$.  Such cycles bound each face using the dimer
model, which was associated to a gauge group in the dimer model rules.
After untwisting to construct the surface $\Sigma$, one may trace
these paths in the tiling of $\Sigma$ and verify that they
generically turn into 1-cycles with non-trivial homology class.

What we have done is to start with a dimer model which tiles $T^2$ 
(and whose dual graph is the planar quiver) and simply re-identified the
faces and edges using the untwisting procedure. The result is a new
bipartite graph $\Gamma$, i.e., a new dimer model, which now tiles
the Riemann surface $\Sigma$ of genus $g=I$.

%%%%------------------------
\subsubsection{Summary: Duality between Dimer Models on $T^2$ and $\Gamma
  \subset \Sigma$}
\label{s:duality}

In summary, we find a kind of duality between dimer models in $T^2$
and the new bipartite graph $\Gamma$ in $\Sigma$: {\it the winding
cycles and boundary of faces of one object are mapped to the boundary
of faces and the winding cycles of the other}\footnote{This duality
was called the {\it antimap} in \cite{Lins:1980}.  We thank D.~Vegh
for pointing out this reference to us.}.

In fact, it is clear that apart from re-identifying the nature of the
faces of the dimer graph, (i.e.~considered as an operation on the abstract
graph itself), the twisting operation is a {\bf graph isomorphism}
since we do not change the adjacency of the edges or the vertices.  In
particular, the dimer model on $\Gamma$ is the same as the dimer model
on its twisting\footnote{It is interesting to note that since
  these two dimer models are isomorphic, they have the same
  characteristic polynomial, and we have produced a dimer model that is
  {\it defined on its own spectral curve} $\det K = 0$.}
to $T^2$.

Note that this construction is not limited to the case where $\Sigma$
has genus 1 as in \cite{Hanany:2001py}.  For $g>1$ the wrapped branes
correspond to some combination of the $A_i$ and $B_j$ cycles of the
curve, and one may use the zig-zag construction to read off their
homology class and intersection numbers.

By construction, we have not changed any of the data of the quiver
theory, which was efficiently packaged in the graph of the dimer
model.  We have simply redrawn the graph of the dimer model as a
certain graph on the curve $P(z,w)=0$, with the same combinatorial
properties.  This graph winds the nontrivial cycles of $\Sigma$, and
certain of these cycles are those wrapped by the D6-branes, which we
will discuss further in the next section.  

For now, let us present 2 illustrative examples.  In
\fref{fig:c3-cycles}, we study the $\IC^3$ geometry encountered in
\fref{f:C3dimer}. Here, $\Sigma$ has $g=0$, and is a 3-punctured
sphere. The dimer model was drawn in \fref{f:C3dimer} and has a single
face. The face is dualised to a single bounding path on the sphere.
\FIGURE[h]{\centerline{\epsfig{file=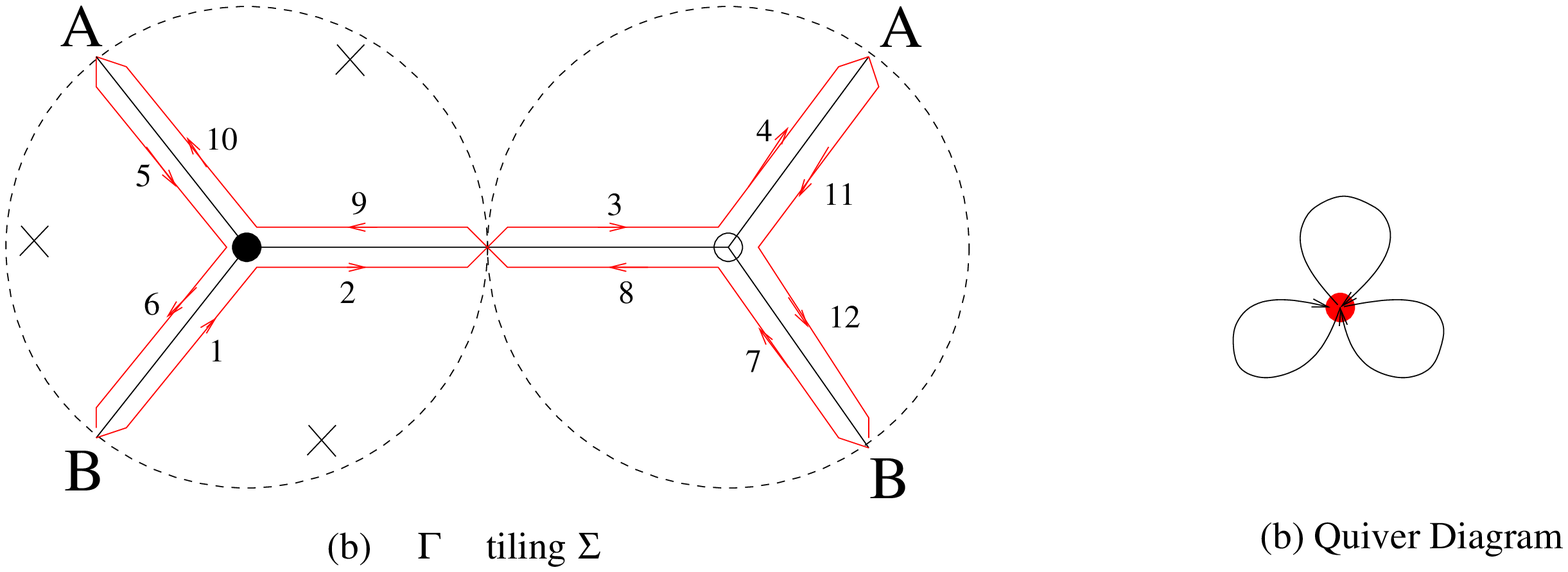,width=13cm}}
\caption{(a) The graph $\Gamma$ obtained by untwisting the dimer model
for $\IC^3$ (see \fref{f:C3dimer}).  The curve $\Sigma$ is obtained by
gluing the two discs back-to-back along their boundary circle so that
the points $A, B$ coincide.  The three punctures on the curve
correspond to the faces of the graph $\Gamma$.  The single face of the
dimer model maps to a single self-intersecting contour (path segments
numbered sequentially) that produces the correct quiver for the
world-volume theory (shown in (b)).  Moreover, the two trivalent
vertices of $\Gamma$ produce the cubic superpotential.}
\label{fig:c3-cycles}}

In \fref{fig:conifold-cycles} we study the conifold. For completeness,
we also included the toric diagram, the quiver (as well as the
periodic planar quiver) diagrams and the dimer models of the theory.
The Riemann surface $\Sigma$ is here again of genus 0, now with 4
punctures. We have labeled the 4 fields $p_{1,2}, q_{1,2}$ explicitly
in part (a). Upon graph-dualising to the dimer model in (b) the fields
become edges and we retain their naming while the $+$ (resp.~$-$) face
becomes the white (resp.~black) node.
\FIGURE[h]{\centerline{\epsfig{file=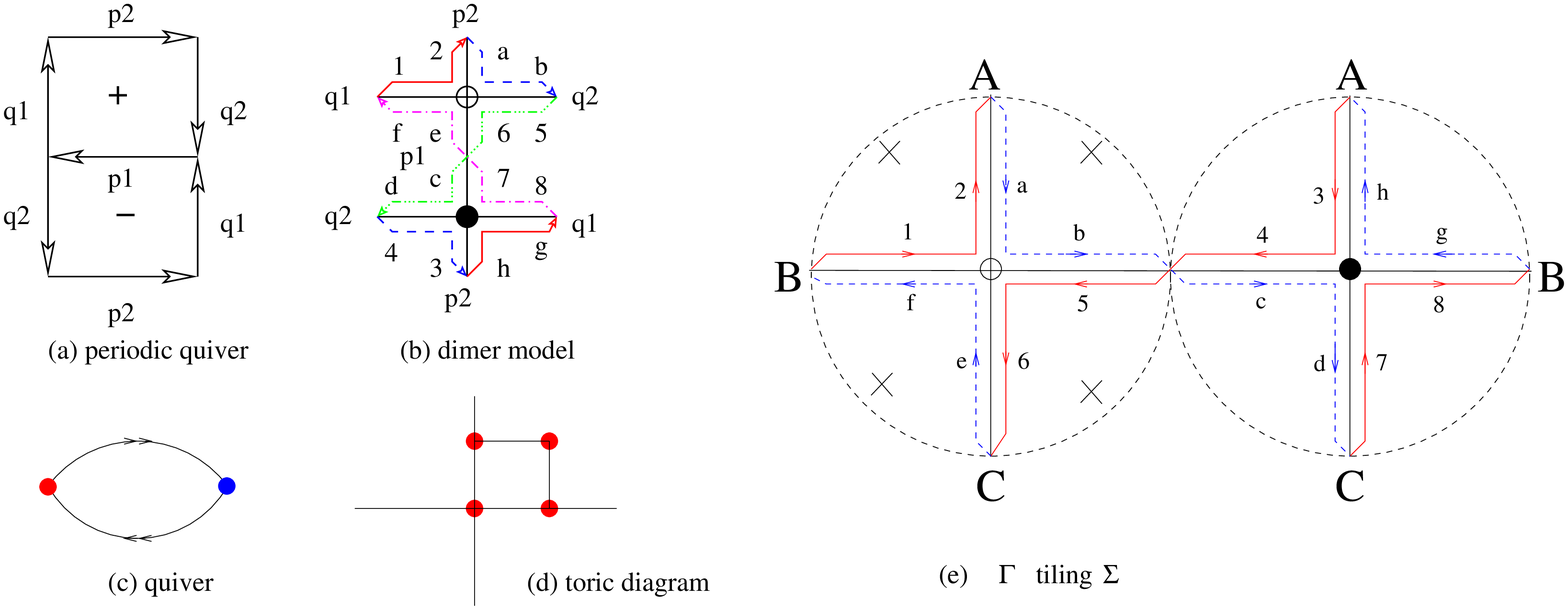,width=18cm}}
\caption{The graph $\Gamma$ obtained by untwisting the dimer model for
the conifold in shown in (e).  For reference, we also draw the
quivers, the dimer model and the toric diagram.
The two faces of the dimer model produce two
intersecting contours on $\Sigma$, whose intersections produce the
quiver diagram for this theory.  The two quartic vertices of $\Gamma$
give rise to the quartic superpotential of the quiver theory.}
\label{fig:conifold-cycles}}
%

%%%%----------
\subsection{The Gluing Locus}
We have seen how dimer models on $T^2$ can be untwisted to give dimer
models on $\Sigma$ (and vice versa). In light of the discussions in
\fref{fig:gluing} and \sref{sec:proposal}, we see that all this takes
place as we glue $\Sigma$ from its semi-infinite cylinders
(punctures). The untwisting procedure thus
furthermore provides a
decomposition of $\Sigma$ into half-infinite cylinders. We show this in
\fref{fig:gluing-graph} for our familiar $\IC^3$ example.
\FIGURE[h]{\centerline{\epsfig{file=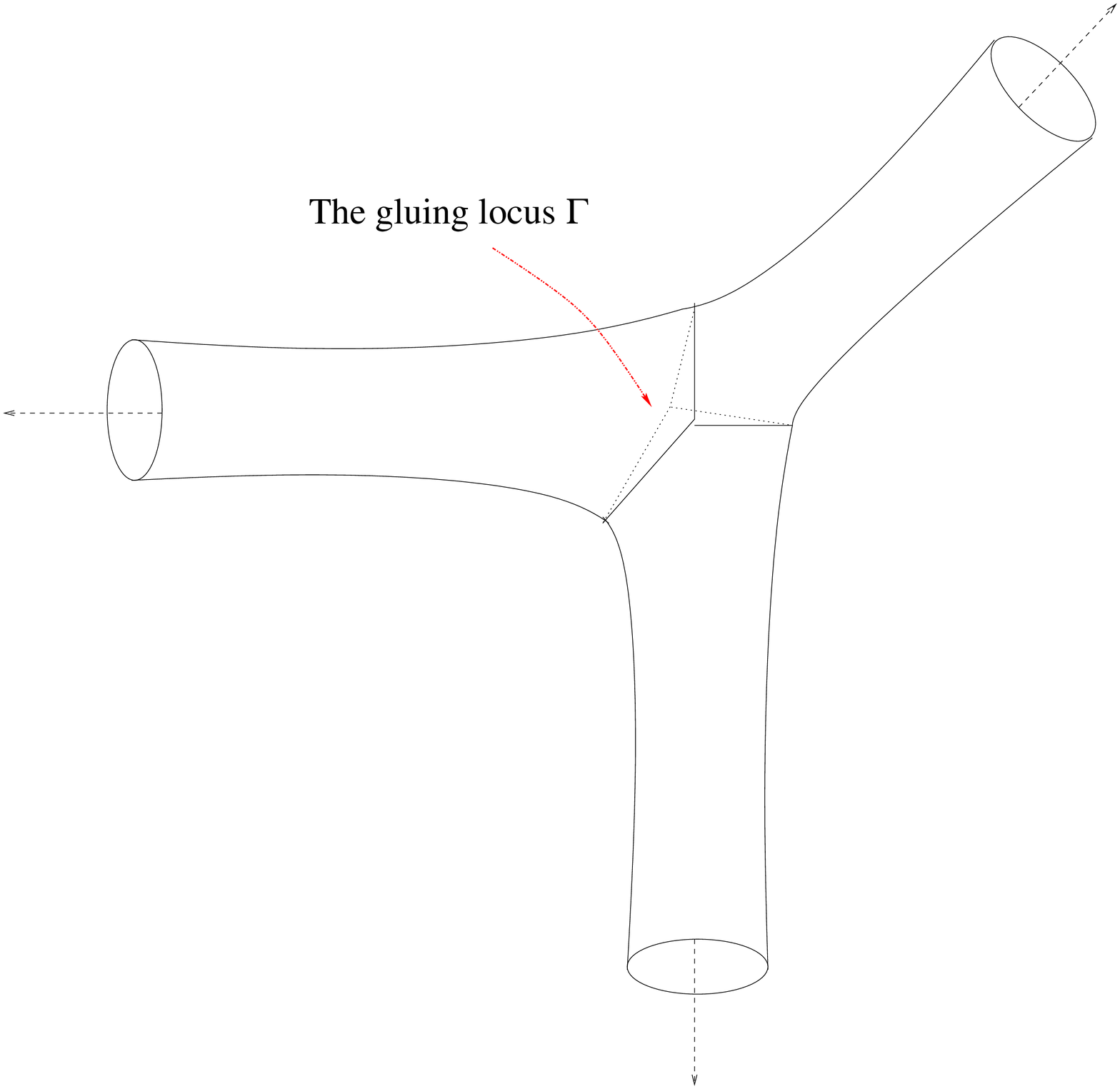,width=5cm}}
\caption{Gluing together three half-infinite cylinders along their
  $S^1$ boundaries to form the curve $1+w+z=0$, which is part of the
  mirror to $\IC^3$.  The gluing locus is as shown in \fref{fig:c3-cycles}.} 
\label{fig:gluing-graph}}

What about the converse operation to this decomposition?  As discussed
in the previous subsection, for each puncture on the curve, we obtain
a half-infinite cylinder.  The curve $P(z,w)=0$ may be recovered by
gluing together the set of deformed cylinders along their $S^1$
boundaries.  The gluing locus corresponds to a graph $\Gamma$
inscribed on the curve $\Sigma$, as shown in \fref{fig:gluing-graph}.
Said differently, we take contours that encircle the punctures of
$\Sigma$ and evolve them continuously into the interior until they
meet.  When they join up completely they do so along the graph $\Gamma$.

The various Seiberg-dual phases of the quiver theory are among the
possible graphs obtained by such gluing.  However, not every such
$\Gamma$ describes a consistent dimer model ($\Gamma$ may not even be
bipartite, although it may always be possible to obtain it as the
limit of a bipartite graph when some edges shrink to zero).  Even for
bipartite graphs, the zig-zag paths on this graph may not produce a
consistent quiver theory (similarly, neither does every dimer model on
$T^2$).  We leave this problem for future study.

%%%%%++++++++++++++++++++++++++++++++++++++++++++++++++
\subsection{Dimer Models from Mirror Symmetry}
\label{sec:dimer-mirror}

We are now able to combine the discussion of the previous sections to
show how the intersecting D6-branes are equivalent to the dimer models
on $T^2$.  As discussed in \sref{s:branes}, each D6-brane is
associated to a disc that is stretched from the vanishing cycle in the
fibre above a critical point in the $W$-plane, to the fibre above
$W=0$, i.e.~the boundary of the disc is attached along an $S^1$ in the
curve $\Sigma$, see \fref{fig:gluing}.

In the previous section we showed that the various $S^1$'s form zig-zag
paths on a certain graph $\Gamma$ inscribed in this curve.  $\Gamma$
defines a tiling of $\Sigma$.  The faces of this tiling correspond to
the punctures on $\Sigma$ (i.e.~the half-cylinders), and the winding
cycles (zig-zag paths) on this curve are the D6-branes.

The intersection of these D6-branes with each other produces the
matter content of the quiver (including any non-chiral matter):
since the zig-zag paths cross along every edge, in string theory we
obtain a massless chiral multiplet localized at the intersection
point, coming from the massless stretched string with consistent
orientation \cite{Berkooz:1996km}.

Furthermore, the vertices of $\Gamma$, where multiple zig-zag paths
form a closed loop around the vertex, give rise to the superpotential
terms.  These are computed by worldsheet disc instantons ending on
this loop.

The discs associated to the D6-branes intersect to form a $T^2$
embedded in the mirror geometry.  To visualize the $T^2$ one may use
the twisting operation described in the previous section.  After
twisting the graph $\Gamma$, we have a tiling of $T^2$, the dimer
model.  The $S^1$ winding paths on $\Gamma$ map to the boundary of the
faces of the tiling, and the discs attached along these $S^1$ map to
the interior of the faces.  The vertices of $\Gamma$ {\it remain}
vertices of the dimer model on $T^2$.

This explains the origin of the dimer models on $T^2$ and their
relevance for describing the physics of the quiver gauge theory.  It
also explains the physical relevance of the observation
\eref{totfield} about the number of fields of the quiver being counted
by the intersection of $(p,q)$ winding cycles on a $T^2$, since we
obtain such cycles by twisting the D6-branes from $\Sigma$ to $T^2$.

In general, the $T^2$ of the dimer model together with the
intersection along the curve $\Sigma$ can only be embedded in four
dimensions, since in 3 dimensions the discs attached along the $S^1$'s
would have to pass through one another.  This is indeed the case here,
since the curve $\Sigma$ is defined in $(\IC^*)^2$.  Furthermore, we
claim that the twisting operation discussed in the previous section
may be performed {\it continuously} when the system is embedded in the
mirror geometry.  This identifies the $T^2$ of the dimer model with a
$T^2 \subset T^3$ of the world-volume of the D6-branes.
The remaining $S^1$ fibre is given by phase rotations in the $uv$
plane, which is finite over the interior of the faces (away from
$W=0$), and vanishes along the graph of the dimer model (above $W=0$).
This gives a singular $T^3$, which is identified with the mirror to
the D3-brane at the singular point.  Thus, we have obtained a concrete
construction of the dimer models using mirror symmetry.

We can now clarify the status of the NS5-D5 ``brane tilings'' proposed
in \cite{Franco:2005rj} (see also \cite{Aganagic:1999fe}), at least
heuristically.  If we T-dualize along the $S^1$ fibre in the $uv$
plane, we should obtain an NS5-brane at the point where the fibre
becomes singular, i.e.~above $W=0$.  Thus, the NS5-brane will wrap the
curve $\Sigma$.  Since the D6-branes were extended along this fibre
direction, they become D5-branes with topology of a disc, with the
boundaries of these discs intersecting along the $NS5$-brane.
Together these D5-branes form a $T^2$ embedded in the geometry, and
the graph of the dimer model is again the locus where the $T^2$
intersects the curve wrapped by the NS5-brane.

Therefore, one may indeed hope to obtain the dimer model from an
intersecting NS5-D5 system, although it is difficult to be clear about
the geometry of this system since it is not prescribed directly by
local mirror symmetry.  In particular it clarifies the relation of the
$T^2$ of the dimer model to the geometry of the NS5-brane, which turns
out to have been not quite correctly specified in previous literature.

%%%%%%%%%%%%%%%%%%%%%%%%%%%%%%%%%%%%%%%%
\subsection{Seiberg duality}
\label{sec:sd}
%%%%%%%%%%%%%%%%%%%%%%%%%%%%%%%%%%%%%%%%
Now we discuss how to understand Seiberg Duality in our
picture. Again we illustrate with the above $F_0$ example.
In \fref{f:F0seiberg}, we have redrawn the
two phases presented in \fref{fig:zigzag-f0-2}, but now
as a dimer model in $\Sigma$.
Here, we have used
$L_i$ to denote the boundary to be glued together by our lifting
procedure. The first thing we need to notice is that while the
gluing of the left figure is a straight-forward rectangular one, the
gluing of the right figure has some shifting. That is, these
two figures have different complex moduli-parameters $\tau$ of the
resulting $\Sigma$, which here happens to be a (punctured) torus.
\FIGURE[h]{\epsfxsize = 4in \epsfbox{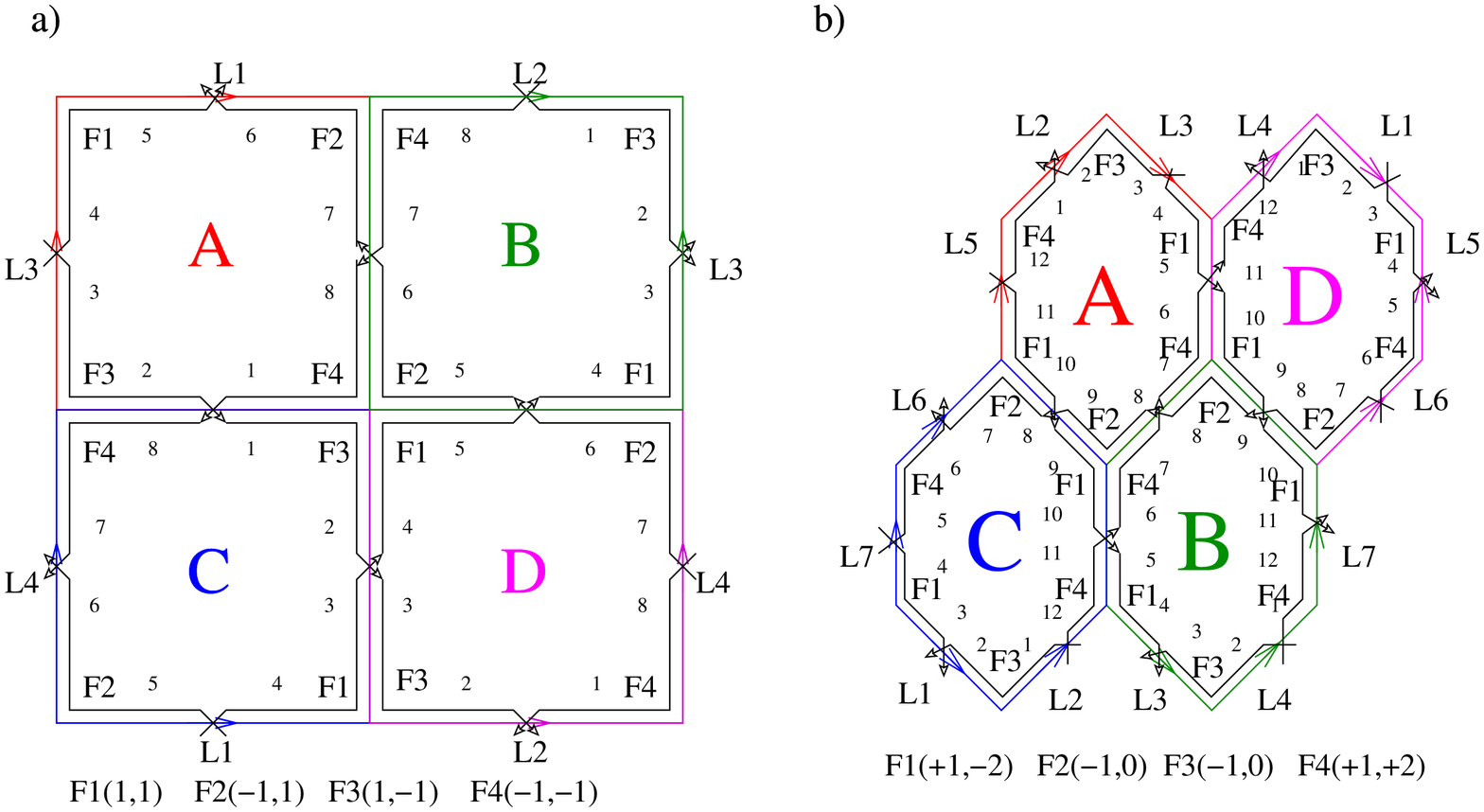}
\caption{The two Seiberg dual phases of $F_0$ from
  \eref{fig:zigzag-f0-2}, here drawn as dimer model in $\Sigma$ using
  the twisting procedure. We have labeled the segments and direction
  of the zig-zag paths explicitly.}
\label{f:F0seiberg}}

The second thing we need to notice is that,
as we have discussed before, the boundary of faces in the dimer model
lifts to nontrivial cycles in $\Gamma$. Here we have redrawn these
liftings carefully and recorded their non-trivial homology classes in
$\Gamma$.
We have written down these $(p,q)$-cycles in $\Sigma$ in the Figure. 
They are
\beq\label{pqF0seiberg}
\begin{array}{ccc}
(a)~(F_0)_{II} & ~~~~& F1(1,1),~~~F2(-1,1),~~~F3(1,-1),~~~F4(-1,-1) \\
(b)~(F_0)_{I}& ~~~~ & F1(1,2),~~~F2(-1,0),~~~F3(-1,0),~~~F4(1,-2) \ .
\end{array}
\eeq
It is easy to
check that the intersection number given by $\chi(i,j)={\rm det}
\left( \begin{array}{cc} p_i & q_i \\ p_j & q_j
\end{array}\right)$, which we recall from \eref{pqdet},
gives the right matter contents for both phases.

In fact these two different sets of cycles have been observed in
\cite{Franco:2002ae} on the discussion of the relation between the
$(p,q)$-web and quiver theory. Such a relation has been clarified
further in \cite{Feng:2004uq} where two different terminologies, viz.,
the toric $(p,q)$-web and the quiver $(p,q)$-web, have been
distinguished.  Now, from our construction, it is clear that the toric
$(p,q)$-web corresponds to zig-zag paths on $T^2$ while the quiver
$(p,q)$-web corresponds to cycles in $\Sigma$. The latter, of course,
only makes sense if $\Sigma$ were a torus as well, which only happens
if there is one internal point in the toric diagram.  Indeed, as we
have emphasized, in general $\Sigma$ can be of arbitrary genus. In
this case, a single pair of $(p,q)$-charges (corresponding to the
quiver $(p,q)$-web) no longer makes sense. However, it still has
meaning in the dimer model on $T^2$.

Now, we would like to show that these two sets of cycles in
\eref{pqF0seiberg} are related to
each other by Picard-Lefschetz (PL) transformations, which we recall from
\cite{Hanany:2001py} and \cite{Feng:2002kk}.
Let us start from
the set
\beq 
F3(1,-1)~~~~~ F1(1,1) ~~~~~ F2(-1,1)~~~~~F4(-1,-1) \ ,
\eeq
where we have reordered these cycles cyclically
according the rules in \cite{Feng:2002kk}. Now we move $F1$ to the
right of $F2$. The new cycle of $F2$ is given by
$(-1,1)+ \chi(1,2) (1,1)=(1,3)$. However, since the new
$n_{F1}=1-\chi(1,2)=-1$ we need to add extra sign for the charge
of $F1$. In other words, we have new cycles:
\beq\label{F0-PL}
F3(1,-1)~~~~~ \widetilde{F2}(1,3)
~~~~~\widetilde{F1}(-1,-1)~~~~~F4(-1,-1) \ .
\eeq
It is then easy to check that \eref{F0-PL}, when acted upon by
an $SL(2,Z)$ transformation  $\left( \begin{array}{cc} 1 &
0 \\ -1 & 1 \end{array}\right)$, gives us:
\beq
F3(1,-2)~~~~~\widetilde{F2}(1,2)
~~~~~\widetilde{F1}(-1,0)~~~~~F4(-1,0) \ .
\eeq
This we instantly recognise to be the set (b) (up to cyclic
permutation).
We conclude, that the sets (a) and (b), obtained from our two
different gluings, indeed are Picard-Lefschetz dual, and, hence,
Seiberg dual, to each other. It is for this reason that we have
judiciously labeled in \eref{pqF0seiberg}, the two phases as
$(F_0)_I$ and $(F_0)_{II}$, in the convention of the literature
(cf.~e.g.,~\cite{Feng:2002zw}).

%----SUMMARY
\subsection{Summary of the Various Correspondences}

We have introduced many concepts in this section, so before proceeding
to the next section where we will see how in some cases one may
concretely realise the $T^2$ in the geometry, we find it expedient to
summarise some key results discussed above by itemising the
correspondences amongst the various objects:

\begin{itemize}
\item
The $(p,q)$-web is the graph dual of the toric diagram $D$ of $\cM$,
while the dimer model on $T^2$ is the graph dual of the quiver
diagram, when drawn as a planar graph on $T^2$.
%In the case of $D$
%having one internal point, the antisymmetric part of the quiver can be
%obtained from the $(p,q)$-web by the intersection formula
%\eref{pqdet}.

\item
The mirror geometry $\cW$ of $\cM$
is given by a double fibration over a $W$-plane, consisting
of a $\IC^*$ fibre ($uv=W$) and a (punctured)
Riemann surface $\Sigma$ defined by $P(z,w)=W$. The expression
$P(z,w)$ is the Newton
polynomial $P(z,w)$ of $D$ (cf.~\fref{f:W} and eq.~\eref{Wuv}).
The genus of $\Sigma$ is
equal to the number of internal points of $D$. Moreover, $\Sigma$ is a
thickening of the $(p,q)$-web, while its punctures, which tend to
cylinders at infinity, are aligned with the $(p,q)$-legs
(cf.~\fref{f:F0pq}).

\item
The $(p,q)$ winding cycles on a $T^2$ can be deformed into zig-zag
paths on the dimer model on $T^2$, or, dualistically, to a bounding of
the planar quiver (cf~\fref{f:pq-deform}). The direction (clock-wise
or counter-clockwise) of the loops formed by the zig-zag paths around
each vertex of the dimer model gives the bipartite nature of the
dimer model (cf.~\fref{f:doubleline}).
\item
The dimer model on $T^2$ can be mapped, using an untwisting procedure,
to an isomorphic bi-partite graph $\Gamma$ on the curve $\Sigma$,
which is part of the mirror geometry. In particular, a zig-zag path
with $(p,q)$ winding maps to an $S^1$ that winds around the puncture
along the $(p,q)$ direction in $\Sigma$ (cf.~Figures
\ref{fig:c3-cycles}, \ref{fig:conifold-cycles}).  Conversely, we can
glue cylinders to form $\Sigma$; then, the $S^1$ winding the punctures
(cylinders) join along $\Gamma$ (cf.~\fref{fig:gluing-graph}).

\item
Dualistically, a closed loop formed by segments of different zig-zag
paths, whereby bounding a face in the dimer model
(cf.~\fref{fig:zigzag-f0-2}), lifts to a winding path around a
non-trivial homology cycle in $\Sigma$ and forms a zig-zag path on
$\Gamma$.

\item
The zig-zag paths on $\Gamma$ are the intersection of the D6-branes
with $\Sigma$; their intersection with one another on $\Sigma$
dictates the quiver theory.  Thus, $\#$(faces of dimer model)
$=\#$(gauge groups in quiver) $=\#$(critical points of $P(z,w)$)
$=\#$(non-trivial 3-cycles in $\cW$) = 2 Area($D$).  Vertices of
$\Gamma$ give superpotential terms via string worldsheet disc
instantons bounded by the parts of the D6-branes that combine to form
a loop around this vertex.

\item
These intersecting D6-branes span a singular $T^3$ according to mirror
symmetry, and there is a distinguished $T^2$, which is the complement
of the $S^1$ (defined by $uv=0$) that vanishes at $P(z,w)=0$.
This is the $T^2$ on which the dimer
model is defined.  The $S^1$ fibre vanishes along the graph of the
dimer model, which is where the D6-branes (and the $T^2$) intersects
$\Sigma$ defined by the curve $P(z,w)=0$.  
The twisting map is a convenient way to map the
$T^2$ spanned by the D6-branes in the mirror geometry to an abstract
$T^2$, in order to visualize this dimer model. Below, we will see
another concrete way to visualise this $T^2$.
\end{itemize}

%%%%%***************
%%%  AMOEBA & ALGA
%%%%%****************************************

\section{Amoeb\ae\ and Alg\ae}
\label{s:amoeba}
In the previous sections we discussed a topological map that allowed us
to go between the dimer models on $T^2$ and the intersection locus of
the D6-branes on the curve $\Sigma \subset (\IC^*)^2$.  We now discuss
how one may also obtain this $T^2$ from a certain projection of the
geometry.  This is the concrete realisation mentioned in
\sref{sec:proposal}.
This projection is a counterpart of the so-called amoeba projection
used in algebraic geometry, which we now review.

%%-------------- amoebae
\subsection{The Amoeba map}

Let us parametrize the co\"ordinates $(z,w) \in (\IC^*)^2$ as
\begin{equation}
(z,w) = (\exp(s + \imath \theta),\exp(t + \imath \phi))
\end{equation}
where $s,t \in \IR$ while $\theta, \phi \in [0, 2\pi)$. Thus, the
space $(\IC^*)^2$ has topology $\IR^2
\times T^2$.  It is difficult to visualize this four-dimensional space
directly, however two projections will be of vital importance.
The first is the projection of the curve onto $\IR^2$:
\begin{equation}\label{amoebapr}
(\exp(s + \imath \theta),\exp(t + \imath \phi)) \mapsto (s, t)
\end{equation}
This projection is known in the literature as the {\bf amoeba} of
the curve, because the resulting shape of the projected curve is
reminiscent of the mischievous microbe (see \fref{f:F0eg} (a)).
\FIGURE[h]{
\begin{tabular}{ll}
$\ba{c} \epsfxsize = 4cm \epsfbox{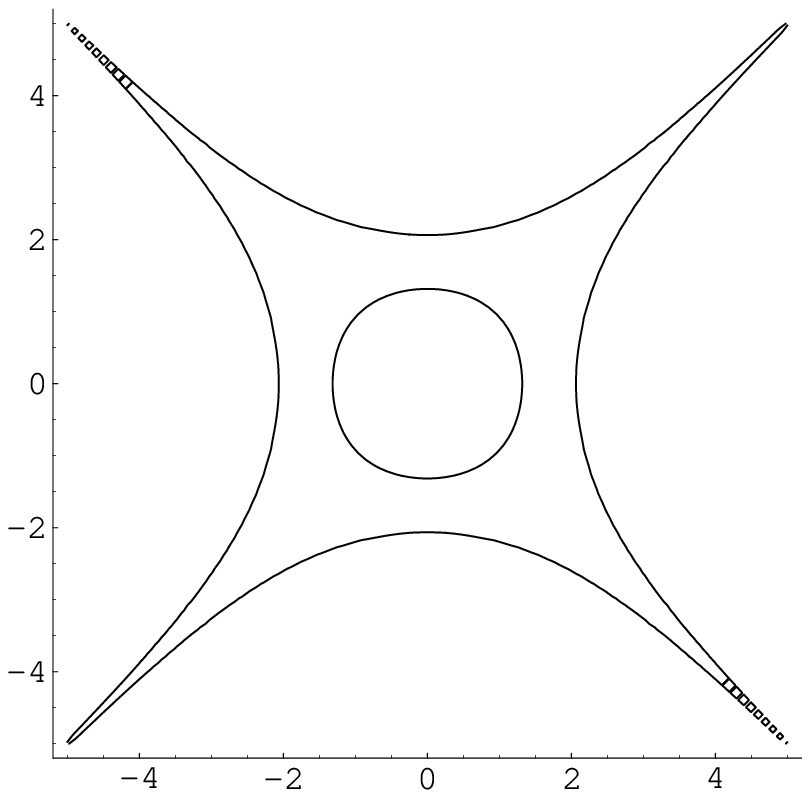} \ea$
&
$\ba{c} \epsfxsize = 6cm \epsfbox{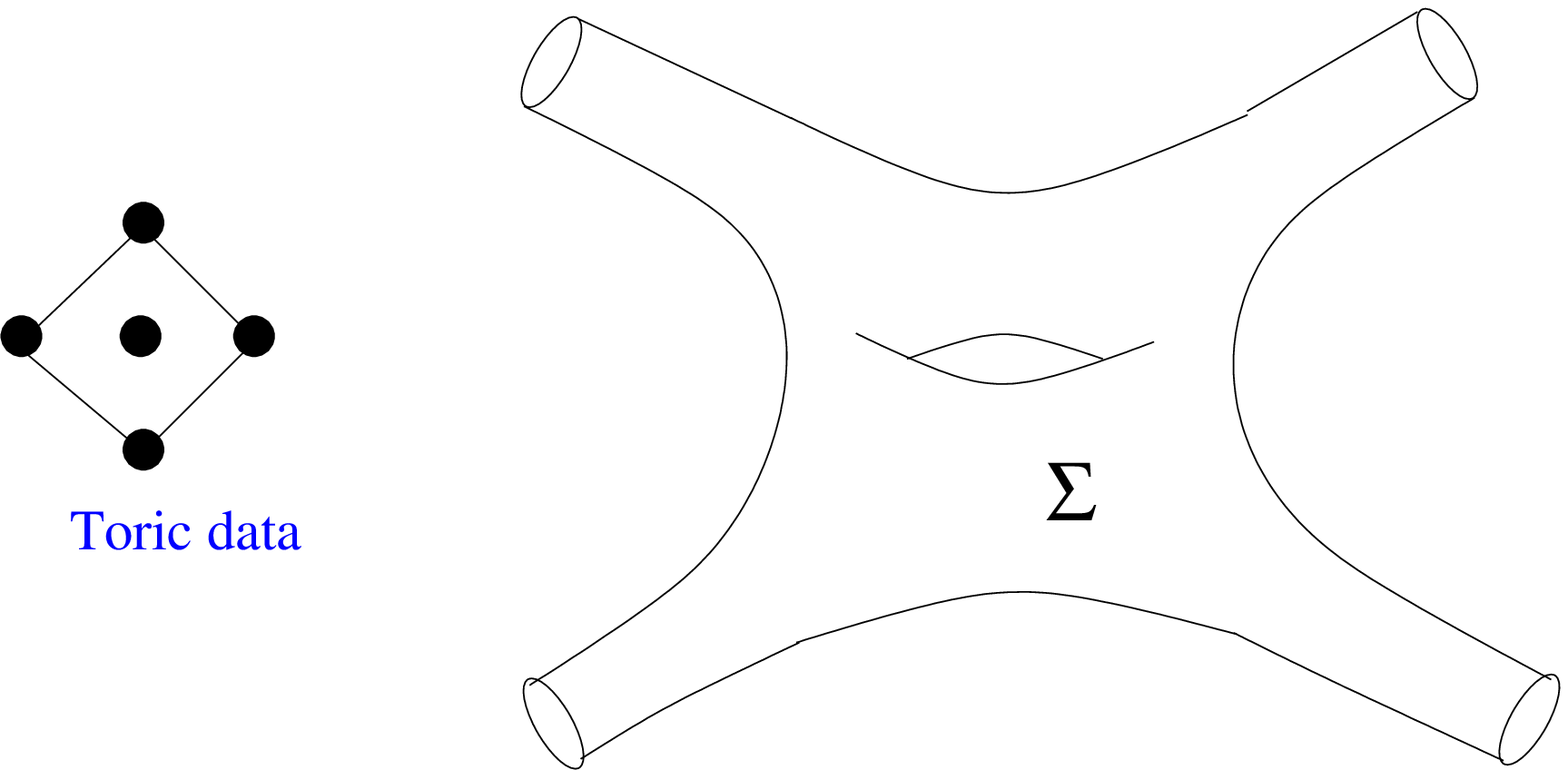} \ea$ \\
(a) Boundary of Amoeba & (b) Toric Diagram and Riemann
    Surface
\end{tabular}
\caption{The amoeba of the Riemann surface $P(z,w) = -w-z-1/w-1/z-6$, which
  corresponds to the zeroth Hirzebruch surface introduced in
  \fref{f:F0pq} in drawn in (a) and the boundary, in (b).
  The amoeba has tentacles which asymptote to lines which are called
  spines, which is the dual graph to the toric diagram, i.e.,
  the $(p,q)$-web. For reference we include the toric diagram and the
  Riemann surface $\Sigma$ in (c).}
\label{f:F0eg}
}

Let us discuss the projection \eref{amoebapr} in detail first. The
reader is referred to
\cite{mikhalkin-2000-2,mikhalkin1,mikhalkin2,rullgard,Kenyon:2003uj}.
The formal definition applies to any variety. Let $V \subset
(\IC^*)^n$ be an algebraic variety and let
\beq\label{Logmap}
Log : (\IC^*)^n \to \IR^n \mbox{ be the map } Log(z_1, \ldots,
z_n) \to (\log|z_1|, \ldots, \log|z_n|)
\eeq
for co\"ordinates $z_{1,\ldots,n}$ of $(\IC^*)^n$. Then the amoeba
of $V$ is a real algebraic set
\beq
Amoeba(V) := A = Log(V) \subset \IR^n.
\eeq

Our focus is on the curve $P_W(z,w) = 0$ and the amoebae will be
regions in the plane $\IR^2$. The map \eref{Logmap} is, of course,
the same as the projection \eref{amoebapr} stated above.

%%--------------------
\subsubsection{Amoeb\ae\ and the $(p,q)$-Web}
\label{s:amoebapq}
One of the first attractive qualities of the amoeba, as one can see
from \fref{f:F0eg}(a), is its {\bf tentacles} which extend
exponentially to infinity, asymptoting to a straight line. Such a line
is called a {\bf spine} of the amoeba.  We can determine the
directions of these spines readily and will show the important fact
that they are simply given by orthogonal directions of the external
edges in the toric diagram.  That is, the tentacles line up with the
dual graph to the toric diagram.

Using an $SL(2,\IZ)$
transformation plus a shift we can always rotate the toric diagram
so that the Newton polynomial is given by 
\[ 
P(z,w)= \sum_{i=0}^n c_i z^i + w G(z,w) 
\] 
with $G(z,w)$ a polynomial of $z,w$ with
only non-negative powers. The number $(n+1)$ is the total number of
lattice points along the external line that is now aligned with the
$z$-axis. The limit $w\to 0$, i.e., as $\log|w| \to -\infty$, we have
the tentacles tending to negative vertical direction.

Rewrite $P(z,w)=0$ as
\[
\sum_{i=0}^n c_i z^i = c_n \prod_{i=1}^n (z- z_i) =- w G(z,w) \ ,
\] 
where we have factorised the polynomial
in $z$ into its roots $z_i$. Therefore, as $w \to 0$, we have that
$\prod_{i=1}^n (z- z_i) \to 0$ and we get $n$ asymptotes along the
vertical negative (i.e., $\log|w|$) direction located at positions
$\log|z_i|$. In other words, we have that the asymptotic
spines are orthogonal directions to (i.e., dual
graph of) the toric diagram. Recalling from the above discussions in 
\sref{s:pqweb}, that this dual
graph is precisely the $(p,q)$-web, we conclude and
summarise\footnote{In the mathematics literature
  (cf.~e.g.,\cite{rullgard}), the spine is considered a deformation
  retract of the amoeba and one usually shows this fact using the
  points of non-differentiability of a so-called Ronkin
  function.
}:
\begin{quote}
{\it The spine of the amoeba of $P(z,w)$ is the $(p,q)$-web
associated to the toric diagram which is the Newton polygon of
$P(z,w)$.} Conversely, the amoeba is a thickening of the
$(p,q)$-web.
\end{quote}

The origin of these deformations from the $(p,q)$-web to the curve
$\Sigma$ are the torus-invariant world-sheet instantons, which
localize near the trivalent vertices of the $(p,q)$ web
($T^3$-invariant points) prior to performing mirror symmetry (recall
from \sref{s:pqweb} that the web describes a $T^2 \times \IR$ special
Lagrangian fibration of the geometry).  Mirror symmetry amounts to
summing these instanton corrections, and produces the effective
geometry described by the local mirror.

The above observation gives another interesting result. 
Notice that for
general moduli $c_i$, the tentacles are parallel along
different locations $\log|z_i|$. However, if some of $\log|z_i|$ are
the same, then the corresponding tentacles will merge. 
The most degenerate case is if all the $\log|z_i|$ are same. 
In this case, we have one and only one tentacle. 
Furthermore, if we assume that all $z_i$ are real and
positive, as was needed in \cite{Hanany:2005ve} to compare with the
linear-sigma model, we immediately have that
$$
(z-a)^n= a^n (z'-1)^n= a^n \sum_{k=0}^n
(-1)^k {n!\over k! (n-k)!} z^{'n} \ ,
$$
where $z' = z/a$.
We  see that the
coefficients ${n!\over k! (n-k)!}$ are nothing but the linear sigma
model field multiplicities conjectured in \cite{Feng:2002zw}, for the case of colinear points on the boundary of the toric diagram.

%%--------------------
\subsubsection{Amoeb\ae\ and Dimer Models}
\label{sec:harnack}

Not only are amoeb\ae\ related to $(p,q)$-webs, they have appeared in
relation to dimer models in previous work 
\cite{Kenyon:2003uj,Kenyon:2002a}. 
For our purposes one especially pertinent issue
is that of {\em Harnack curves}.

Harnack curves date back to Hilbert's 16th Problem concerning the
possible topologies of degree $d$ real algebraic curves $C$ in
$\IR\IP^2$. Harnack proved \cite{Harnack}
that the number of components of $C$ can not exceed
$\frac{(d-1)(d-2)}{2}+1$ and when this maximum is attained, $C$ is
now known as a {\bf Harnack curve}. These are in some sense the
``best'' or most generic real curves. Recalling the definition of
the spectral curve an edge weights of a dimer model, a main result
of \cite{Kenyon:2003uj} is that
\begin{quote}
{\em For any choice of real edge weights subject to a certain sign
  constraint, the spectral curve of a dimer model is a Harnack
  curve. } 
\end{quote}
Harnack curves have very nice amoeba projections, in particular,
their amoeb\ae\ can be analytically described by a single
inequality (with the boundary satisfying the equality):
\beq
P_W(z,w) \mbox{ Harnack } \Rightarrow Amoeba(P_W(z,w)) =
\left\{(x,y)\in \IR^2 :
\prod_{\alpha, \beta = \pm 1} P_W(\alpha e^x, \beta e^y) \le 0
\right\} \ .\label{eq:harnackbdy}
\eeq

In this paper, we shall however relax the constraint of reality on
the coefficients in $P_W(z,w)$ and hence will not deal much with
Harnack curves, or, for that matter, real curves. This means that
in the dimer models we relax the constraint on the signs and
reality of edge weights. This turn out to be necessary for the
dimer model to explore the full moduli space of the Calabi-Yau
geometry (cf.~\cite{CYmoduli}). 
The price is that even though the Kasteleyn determinant is
still a sum over perfect matchings of the graph, they are no
longer counted with uniform sign. As we learned from
\cite{Hanany:2005ve}, the Harnack nature of the spectral curve was
good for combinatorics and unveiled the nature of GLSM field
multiplicities, but for now this is not relevant for us. We can
still recover the special Harnack case if we choose edge weights
to satisfy appropriate constraints.

%------------------------
%---- ALGA
\subsection{The Alga Map}
The map \eref{amoebapr}, which
leads to the amoeb\ae\ and has been widely studied by
mathematicians,  is but one of 2 natural projections.
Whereas \eref{amoebapr} projected onto the real parts of $(z,w)
\in (\IC^*)^2$, we now project to the imaginary (angular) part:
\begin{equation}
(\exp(s + \imath \theta),\exp(t + \imath \phi)) \mapsto (\theta,
\phi)
  \ .
\end{equation}
This is a projection onto a $T^2$ component of $(\IC^*)^2$ and by
projecting every point on the curve $P_W(z,w)$ to its angular
component, we obtain a doubly periodic image of the curve. In keeping
with the microbial naming convention, we call these doubly periodic
projections {\bf algae} of the curves, after the microscopic plant
species which like to tile surfaces in bodies of water. We illustrate
this with an example in \fref{f:F0alga}.
\FIGURE[h]{
\begin{tabular}{ll}
$\ba{c}\epsfxsize = 6cm \epsfbox{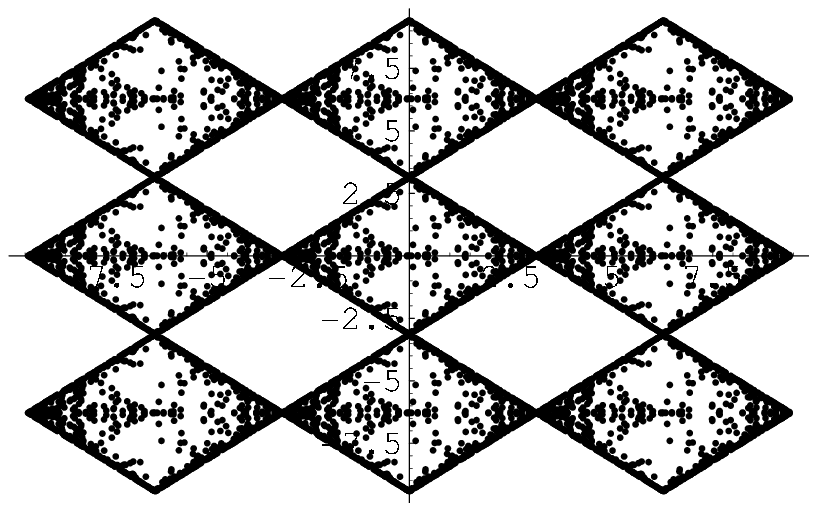} \ea$
&
$\ba{c}\epsfxsize = 7cm \epsfbox{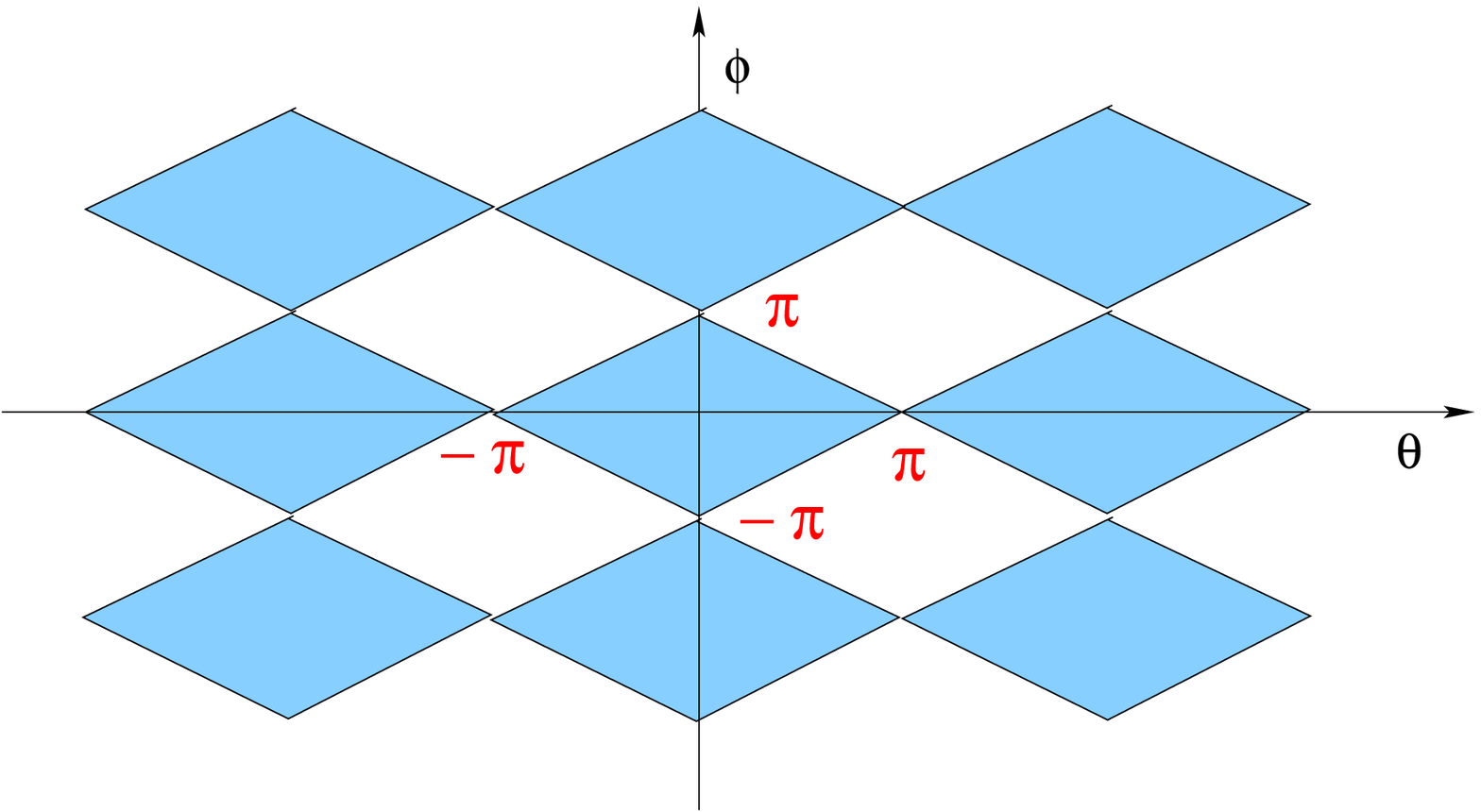} \ea$ \\
(a) & (b)
\end{tabular}
\caption{
(a) The alga of $F_0$ (cf.~\fref{f:F0eg}) draw using
  Monte-Carlo and redrawn in (b)
%The alga of $F_0$ (cf.~\fref{f:F0eg}), redrawn here from a Monte-Carlo
%  algorithm in order
  to show the boundaries more clearly. We have indicated $3 \times 3$ of
  the fundamental region to show the periodicity.
  Not all alg\ae~are as regular
  as this.}
\label{f:F0alga}
}
We will show below that the $T^2$ of the dimer models can be
identified with the above $T^2$ inhabited by the algae, at least in
certain cases.

As far as we are aware, the properties of this projection have not
been studied by mathematicians.  For the amoeba map, the boundary may
be parametrized explicitly when the curve is Harnack (see
\eref{eq:harnackbdy}).  In this case, the curve admits the
antiholomorphic involution $(z, w) \mapsto (\overline z, \overline
w)$.  This fixes the boundary, and shows that the curve is 2-1 over
the interior of the amoeba, with those points related by complex
conjugation.

Therefore, under the alga projection the boundary of a Harnack curve
maps to the points $(0, 0), (0, \pi), (\pi, 0), (\pi, \pi)$ since it
is real.  The rest of the alga is symmetric under the map $(\theta,
\phi) \mapsto (-\theta, -\phi)$ which descends from the involution on
$\Sigma$.  Beyond this, we do not know how to explicitly parametrize
the alga projection, let alone in the non-Harnack case.  In practice,
we therefore resort to Monte-Carlo simulation to plot its interior.
Furthermore, as we discuss later, to get a non-degenerate alga
projection of the dimer models we really need generic complex moduli
in the curve $P(z,w)$.

%%
%+++++++++++++++++++++++++++++++++++++++++++++++++++++++
%
\section{Dimer models from Alg\ae}
\label{sec:alga}
In this section we describe how the graph of the dimer models may be
obtained by the alga projection of the intersecting D6-brane system
discussed in \sref{sec:dimer-mirror}.

Recall that the dimer models on $T^2$ may be obtained by ``twisting''
the intersection locus of the D6-branes with the curve $P(z,w)=0$.
Key to this twisting procedure was the fact that the graph $\Gamma$ on
which the D6-branes are zig-zag paths admits a decomposition into
contours encircling the $(p,q)$ spines of the curve $\Sigma$.  After
the twisting map these became the $(p,q)$ winding cycles of the torus.

We will now show that the same is true when we project onto the $T^2$
defined by the alga projection, and furthermore that in certain
situations the $T^2$ obtained by twisting may be identified with the
projection onto the angular variables.

\subsection{The $(p,q)$ winding cycles}
\label{s:pq}

Recall from our above discussions in
\sref{s:amoebapq} that the spine of the amoeba aligns with the
$(p,q)$-web. In particular, consider the simple case where
we have only 2 lattice points for a given external edge in the toric
diagram whose normal direction is $(p,q)$. 
Then we can normalize $P$ to the form

\begin{equation}
P(z,w) = c_1 + c_2 z^{-q} w^p + \sum_i c_i z^{-q_i} w^{p_i}
\end{equation}
Rescaling $z \mapsto \lambda^p z$, $w \mapsto \lambda^q w$, the curve becomes
\begin{equation}
P(z,w) = \frac{c_1}{c_2} + z^{-q} w^p + \sum_i \frac{c_i}{c_2} \lambda^{(p,q) \cdot (-q_i, p_i)} z^{-q_i} w^{p_i} = 0
\end{equation}
Since the toric diagram is convex, in the limit $\lambda \rightarrow \infty$ only the first two terms survive, and the curve becomes
\begin{equation}
P(z,w) = z^{-q} w^p + \frac{c_1}{c_2} = 0
\end{equation}
i.e.~in the neighbourhood of a puncture of $P(z,w)=0$, the curve approaches a flat cylinder 
\be\label{pqK}
z^{-q} w^{p} = -\frac{c_1}{c_2}
\ee
The constant is given by the ratio of the coefficients of the two
vertices forming the edge of the Newton polygon that is orthogonal to
the $(p,q)$ spine.

This cylinder admits the $\C^*$ action
\begin{equation}
z \mapsto \lambda^p z,\quad w \mapsto \lambda^q w
\end{equation}
and so the cylinder is described by a trivial $S^1$ fibration over a
line.  Under the projection to the alga, this $S^1$ maps to a
straight-line with $(p,q)$ winding and with offset given by
$\arg(-c_1/c_2)$, i.e.~determined by two of the moduli of the curve. In other words, \eref{pqK} projects to
\beq  
q \theta-p \phi= \arg(-c_1/c_2)
\label{eq:straight}
\eeq 
which is a straight line with winding number $(p,q)$ in the $T^2$
defined by the alga projection.  On the other hand, the base of the
fibration is the spine and is a line aligned in the $(p,q)$ direction
in $\IR^2$ under the amoeba projection.  Thus, {\it just as the amoeba
can be viewed as a thickening of the $(p,q)$ web whose semi-infinite
legs (spines) are straight lines of slope $(p,q)$, so too can the alga
be viewed as a thickening of the straight line winding cycles with
slope (or winding number) $(p,q)$.}

The above observations are for $S^1$ cycles far away along the
spines. Now if we deform these contours, the projection to the alga
will also deform. Suppose that there exists a family of contours
encircling the punctures on $\Sigma$ such that under the projection to
$T^2$ the boundary of the $n$-valent polygons (i.e., polygons bounded
by consistent orientation) retracts onto $n$-valent vertices in the
interior, as shown in \fref{fig:dimerc3}.  Then one obtains a doubly
periodic bipartite graph in the $T^2$, which is a dimer model for the
quiver theory.  One recognises the retraction of the straight-line
paths as the zigzag paths introduced in \fref{f:doubleline}.

\FIGURE[h]{\epsfig{file=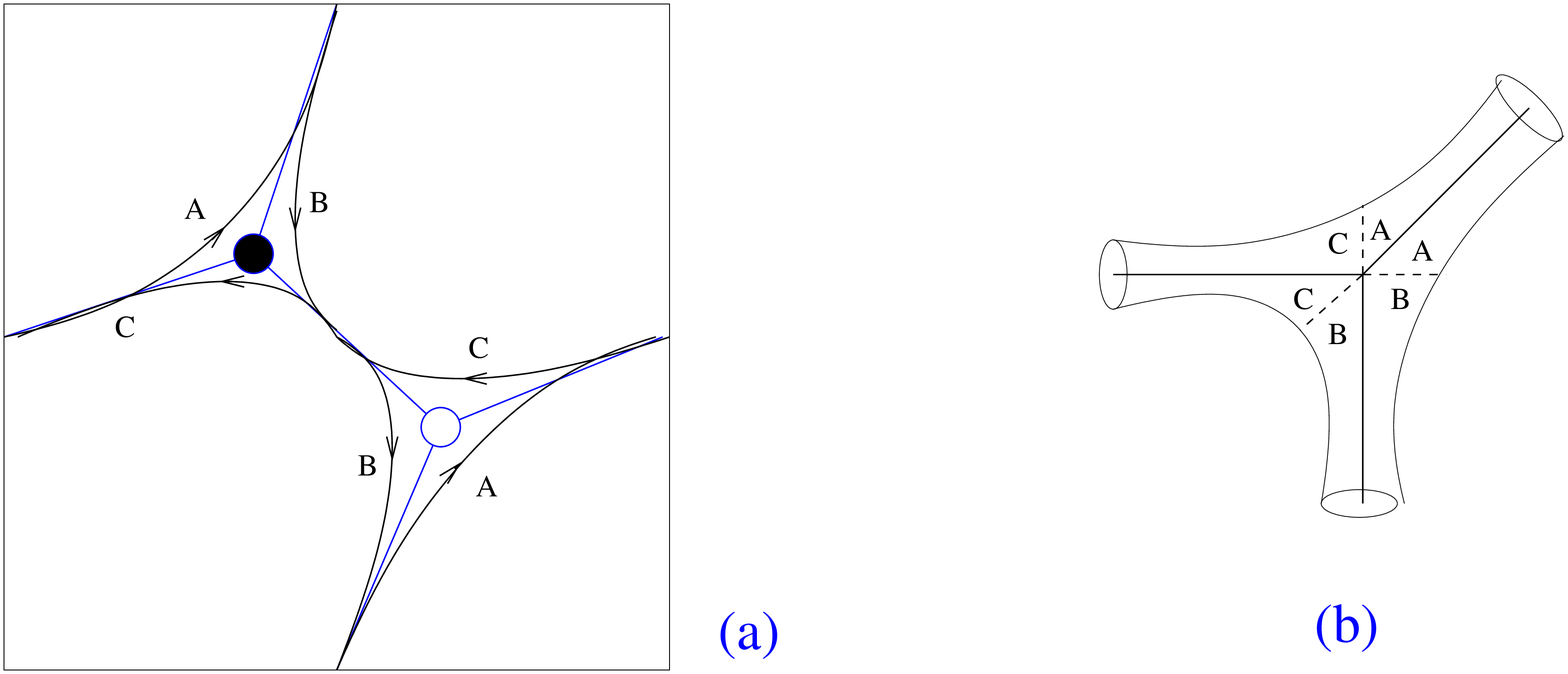,width=12cm}
\caption{(a) 
  The projection of the gluing locus for $\IC^3$ onto the alga.
  The 3 punctures on the curve define cylinders along the $(1,1)$,
  $(0,-1)$ and $(-1,0)$ directions which project to cycles with the
  same winding number, labeled by A, B and C.  They are shown as
  smooth curves, and in the limit where they glue together they meet
  along the locus in blue, which is the graph of the dimer model for
  $\IC^3$.  Note that the bipartite nature of the graph follows
  automatically, since neighbouring vertices are surrounded by
  opposite orientations.  This choice of $(p,q)$ winding cycles bound
  polygonal regions on the torus, which are the faces of the dimer
  model and will later be identified with the wrapped branes.  Note
  that each ``edge'' of the faces is produced by two intersecting
  $(p,q)$ cycles, which will each give rise to a bifundamental chiral
  multiplet.  (b) The dashed lines on the curve $P(z,w)=0$ shows the
  gluing locus, i.e.~the graph $\Gamma$.}
\label{fig:dimerc3}} 

This points toward the geometrical origin of the dimer models as
winding cycles on the curve $P(z,w)=0$.  However, it is not the end of
the story, because in string theory we obtain the massless matter of
the quiver theory from the intersection of D6-branes, so we must
choose a set of contours that intersect with the correct properties.
Clearly, the straight-line contours do not satisfy this, since they do
not intersect one another on $\Sigma$ because they are far away along
the spines.  

However, we saw in \sref{sec:untwisting} that there indeed
exists a suitable deformation of these contours on $\Sigma$ to produce
the dimer model as above, namely the graph $\Gamma$ introduced
therein.  It may be constructed by suitably gluing together a set of
contours that encircle each of the punctures on the curve.  We
constructed this graph using the untwisting map from the dimer models
on $T^2$, and observed that the resulting graph has the correct
properties to support the wrapped D6-branes that produce the desired
quiver gauge theory on their world-volume.

Given a suitable embedding of this graph $\Gamma \subset \Sigma$,
which was obtained by a map from an {\it abstract} $T^2$, we will show
that the graph of the dimer model is again reproduced by projecting to
the {\it particular} $T^2$ defined by the angular parts of $(z,w)$,
and argue that these $T^2$ may be identified in such cases.  However,
it is important to keep in mind that the image in this alga $T^2$ is a
projection of the curve in which we have discarded half of the
information (by projecting onto a half-dimensional subspace of
$(\IC^*)^2$).  It may sometimes be the case that the projection to the
$T^2$ is not faithful, e.g. contours that cross after the $T^2$
projection may not really cross on the curve $\Sigma$.  It is
therefore better in general to consider the contours on $\Sigma$
rather than their projection; we discuss this further in
\sref{sec:deg}.

%--------------------
\subsection{Projection of the Intersection Locus}

If we assume that the embedding of the graph $\Gamma$ is such that the
projection to the $T^2$ is an isomorphism (i.e.~there do {\it not}
exist two distinct points on $\Gamma$ with the same angular parts),
then the alga projection of $\Gamma$ is identified with the graph of
the dimer model obtained by the twisting procedure of
\sref{sec:untwisting}, up to homotopy.  We conjecture that it is always
possible to arrange this, and we give several examples in
\sref{sec:examples}.

Even in non-degenerate cases, identifying the $T^2$ of the dimer model
with that of the angular variables imposes additional restrictions on
the properties of the projection.  Firstly, since the graph $\Gamma$
is inscribed on the curve $P(z,w)=0$, the image of this graph is
restricted to the interior of the alga projection of this curve.
Thus, the graph of the dimer model cannot be embedded arbitrarily
within the $T^2$, but must lie within the subset of $T^2$ defined by
the alga projection of the curve.

Secondly, we show that the allowed deformations of the graph in $T^2$
are restricted (in particular, arbitrary deformations of the $(p,q)$
cycles are not allowed).  Let us consider more closely the effect on
the alga projection of deforming the contours.  Above we showed that a
contour very close to a $(p,q)$ puncture projects to a straight line
on the alga with $(p,q)$ winding number.  It is also true that {\it
any} choice of contour encircling a puncture of $P(z,w)=0$ maps to a
cycle with the same $(p,q)$ winding number in the alga, since they are
related by continuous deformations.  Moreover, deforming the choice of
contour on $\Sigma$ does not change the average position of the
projection to the alga (i.e.~does not produce a translation of the
winding path on the torus).  The allowed projections of contours to
the alga are therefore constrained by the moduli of the curve, see
\fref{fig:contour-deform}.  Let us now show this fact.

\FIGURE[h]{\centerline{\epsfig{file=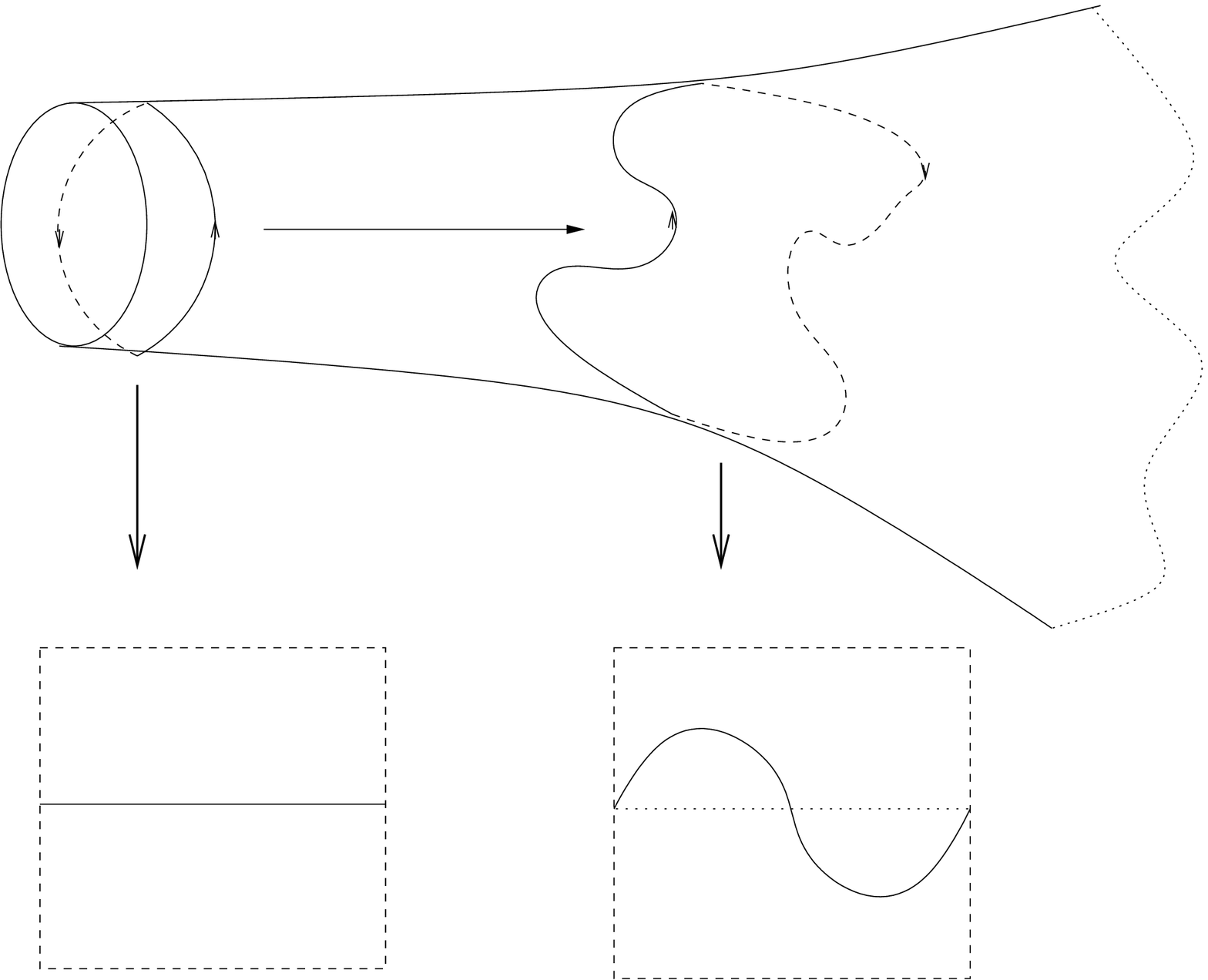,width=7cm}}
\caption{Two choices of contour encircling a puncture on the curve
  $\Sigma$.  Far along the cylinder, i.e.~close to the puncture, the
  contour projects to a straight line in the angular $T^2$.  When this
  contour is deformed, the projection to the $T^2$ is also deformed,
  but it is not translated in the torus and has the same average
  position.}
\label{fig:contour-deform}}

Without loss of generality, we may consider the puncture to be a
cylinder aligned in the $(-1,0)$ direction (any other choice of
orientation is related by an $SL(2,\IZ)$ transformation).  The location
of the puncture is therefore at the point $z=0$ on $\Sigma$ (i.e.~$\log |z|
\rightarrow -\infty$), and a contour very close to the puncture
projects to a horizontal straight line in the alga given by $\phi =
\mbox{Arg}(w) \equiv \phi_0$, a constant determined by the curve moduli, as in \eref{eq:straight}.  For any other choice of contour, the average value of
$\phi$ along the contour is given by 
\beq
\begin{array}{rcl}
\overline \phi &=& \frac{1}{2 \pi \imath} \oint_{z=0}
\mbox{Arg}(w(z)) d(\log z) \\
&=& \frac{1}{2 \pi \imath} \oint_{z=0}  \mbox{Arg}(w(z)) \frac {dz}{z}
\\
&=& \mbox{Arg}(w(0)) \\
&=& \phi_0
\end{array}
\eeq
as long as the puncture is isolated, i.e.~two punctures do not collide. 

%%%%%%%%%%%%%%%%
%------------------------=====================-------------------
%%%%%%%%%%%%%%
\subsection{Projection of the D6-branes}

As we discussed in \sref{sec:proposal}, the wrapped D6-branes that
provide a basis for the quiver correspond to 1-cycles on the curve
$\Sigma$ that wind the cycles which vanish at critical points in the
$W$-plane.  Under the untwisting map the discs attached with boundary
along these 1-cycles map to the faces of the dimer model on $T^2$.  We
will later show in several examples that under the (alga) $T^2$
projection to phases of $(z,w)$ these vanishing cycles also map to the
faces of the dimer model.

Taking $W=P(z,w)$, critical values of $W$ define points where the
curve fibre degenerates, i.e.~a certain homology
1-cycle of the curve $P(z,w) =
W$ vanishes.  At this point, the vanishing 1-cycle collapses to a
point, and therefore the projection of the cycle to the alga is also a
point.  As we move away from this point by changing $W$, this point
will resolve into a circle, which will project to a small closed
circle in the alga also.  Along the straight-line path from the
critical point $W=W_*$ toward $W=0$, the vanishing cycle sweeps out a
closed disc in the angular $T^2$.

Multiple such straight-line paths all meet at $W=0$.  Thus, near
$W=0$, where the discs are attached along the $S^1$ paths along the
graph $\Gamma$, which projects to the graph of the dimer model, the
disc swept out by the vanishing cycle in the $T^2$ approaches the
piecewise linear polygonal face of the dimer model.  Hence, the image
of the vanishing cycles along the straight-line paths projects to the
{\it closed} polygonal faces of the dimer model. Let us now enlighten
the reader with some illustrative examples.
%%%%%%%%%%%%%%-----------
\subsection{Examples}
\label{sec:examples}
%--------
\subsubsection{$\IC^3$}

If one prescribes an embedding of the graph of the dimer model within
the alga projection of the curve $P(z,w)=1 - w - z = 0$
(i.e.~prescribing the angular parts of $z,w$), it is straightforward
to solve for the absolute values to lift this to an embedding of the
graph $\Gamma$ on the curve.

Writing 
\begin{equation}
z = A e^{\imath \theta}, \qquad
w = B e^{\imath \phi}
\end{equation}
and setting the real and imaginary parts of $P(z,w)$ to zero, we find

\begin{equation}
A = -\frac{\sin(\phi)}{\sin(\theta - \phi)}, \qquad
B = \frac{\sin(\theta)}{\sin(\theta - \phi)}
\label{eq:ab-c3}
\end{equation}
If one chooses the two vertices of the dimer model to be located at
$\pm(-\pi/3, \pi/3)$ and connected by straight lines (see
\fref{fig:c3-embed}), the graph of the dimer model on $T^2$ may be
parametrized by the 3 paths:

\FIGURE[h]{\centerline{\epsfig{file=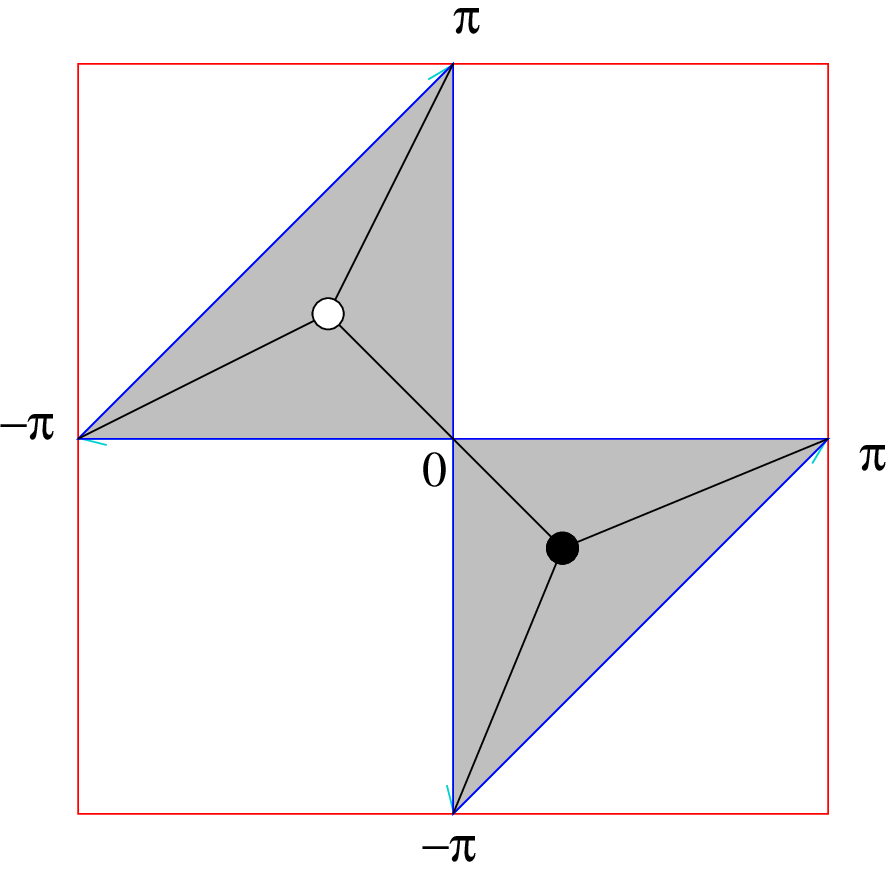,width=5cm}}
\caption{An embedding of the dimer model for $\IC^3$.  The shaded
  region is the alga projection of the curve $P(z,w)=1-z-w = 0$, and
  in this case it is bounded by the straight-line winding cycles with
  $(p,q) = (1,1), (-1,0),(0,-1)$.}
\label{fig:c3-embed}} 

\begin{eqnarray}
\theta_1(t) = -\frac{\pi}{3} + \frac{2\pi}{3} t, &\quad&
\phi_1(t) = \frac{\pi}{3} - \frac{2\pi}{3} t; \nonumber \\
\theta_2(t) = \frac{\pi}{3} + \frac{4\pi}{3} t, &\quad&
\phi_2(t) = -\frac{\pi}{3} + \frac{2\pi}{3} t; \nonumber \\
\theta_3(t) = -\frac{\pi}{3} + \frac{2\pi}{3} t, &\quad&
\phi_3(t) = \frac{\pi}{3} + \frac{4\pi}{3} t,
\end{eqnarray}
where $t \in [0,1)$.  Substituting these paths into (\ref{eq:ab-c3})
parametrizes an embedding of the graph $\Gamma$ into the curve
$P(z,w)=0$.  Since this curve has real coefficients, it is Harnack,
and the amoeba projection is simple (see \sref{sec:harnack})
(recall each point on the interior of the amoeba lifts to two points
on the curve).  The amoeba projection of $\Gamma$ is shown in
\fref{fig:c3lift}.  We see that this graph agrees with the one obtained
in \sref{s:duality} by the untwisting method (see also
\fref{fig:dimerc3}).

\FIGURE[h]{\centerline{\epsfig{file=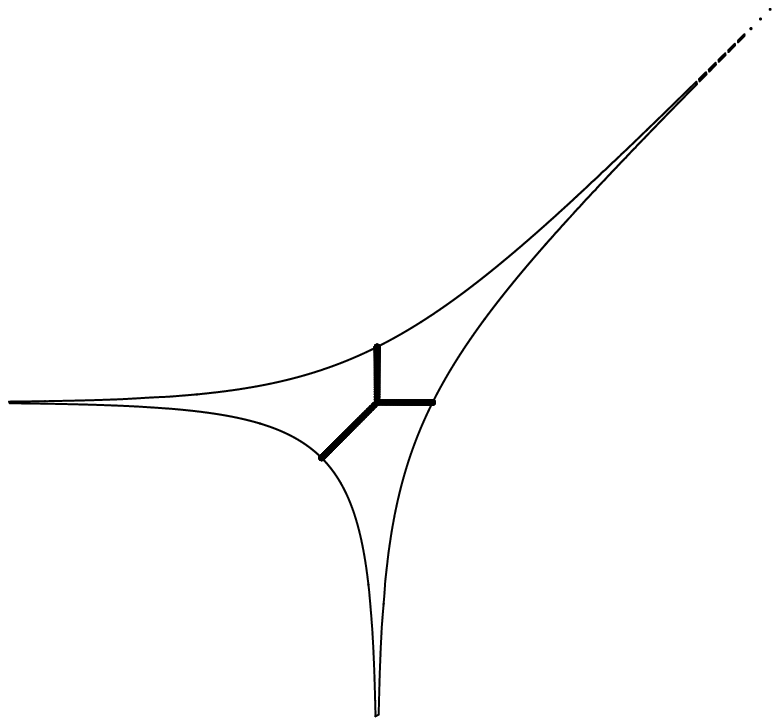,width=5cm}}
\caption{The amoeba projection of the graph $\Gamma$ on the curve
$P(z,w)=1-z-w=0$ for $\IC^3$, 
obtained by solving for the absolute values of $z,w$ along
the locus prescribed in \fref{fig:c3-embed}.  This $\Gamma$
agrees with the graph obtained by untwisting, see
\fref{fig:gluing-graph}.}
\label{fig:c3lift}}

Note that even though we cannot be completely explicit about the
topology of the D6-branes in the full CY geometry (since $\Sigma$ has
genus 0 and the corresponding $W=P(z,w)$ does not have enough critical
points to use the method described in \sref{s:branes}), we see
that we still obtain the full quiver theory from self-intersections of
a closed path on the curve $P(z,w)=0$ (the zig-zag path on the graph
$\Gamma$, as shown in \fref{fig:c3-cycles}).  This 1-cycle
should be thought of as the intersection of the D6-brane with the
curve $P(z,w)=0$.

%%%%%%%%%%---------
\subsubsection{Conifold}

For simplicity we take $P(z,w) = 1 - z - w - e^t z w$ with $t$ real.
One may again solve for the absolute value of $z,w$ in terms of the
angular parts, but the expressions are more complex since $P$ is
quadratic.  Instead of writing these out we simply plot the result in
\fref{fig:conifold-lift}: since we have chosen $t$ real, the
amoeba projection again gives a nice representation of the data, and
one can verify that the lift of the graph of the dimer model is again
isomorphic to the graph obtained by untwisting in
\sref{s:duality}, see \fref{fig:conifold-cycles}.  As in the
previous example, we obtain the full quiver theory from the
intersection of the D6-branes with the curve $P(z,w)=0$.

\FIGURE[h]{\centerline{\epsfig{file=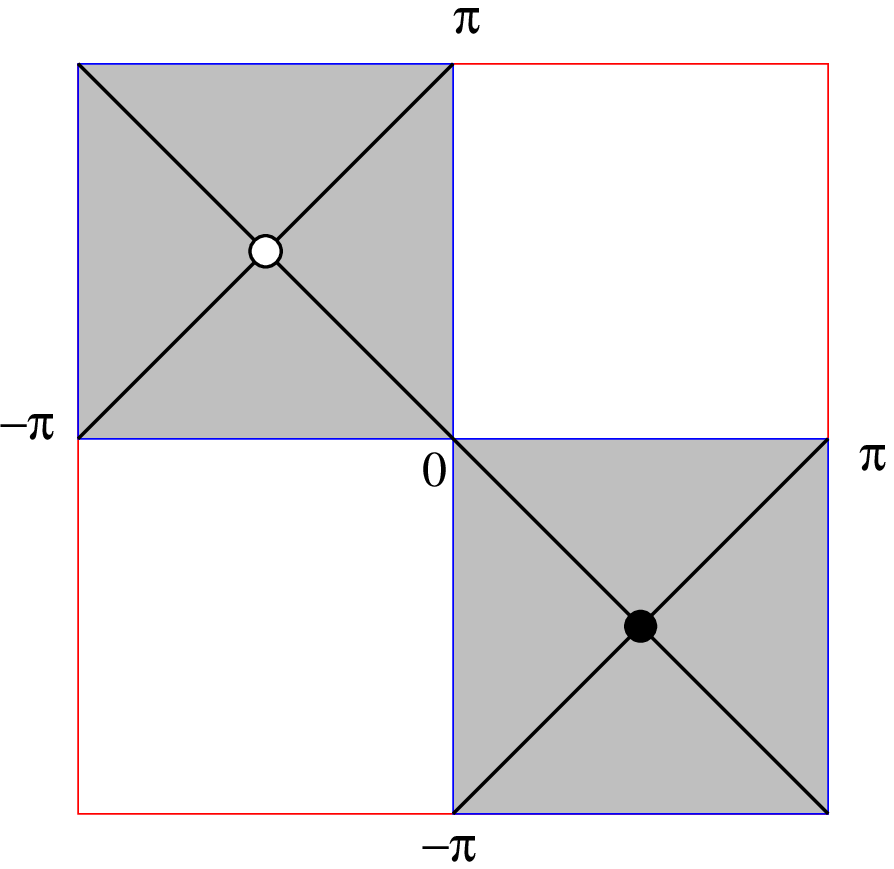,width=5cm}}
\caption{An embedding of the dimer model for the conifold.  The shaded
  region is the alga projection of the curve $P(z,w)=0$, and
  in this case it is bounded by the straight-line winding cycles with
  $(p,q) = (1,0), (-1,0), (0,1), (0,-1)$.}
\label{fig:conifold-embed}} 

\FIGURE[h]{\centerline{\epsfig{file=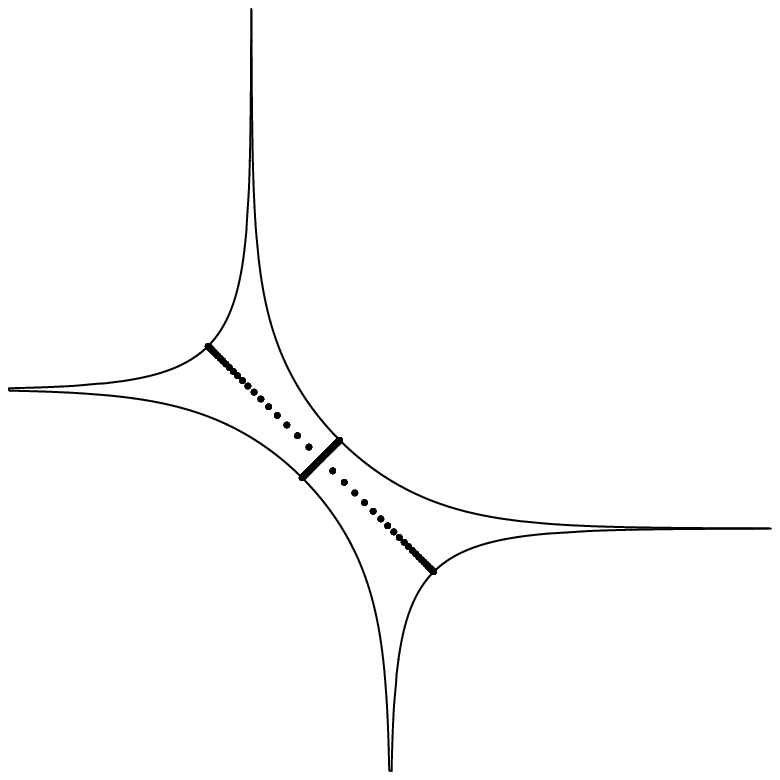,width=5cm}}
\caption{The amoeba projection of the graph $\Gamma$ on the curve
$P(z,w)=0$, obtained by solving for the absolute values of $z,w$ along
the locus prescribed in \fref{fig:conifold-embed}.  This $\Gamma$
agrees with the graph obtained by untwisting, see
\fref{fig:conifold-cycles}.}
\label{fig:conifold-lift}}

For more general curves $P(z,w)$, it is necessary to take complex
moduli in order to inscribe a non-degenerate version of the dimer
model graph.  One may in principle still solve for the absolute values
of $z,w$ to lift the dimer model graph to a graph on the curve, but
since $P$ is no longer Harnack, the amoeba projection does not give a
nice projection of this graph.

We now show in several higher-genus examples that the critical points
of $W$ correspond to the faces of the dimer model, in agreement with
the general proposal.

%%%%%%%%%%%%%%--------
\subsubsection{$\IC^3/\IZ_3$}

\begin{equation}
P(z,w) = 1 + z + w + \frac{e^t}{z w}
\end{equation}
Critical points of $W = P(z,w)$ are obtained when $z = w = \lambda
e^{t/3}$, with $\lambda^3 = 1$.  The critical values are $W_* = 1 + 3
\lambda e^{t/3}$. We show this in part (a) of \fref{fig:c3z3-2}.
 
Thus, there are three wrapped D6-branes, corresponding to the 3 gauge
groups in the quiver.  The vanishing cycles are located at $z = w =
\lambda e^{t/3}$, which project to the points $(0,0)$, $(2 \pi/3,
2\pi/3)$, $(4 \pi/3, 4\pi/3)$ in the alga.  These are indeed the
center of the hexagonal faces in the graph of the dimer model, see
\fref{fig:c3z3}. As one proceeds from $W=0$ to critical points in a
straight line, faces of the dimer model get filled in, see
parts (b) and (c) of \fref{fig:c3z3-2}.

\FIGURE[h]{\centerline{\epsfig{file=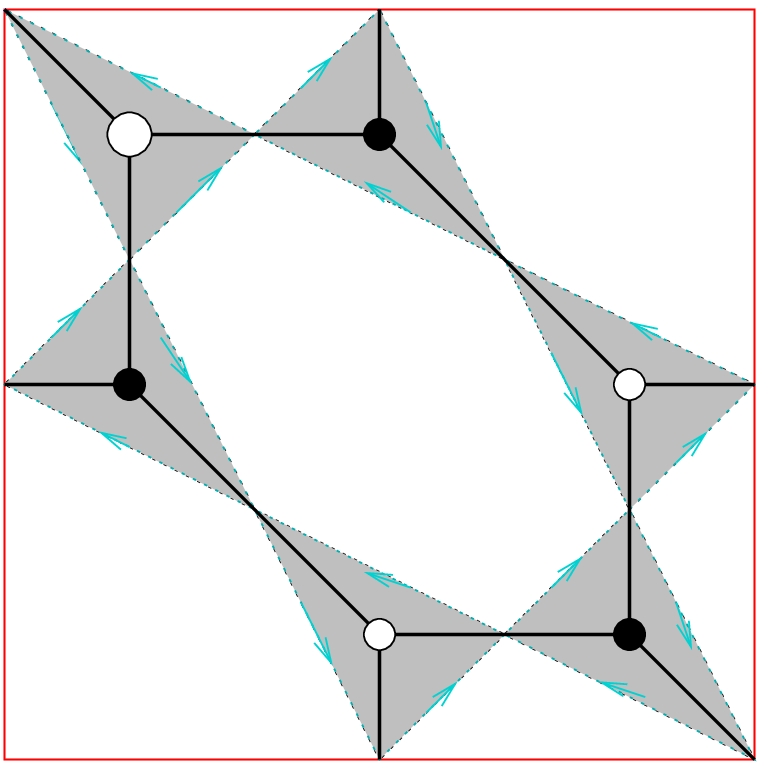,width=4.5cm}}
\caption{The dimer model for $\IC^3/\IZ_3$.  The shaded region is the
  alga projection of the curve, and in this case it is bounded by the
  straight-line winding cycles with $(p,q) = (1,1), (-2,1),(1,-2)$,
  which form triangular regions.  The graph of the dimer model
  describing the quiver theory of $\IC^3/\IZ_3$ may be inscribed
  within these triangular regions, as shown.}
\label{fig:c3z3}} 

\FIGURE[h]{\centerline{\epsfig{file=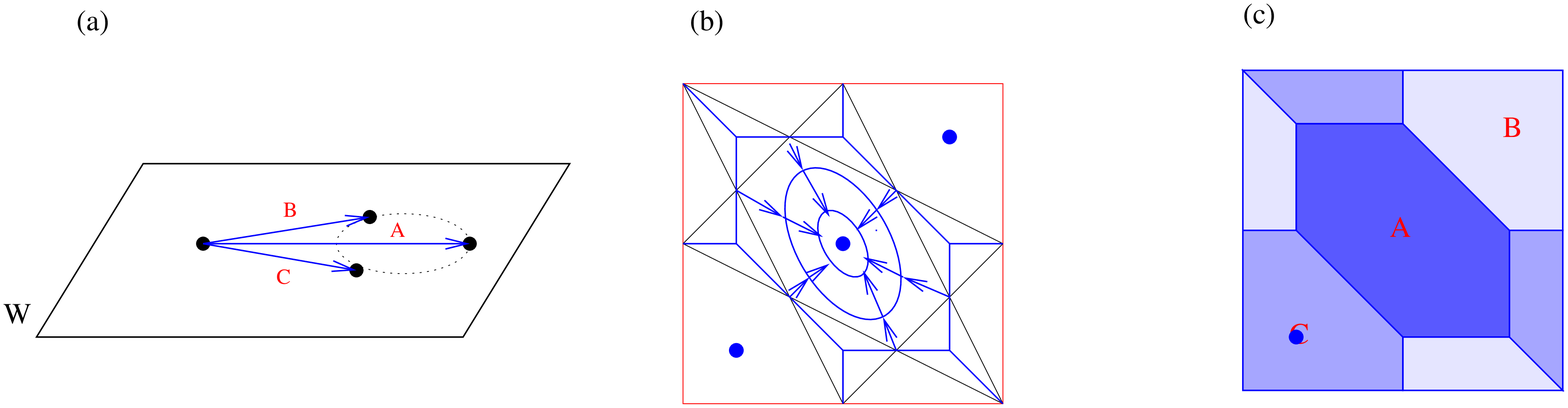,width=14cm}}
\caption{
(a) For $\IC^3/\IZ_3$, there are 3 critical points in the $W$-plane;
(b) The alga projection of the critical values of $z, w$ at the
  critical points of $W$ are in the interior of the three hexagonal
  faces of the dimer graph.  Along the straight-line paths from $W=0$
  to a critical point, the corresponding face of the dimer model fills
  in.  
(c) The combination of these paths fills the entire $T^2$ in the
  angular parts of $z, w$.  The $S^1$ fibre in the $u,v$ plane
  vanishes over the boundary of each face and is non-vanishing in the
  interior, which exhibits these faces as the $S^3$ wrapped branes.}
\label{fig:c3z3-2}}

%%%%%%%%%%%%----------
\subsubsection{$Y^{3,1}$}
%\FIGURE[h]{\centerline{\epsfig{file=y31.eps,width=3cm}}
%\caption{The Newton polytope of the toric space $Y^{3,1}$.}
%\label{fig:y31}}
%The Newton polygon of the toric space is shown in
%\fref{fig:y31}.  
The spaces $Y^{p,q}$ have recently been much studied. Let us focus on
$Y^{3,1}$. The toric
diagram and the corresponding $P(z,w)$ are:
\begin{equation}
\begin{array}{cc}
\epsfxsize = 4cm \epsfbox{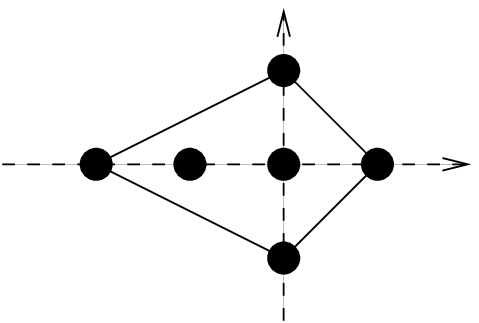}
& P(z,w) = 1 + z + \frac{e^{-a}}{z} + \frac{e^{-b}}{z^2} + w +
  \frac{e^{-c}}{w}
\end{array}
\end{equation}
Indeed, the toric diagram contains 2 internal points, so the Newton
polynomial defines a genus 2 curve in the mirror geometry. 

There are generically 6 critical points of $W=P(z,w)$ by the arguments
of appendix \ref{app:counting}.  In \fref{fig:ex-ypq} we plot
the alga of $P(z,w)$ with randomly chosen values of the moduli
($e^{-a}=1+\imath, e^{-b}=3-2\imath, e^{-c}=-1-4\imath$), and the
location of the corresponding critical values of $z,w$.  We may
inscribe a dimer model within the alga projection of the curve such
that these points lie within the faces of this dimer model.  This
dimer model is also consistent with the straight-line $(p,q)$ winding
paths, which are shown.  In this case the straight line paths bound
regions enclosing the vertices with alternating orientations, so the
dimer model is easy to read off.  In general this may not be true,
although based on the argument in \sref{s:pq} (see also
\sref{s:interpolate}) one can always deform these straight-line paths
to meet at vertices while keeping the same average value.

\FIGURE[h]{\centerline{\epsfig{file=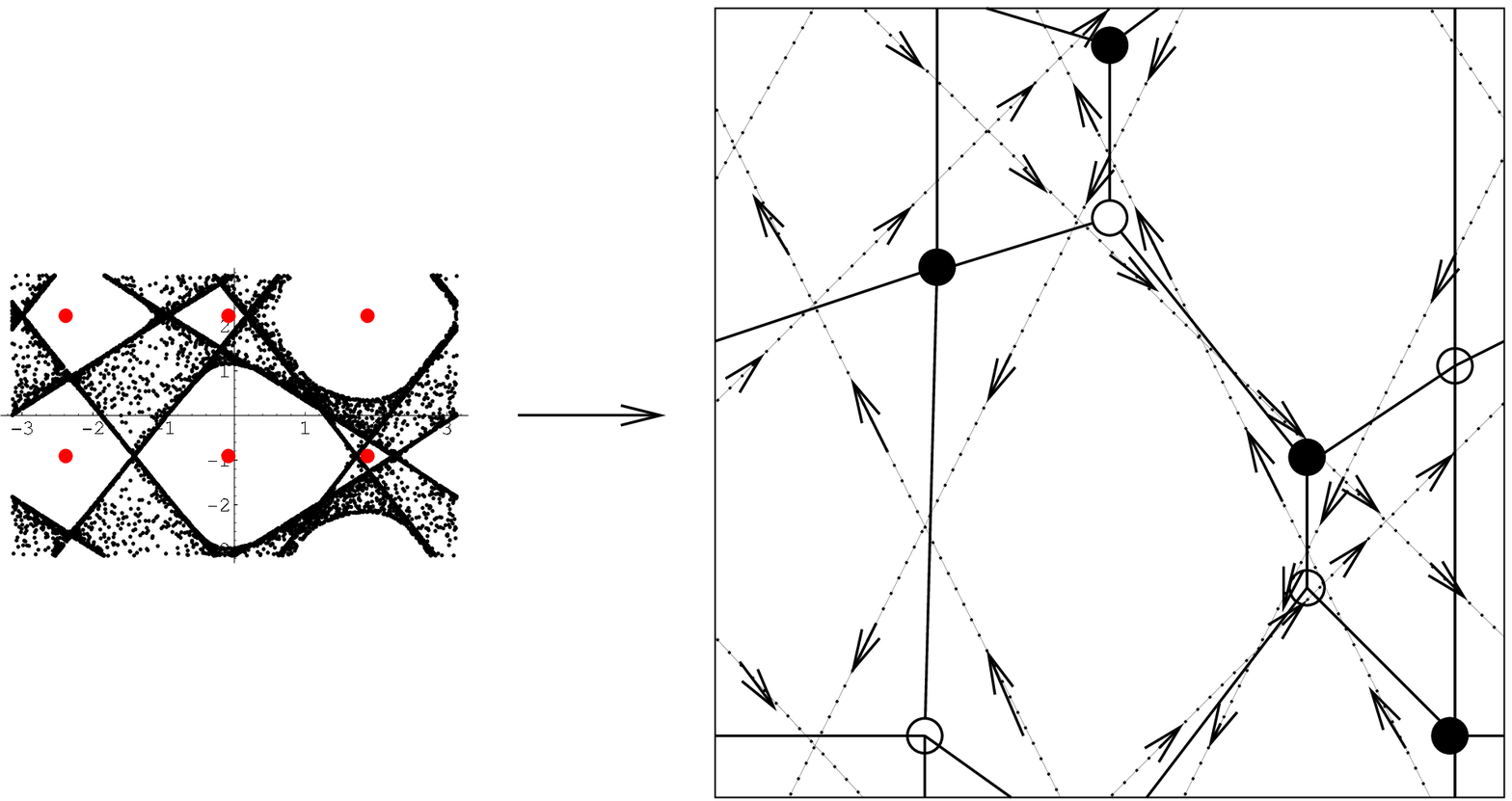,width=10cm}}
\caption{The alga for $P(z,w) = 1 + z + \frac{1+\imath}{z} +
  \frac{3-2\imath}{z^2} + w + \frac{-1-4\imath}{w}$.  Critical values
  of $z, w$ are marked by red dots in the alga projection.  The
  straight line $(p,q)$ winding cycles are visible as accumulation
  points, the projection of points lying along the half-cylinders.
  This geometrical data is consistent with the dimer model shown.}   
\label{fig:ex-ypq}} 

\FIGURE[h]{\centerline{\epsfig{file=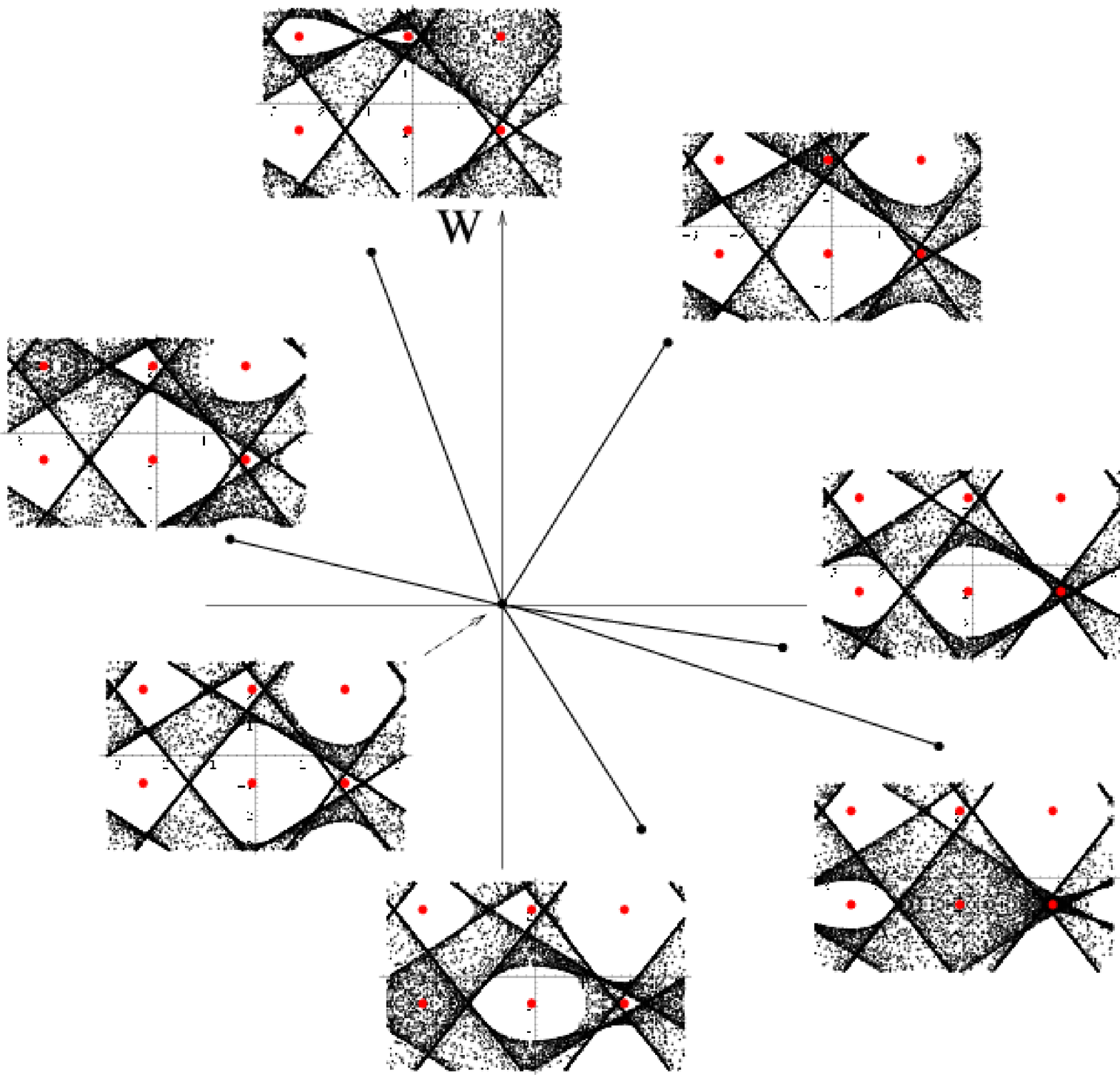,width=10cm}}
\caption{By plotting the alga projection of the curve $P(z,w)=W$ we
  see that near the critical points $W_*$ the alga fills in the region
  of the vanishing cycle (critical values of $z,w$, marked with red
  dots).  Thus, the vanishing cycles sweep out the faces of the dimer
  model shown in \fref{fig:ex-ypq}.} 
\label{fig:ex-ypq-1}.}

In \fref{fig:ex-ypq-1} the straight line paths from $W=0$ to the
critical points $W=W_*$ are shown.  The alga projection of $P(z,w) =
W_*$ shows the vanishing cycle located at the critical values, since
the corresponding face of the dimer model fills in along this path. 

%%%%%%%%%%%%%%%%%%%%%%%%%%%%%%%%%%%%%%%%
%%%%%%%%%%%%----------------------=============================
%%%%%%%%%%%%%%%%%%%%%%%%%%%%%%%%%%%%%%%%
\subsection{Degenerations}
\label{sec:deg}

Since the $T^2$ is only a projection onto a half-dimensional subspace,
various things may go wrong with the image of this projection.  We
discuss some of them, although we do not study these degenerations in
detail.

When $P(z,w)=0$ is a Harnack curve, i.e.~when the coefficients of $P$
are real, the amoeba map has the property that the boundary of the
amoeba is also real.  This is maximally nice from the point of view of
the amoeba, but maximally bad from the point of view of the alga,
since it means that on the boundary of the fundamental domain the
image of the curve contracts to the points $(0, 0), (0, \pi), (\pi,
0), (\pi, \pi)$.

Thus, {\it any} contour that touches the boundary of the amoeba (in
particular, any contour that encircles a puncture on $\Sigma$, which
projects in the amoeba to a line segment extending between two
boundary components) passes through these degenerate points.  This
obscures the projection of $\Gamma$ and does not produce an isomorphic
graph.  Taking the moduli of the curve to be generic complex numbers
avoids this problem. 

In principle different embeddings of $\Gamma$ may produce degenerate
projections to $T^2$, i.e.~if they contain two points with the same
angular parts, they will project to the same point and create
``spurious vertices'' in the projection of $\Gamma$. 

However, we conjecture that it is always possible to find situations
where the projection is non-degenerate (by adjusting the moduli of the
curve $\Sigma: P(z,w)=0$ and/or by deforming the embedding of $\Gamma
\subset \Sigma$).  As in the examples discussed in the previous
section, when the alga is suitably non-degenerate one may inscribe the
desired dimer graph in the interior of the alga projection (which
prescribes the angular parts of $z, w$ along $\Gamma$) and solve for
$|z|, |w|$ to map the dimer graph to $\Gamma$ (by definition of the
alga projection, a solution exists on the curve).  This constructs a
$\Gamma$ that is graph-isomorphic to the dimer graph $T^2$.

In such cases, the dimer models in $T^2$ are obtained by projection of
$\Gamma$.  In general it is simpler to consider the dimer model on
$\Gamma$ itself, or to use the twisting map to relate it topologically
to a graph on $T^2$ without having to worry about a choice of
embedding.  Considering the specific projection to angular variables
imposes restrictions on the image in $T^2$ and may provide further
information.  This problem should be studied in more detail.

%
%++++++++++++++++++
\subsection{Seiberg Duality from Alg\ae}
Now we discuss Seiberg duality from the point of view of alga
projections. As we have shown \cite{Hanany:2001py,Feng:2002kk} 
and discussed in \sref{sec:sd}, 
Seiberg duality is the result of
PL transformations, where one cycle passes over another. 
More concretely, $P(z,w)-W=0$ gives various critical
points in the $W$-plane and the bases $S^3$ of the third-homology
are given by straight lines in the
$W$-plane connecting the origin to these critical points. 

As we
vary moduli (coefficients in $P(z,w)$), the positions of critical points
change correspondingly. Hence, it is clear that sometimes one straight
line will pass another straight line, whereby inducing the PL
transformation.
From the point of view of the actual geometry, such a transformation is
non-trivial. We would thus like to see that after the alga projection,
how this effect shows up in the dimer model. One possible way
is illustrated by the following.

Let us again exemplify using the Hirzebruch-zero geometry $F_0$.
The curve, rescaling \eref{F0toric}, is given by  
\bean 
P_{F_0}(z,w)= k -z-1/z-w -e/w 
\eean
where $k,e$ are complex moduli. 
Now let us study how phases change when
we vary the moduli. To do this we have chosen $k=3+3i$ and $e=2+
i\alpha $ with $\alpha$ varying. 
We choose four cycles, corresponding to the 4 legs (spines), parametrized 
as $z=r_0 e^{it}$ with  $t \in [-\pi,+\pi]$ and such that
$r_0=e^3$ for cycles $(1,\pm 1)$ and $r_0=e^{-3}$ for cycles
$(-1,\pm 1)$. 

\FIGURE[h]{
$\begin{array}{cc}
\epsfxsize = 6cm \epsfbox{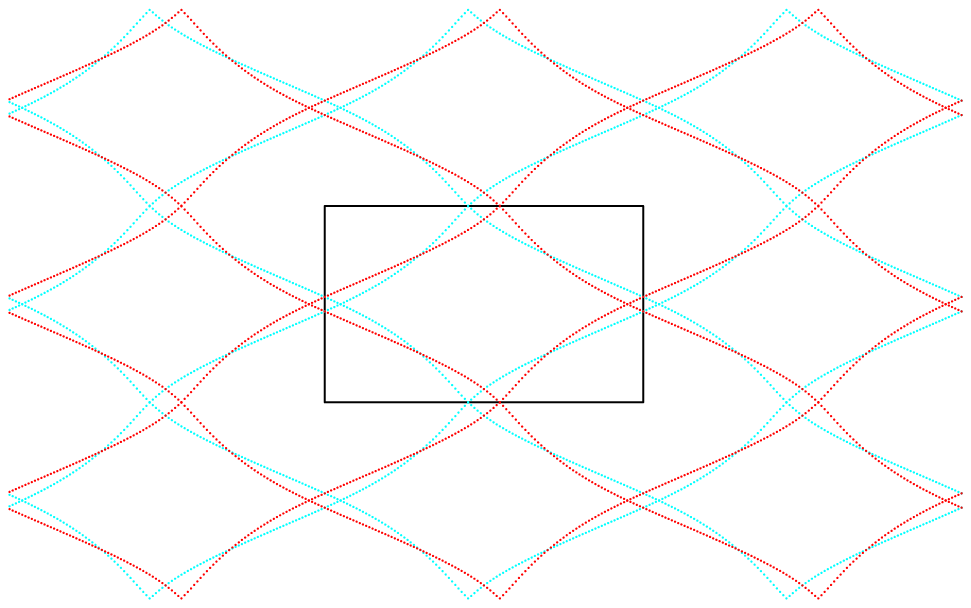}
&
\epsfxsize = 6cm \epsfbox{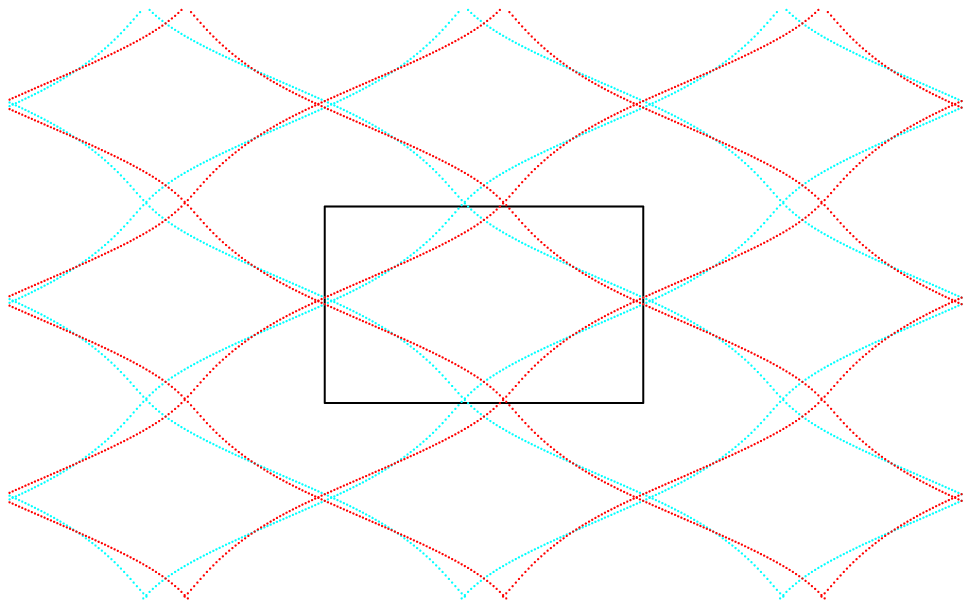} \\
\alpha=0 & \alpha=0.5 
\\
\epsfxsize = 6cm \epsfbox{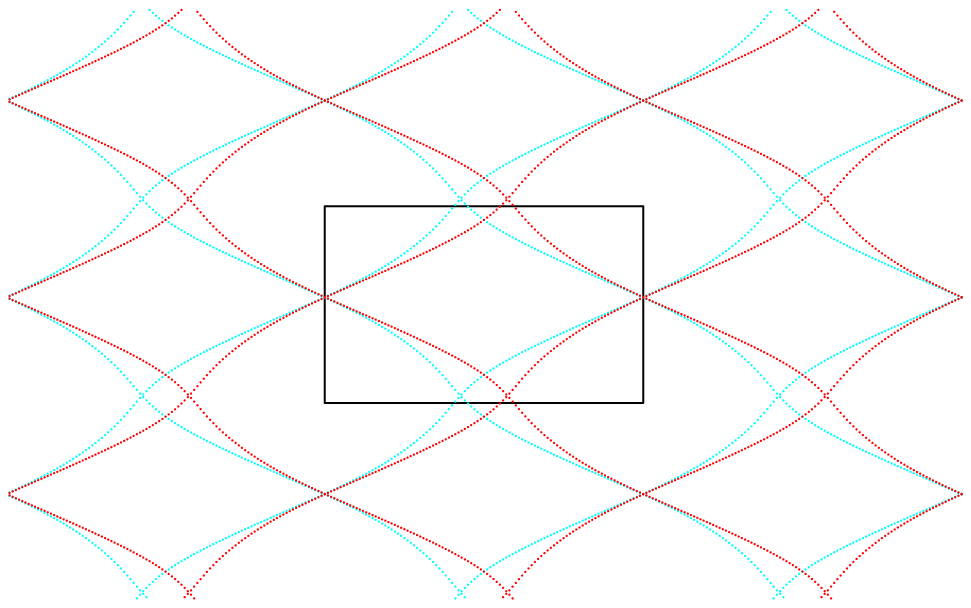}
&
\epsfxsize = 6cm \epsfbox{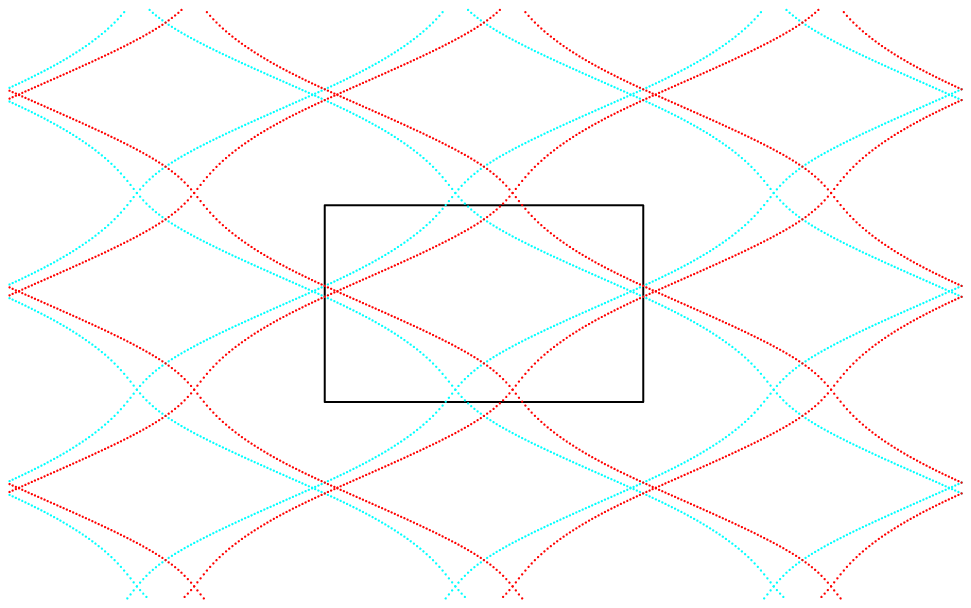} \\
\alpha=1 & \alpha=2 \\
\end{array}$
\caption{The 4 $S^1$'s winding the 4 cylinders of $F_0$ under the alga
projection. We vary the parameter $\alpha$ which represents a complex
modulus of $\Sigma$ and see how the intersection varies for 4
different values of $\alpha$ as indicated. The fundamental domain
$[-\pi,\pi]$ is indicated by the black box.}
\label{f:F0chaning}
}

We now follow how these 4 cycles project in the alga and
the results are given in parts (a), (b), (c) and (d) of
\fref{f:F0chaning}.
To see the periodicity of the alga projection more clearly, 
we draw in $3\times 3$ times the fundamental unit, which is
indicated in the black square. 

Now we can see how the intersection of the four cycles
changes from $\alpha=0$ to $\alpha=2$.
The main difference is that
the zig-zag cycles $(1,1)$ and $(-1,-1)$ do not intersect at
$\alpha=2$,
but do have two extra intersections at $\alpha=0$,
one with plus and one with minus sign,
so the topological intersection number is still zero from
the perspective of dimer. However, such ``superficial'' extra intersections
in the dimer model is crucial for Seiberg duality because they come from the
intersection in the real geometry. In other words, they are actual
intersections in $\Sigma$.

It is worth to notice that at
$\alpha=1$, all four cycles intersect at the same point. Thus these
two phases interpolate to each other through this degenerate point.
We see that indeed for certain choices of the moduli, the projection
onto the alga precisely reproduces the zig-zag paths. We conjecture
similar procedures may be applied in general.

We must emphasize that the above results, though highly suggestive, are
not fully right because the four contours we have chosen do not
actually intersect each other in the real geometry $\Gamma$.
However, we think it may give the
right topological picture about Seiberg duality in the dimer model.

%%%%%%%%%%%%%%%%%%%%%%%%%%%%%%%%%%%%%%%%
%%%%%%%%%%%%----------------------=============================
%%%%%%%%%%%%%%%%%%%%%%%%%%%%%%%%%%%%%%%%
\section{Discussion}\label{s:conc}

In this paper we have studied the intersecting D6-brane systems that
give rise to toric quiver gauge theories on their world-volume.  We
found that the intersection of the D6-branes with the curve
$P(z,w)=0$, which is part of the CY geometry $\cW$ mirror to the
original toric CY $\cM$, encodes the matter content and tree-level
superpotential of the quiver theory.

The D6-brane world-volumes form a singular $T^3$.  In principle the
singularities of this $T^3$ could have been of a complicated form.
However, we found that for the quiver theories coming from D3-branes
at the tip of toric CY cones, the only singularity of the mirror
D6-branes comes from a vanishing $S^1$ fibre over a $T^2$, where the
fibre vanishes along a bipartite graph\footnote{This may
  have implications for the special Lagrangian fibration structure of
  the mirror geometry, cf \cite{Gross:2000}.}
in $T^2$ .  
This realizes the dimer
models on $T^2$ as living on a subset of the $T^3$ world-volume of the
D6-branes, and provides a concrete string theoretical framework for
the previous combinatorial results relating dimer models and quiver
gauge theories \cite{Hanany:2005ve,Franco:2005rj,Vegh:2005}.

Here we have only discussed the {\it topology} of the D6-branes.  From
general arguments the supersymmetric D6-branes should wrap special
Lagrangian cycles.  Since they have topology $S^3$ (i.e.~$b_1=0$) these
cycles should be rigid.  It would be interesting to see whether the
ideas of isoradial embeddings of the dimer model graph, and the
interpretation of R-charges as angles in such embeddings
\cite{Vegh:2005}, can be fit into the present framework.
Specifically, the $a$-maximization of angles of the graph in $T^2$
selects one embedding of the dimer model; is this related to the
selection of the special Lagrangian cycle?  Alternatively, we may be
closer to the $Z$-minimization of \cite{Martelli:2005tp}, which is
also formulated in terms of toric geometry.

It should be possible to understand these results on dimer models from
intersecting D6-branes from a world-sheet perspective using linear
sigma models with A-type boundary conditions.  This problem is
currently under investigation \cite{KK-KH}.  From the target space
point of view one should also be able to relate the D6-branes and
associated dimer models on $\cW$ to the mirror branes on $\cM$, which
form an exceptional collection of sheaves.

The structure governing the intersection of the D6-branes with the
curve $P(z,w)=0$ is that they form zig-zag paths on a bipartite graph
$\Gamma$ on the curve.  This graph $\Gamma$ also produces a
decomposition of the curve into half-cylinders.  One might imagine
constructing different possible $\Gamma$ by the converse gluing
operations; in general there is an ambiguity in how $\Gamma$ may be
constructed by gluing of half-cylinders (this ambiguity admits the
various Seiberg-dual phases of the quiver theory).  It should be
possible to clarify the consistency conditions for the resulting
quiver theories.  It would also be interesting to understand this
structure geometrically.

We were not able to be completely explicit about the D6-branes when
the toric CY contains only a vanishing 2-cycle.  While we still obtain
the full quiver theory from intersections of 1-cycles on the curve
$P(z,w)=0$, the function $W=P(z,w)$ does not have enough critical
points to treat these cases uniformly with the vanishing 4-cycle
cases.  See Appendix \ref{app:counting} for one idea on the resolution
to this puzzle.

We introduced a projection of the mirror geometry (dubbed the alga
projection) that produces the dimer models on $T^2$ in suitable cases.
This projection has not been well-studied by mathematicians, so
further study of the properties of this projection and its relation to
dimer models is needed.

Now that we have understood the role of dimer models in describing the
tree-level superpotential of the quiver theories, can we also use the
dimer models as a basis for studying quantum corrections?  It is
already known \cite{Okounkov:2003sp} (see also
\cite{Stienstra:2005a,Stienstra:2005b}) that these same dimer models
may be used to compute the topological string partition function, so
it is likely that the answer is yes.

%%%%%%%%%%%%%%%%%%%%%%%%%%%%%%%%%%%%%%%%
%%%%%%%%%%%%----------------------=============================
%%%%%%%%%%%%%%%%%%%%%%%%%%%%%%%%%%%%%%%%
\section*{Acknowledgments}
BF gratefully acknowledges Marie Curie Research Training Network under
contract number MRTN-CT-2004-005104 and the IAS, Princeton under NSF
grant PHY-0070928; he extends his thanks to the Universities of
Amsterdam, Durham and Oxford for their hospitality.  YHH is indebted
to the patronage of Merton College, Oxford (especially to the
Rt.~Rev.~Warden Richard
FitzJames, Bishop of London) as well as the
Mathematical Institute of Oxford University; he would also like to
express his sincere gratitude to the Dept.~of Physics
(DE-FG02-95ER40893) and the Math/Physics RG (NSF-DMS0139799 Focused
Grant for ``The Geometry of Strings'') at the University of
Pennsylvania where this work began and for three happy years past.  KK
is supported by NSERC, and thanks Alastair Craw, Jarah Evslin, Manfred
Herbst, Kentaro Hori, Robert Karp, David Page, Christian
R\"omelsberger and James Sparks for enlightening discussions, and the
Fields Institute and Department of Physics at the University of
Toronto for support; he also is grateful to Harvard University, and
the University of Pennsylvania for being warm hosts.  CV is supported
in part by NSF grants PHY-0244821 and DMS-0244464.  BF, YHH and KK
jointly thank Sebastian Franco, Amihay Hanany, David Vegh and Brian
Wecht and the kind hospitality of MIT.  KK and CV thank the Simons
Workshop in Mathematics and Physics, Stony Brook for providing a
stimulating and productive environment.

\appendix
\section{Number of Critical Points of $P(z,w)$ and The Newton Polytope}
\label{app:counting}
In this Appendix, we show that the number of critical points of $W =
P(z,w)$ is equal to twice the area of the toric diagram, which in turn
is equal, by \eref{pick}, to the number of gauge groups in the
quiver.

First, there are some nice theorems on the
relationship between the number of
solutions of polynomial systems and the Newton polytope spanned by the
polynomials, culminating in the Koushnirenko-Bernstein theorem
\cite{Koushnirenko,Bernstein} (for relations to amoeb\ae,
q.v.~\cite{mikhalkin-2000-2,shiff-zeld}):
\begin{quote}
{\em
The number of joint zeros in $(\IC^*)^m$ of $m$ generic polynomials
$\{f_1, f_2, \ldots f_m\}$ with a given Newton polytope $\Delta$ is
equal to $m! {\rm Vol}(\Delta)$.
}
\end{quote}

The case at hand has $m=2$, with the equation system being
\beq\label{crit-2d}
\{\diff{P(z,w)}{z}, \diff{P(z,w)}{w}\} = 0 \ . 
\eeq
Now, the Newton
polytopes of $\diff{P(z,w)}{z}$, $\diff{P(z,w)}{w}$ and $P(z,w)$ are
all different; however, a simple transformation would make the theorem
applicable. We note that the Newton polytope of $z \diff{P(z,w)}{z}$ 
is the same as that of $P(z,w)$ except that
the $w$ axis is deleted.  Similarly $w \diff{P(z,w)}{w}$
has the $z$ axis deleted.
Thus, as long as the origin (common to both axes) is an {\it interior} point
of $P(z,w)$, i.e., when genus is greater than 0, 
$z \diff{P(z,w)}{z}$ and $w \diff{P(z,w)}{w}$ span the Newton polygon of
$P(z,w)$ (the only point missing is the constant term, which is
interior to the polygon and thus included in the convex hull).  
Then, the system
\beq
P1(z,w) = a z \diff{P(z,w)}{z}  + b w \diff{P(z,w)}{w} = 0; \quad
P2(z,w) = a z \diff{P(z,w)}{z}  + b w \diff{P(z,w)}{w} = 0
\eeq
for $a,b,c,d \ne 0$ and $\det\mat{a&b\\c&d} \ne 0$ has the same number
of solutions as the desired \eref{crit-2d} while $P1$ and $P2$ both have
the same Newton polytope as $P(z,w)$.

Now, the Koushnirenko-Bernstein theorem applies and we conclude that
\beq
\#\mbox{ critical points }(P(z,w)) = 
2 Vol(\mbox{ Newton Polytope of } P(z,w)) = 2 \mbox{Area(Toric Diagram)}.
\eeq

We remark, as an aside, that we could have multiplied \eref{crit-2d}
by $z^s w^t$ for some sufficiently large $s,t \in \IN$.  This shift of
the Newton polytope introduces extra critical points $z=0$ or $w=0$, which
are excluded since $z,w \in \IC^*$. Thus we could equally have run our
argument above and come to the same conclusion. However, for the genus
zero case, this shift does introduce valid extra solutions.
Amazingly, now, we actually get the right number of critical
points. However, it is not clear what this means in the
geometry. Indeed, although it seems that the toric data allows us to
liberally shift the origin, when we are calculating locations of
critical points, results do depend on the choice of origin. The
implications of this should be investigated further.

\bibliographystyle{JHEP}
\bibliography{quiver}

\providecommand{\href}[2]{#2}\begingroup\raggedright\begin{thebibliography}{10}

\bibitem{Douglas:1996sw}
M.~R. Douglas and G.~W. Moore, {\it D-branes, quivers, and {ALE} instantons},
  \href{http://xxx.lanl.gov/abs/hep-th/9603167}{{\tt hep-th/9603167}}.

\bibitem{Douglas:1997de}
M.~R. Douglas, B.~R. Greene, and D.~R. Morrison, {\it Orbifold resolution by
  {D}-branes},  {\em Nucl. Phys.} {\bf B506} (1997) 84--106,
  [\href{http://xxx.lanl.gov/abs/hep-th/9704151}{{\tt hep-th/9704151}}].

\bibitem{Hori:2000ck}
K.~Hori, A.~Iqbal, and C.~Vafa, {\it D-branes and mirror symmetry},
  \href{http://xxx.lanl.gov/abs/hep-th/0005247}{{\tt hep-th/0005247}}.

\bibitem{Hanany:2001py}
A.~Hanany and A.~Iqbal, {\it Quiver theories from {D6}-branes via mirror
  symmetry},  {\em JHEP} {\bf 04} (2002) 009,
  [\href{http://xxx.lanl.gov/abs/hep-th/0108137}{{\tt hep-th/0108137}}].

\bibitem{Wijnholt:2002qz}
M.~Wijnholt, {\it Large volume perspective on branes at singularities},  {\em
  Adv. Theor. Math. Phys.} {\bf 7} (2004) 1117--1153,
  [\href{http://xxx.lanl.gov/abs/hep-th/0212021}{{\tt hep-th/0212021}}].

\bibitem{Cachazo:2001sg}
F.~Cachazo, B.~Fiol, K.~A. Intriligator, S.~Katz, and C.~Vafa, {\it A geometric
  unification of dualities},  {\em Nucl. Phys.} {\bf B628} (2002) 3--78,
  [\href{http://xxx.lanl.gov/abs/hep-th/0110028}{{\tt hep-th/0110028}}].

\bibitem{Feng:2001xr}
B.~Feng, A.~Hanany, and Y.-H. He, {\it Phase structure of {D}-brane gauge
  theories and toric duality},  {\em JHEP} {\bf 08} (2001) 040,
  [\href{http://xxx.lanl.gov/abs/hep-th/0104259}{{\tt hep-th/0104259}}].

\bibitem{Feng:2000mi}
B.~Feng, A.~Hanany, and Y.-H. He, {\it D-brane gauge theories from toric
  singularities and toric duality},  {\em Nucl. Phys.} {\bf B595} (2001)
  165--200, [\href{http://xxx.lanl.gov/abs/hep-th/0003085}{{\tt
  hep-th/0003085}}].

\bibitem{Hanany:2005ve}
A.~Hanany and K.~D. Kennaway, {\it Dimer models and toric diagrams},
  \href{http://xxx.lanl.gov/abs/hep-th/0503149}{{\tt hep-th/0503149}}.

\bibitem{Franco:2005rj}
S.~Franco, A.~Hanany, K.~D. Kennaway, D.~Vegh, and B.~Wecht, {\it Brane dimers
  and quiver gauge theories},
  \href{http://xxx.lanl.gov/abs/hep-th/0504110}{{\tt hep-th/0504110}}.

\bibitem{Franco:2005zu}
S.~Franco, A.~Hanany, F.~Saad, and A.~M. Uranga, {\it Fractional branes and
  dynamical supersymmetry breaking},
  \href{http://xxx.lanl.gov/abs/hep-th/0505040}{{\tt hep-th/0505040}}.

\bibitem{Benvenuti:2005cz}
S.~Benvenuti and M.~Kruczenski, {\it Semiclassical strings in
  {S}asaki-{E}instein manifolds and long operators in {$\N = 1$} gauge
  theories},  \href{http://xxx.lanl.gov/abs/hep-th/0505046}{{\tt
  hep-th/0505046}}.

\bibitem{Benvenuti:2005ja}
S.~Benvenuti and M.~Kruczenski, {\it From {S}asaki-{E}instein spaces to quivers
  via {BPS} geodesics: {$L(p,q,r)$}},
  \href{http://xxx.lanl.gov/abs/hep-th/0505206}{{\tt hep-th/0505206}}.

\bibitem{Franco:2005sm}
S.~Franco, A.~Hanany, D.~Martelli, J.~Sparks, D.~Vegh, and B.~Wecht, {\it Gauge
  theories from toric geometry and brane tilings},
  \href{http://xxx.lanl.gov/abs/hep-th/0505211}{{\tt hep-th/0505211}}.

\bibitem{Butti:2005vn}
A.~Butti and A.~Zaffaroni, {\it R-charges from toric diagrams and the
  equivalence of $a$- maximization and {$Z$}-minimization},
  \href{http://xxx.lanl.gov/abs/hep-th/0506232}{{\tt hep-th/0506232}}.

\bibitem{Okounkov:2003sp}
A.~Okounkov, N.~Reshetikhin, and C.~Vafa, {\it Quantum {C}alabi-{Y}au and
  classical crystals},  \href{http://xxx.lanl.gov/abs/hep-th/0309208}{{\tt
  hep-th/0309208}}.

\bibitem{Feng:2002kk}
B.~Feng, A.~Hanany, Y.~H. He, and A.~Iqbal, {\it Quiver theories, soliton
  spectra and {P}icard-{L}efschetz transformations},  {\em JHEP} {\bf 02}
  (2003) 056, [\href{http://xxx.lanl.gov/abs/hep-th/0206152}{{\tt
  hep-th/0206152}}].

\bibitem{Hori:2000kt}
K.~Hori and C.~Vafa, {\it Mirror symmetry},
  \href{http://xxx.lanl.gov/abs/hep-th/0002222}{{\tt hep-th/0002222}}.

\bibitem{claymirror}
K.~Hori, S.~Katz, A.~Klemm, R.~Pandharipande, R.~Thomas, C.~Vafa, R.~Vakil, and
  E.~Zaslow, eds., {\em Mirror Symmetry (Clay Mathematics Monographs, V. 1)}.
\newblock AMS, 2003.

\bibitem{Strominger:1996it}
A.~Strominger, S.-T. Yau, and E.~Zaslow, {\it Mirror symmetry is {T}-duality},
  {\em Nucl. Phys.} {\bf B479} (1996) 243--259,
  [\href{http://xxx.lanl.gov/abs/hep-th/9606040}{{\tt hep-th/9606040}}].

\bibitem{mikhalkin-2000-2}
G.~Mikhalkin, {\it Real algebraic curves, the moment map and amoebas},  {\em
  ANN.OF MATH.} {\bf 2} (2000) 151,
  [\href{http://xxx.lanl.gov/abs/math/0010018}{{\tt math/0010018}}].

\bibitem{Khovanskii}
A.~G. Khovanskii, {\it Newton polyhedra and toric varieties},  {\em Funkcional.
  Anal. i Prilozen} {\bf 11} (1977) 56--64.

\bibitem{Aharony:1997bh}
O.~Aharony, A.~Hanany, and B.~Kol, {\it Webs of {$(p,q)$} 5-branes, five
  dimensional field theories and grid diagrams},  {\em JHEP} {\bf 01} (1998)
  002, [\href{http://xxx.lanl.gov/abs/hep-th/9710116}{{\tt hep-th/9710116}}].

\bibitem{Leung:1997tw}
N.~C. Leung and C.~Vafa, {\it Branes and toric geometry},  {\em Adv. Theor.
  Math. Phys.} {\bf 2} (1998) 91--118,
  [\href{http://xxx.lanl.gov/abs/hep-th/9711013}{{\tt hep-th/9711013}}].

\bibitem{Kenyon:2003uj}
R.~Kenyon, A.~Okounkov, and S.~Sheffield, {\it Dimers and amoebae},
  \href{http://xxx.lanl.gov/abs/math-ph/0311005}{{\tt math-ph/0311005}}.

\bibitem{Kenyon:2002a}
R.~Kenyon, {\it An introduction to the dimer model},
  \href{http://xxx.lanl.gov/abs/math.CO/0310326}{{\tt math.CO/0310326}}.

\bibitem{Kasteleyn}
P.~Kasteleyn, {\it Graph theory and crystal physics},  in {\em Graph theory and
  theoretical physics}, pp.~43--110.
\newblock Academic Press, London, 1967.

\bibitem{CYmoduli}
A.~Craw, R.~Karp, and K.~Kennaway. Work in progress.

\bibitem{Kenyon:rhombic}
R.~Kenyon and J.-M. Schlenker, {\it Rhombic embeddings of planar graphs with
  faces of degree 4},  \href{http://xxx.lanl.gov/abs/math-ph/0305057}{{\tt
  math-ph/0305057}}.

\bibitem{Vegh:2005}
A.~Hanany and D.~Vegh, {\it Quivers, tilings, branes and rhombi},
  \href{http://xxx.lanl.gov/abs/hep-th/0511063}{{\tt hep-th/0511063}}.

\bibitem{Feng:2001bn}
B.~Feng, A.~Hanany, Y.-H. He, and A.~M. Uranga, {\it Toric duality as {S}eiberg
  duality and brane diamonds},  {\em JHEP} {\bf 12} (2001) 035,
  [\href{http://xxx.lanl.gov/abs/hep-th/0109063}{{\tt hep-th/0109063}}].

\bibitem{Pick}
G.~Pick, {\it Geometrisches zur zahlenlehre},  {\em Sitzungber. Lotos} {\bf 19}
  (1899) 311.

\bibitem{Lins:1980}
S.~Lins, {\it Graphs of maps},
  \href{http://xxx.lanl.gov/abs/math.CO/0305058}{{\tt math.CO/0305058}}. PhD
  Thesis.

\bibitem{Berkooz:1996km}
M.~Berkooz, M.~R. Douglas, and R.~G. Leigh, {\it Branes intersecting at
  angles},  {\em Nucl. Phys.} {\bf B480} (1996) 265--278,
  [\href{http://xxx.lanl.gov/abs/hep-th/9606139}{{\tt hep-th/9606139}}].

\bibitem{Aganagic:1999fe}
M.~Aganagic, A.~Karch, D.~Lust, and A.~Miemiec, {\it Mirror symmetries for
  brane configurations and branes at singularities},  {\em Nucl. Phys.} {\bf
  B569} (2000) 277--302, [\href{http://xxx.lanl.gov/abs/hep-th/9903093}{{\tt
  hep-th/9903093}}].

\bibitem{Franco:2002ae}
S.~Franco and A.~Hanany, {\it Geometric dualities in 4d field theories and
  their 5d interpretation},  {\em JHEP} {\bf 04} (2003) 043,
  [\href{http://xxx.lanl.gov/abs/hep-th/0207006}{{\tt hep-th/0207006}}].

\bibitem{Feng:2004uq}
B.~Feng, Y.-H. He, and F.~Lam, {\it On correspondences between toric
  singularities and $(p,q)$- webs},  {\em Nucl. Phys.} {\bf B701} (2004)
  334--356, [\href{http://xxx.lanl.gov/abs/hep-th/0403133}{{\tt
  hep-th/0403133}}].

\bibitem{Feng:2002zw}
B.~Feng, S.~Franco, A.~Hanany, and Y.-H. He, {\it Symmetries of toric duality},
   {\em JHEP} {\bf 12} (2002) 076,
  [\href{http://xxx.lanl.gov/abs/hep-th/0205144}{{\tt hep-th/0205144}}].

\bibitem{mikhalkin1}
G.~Mikhalkin, {\it Amoebas of algebraic varieties and tropical geometry},
  \href{http://xxx.lanl.gov/abs/math.AG/0403015}{{\tt math.AG/0403015}}.

\bibitem{mikhalkin2}
G.~Mikhalkin, {\it Amoebas of algebraic varieties},
  \href{http://xxx.lanl.gov/abs/math.AG/0108225}{{\tt math.AG/0108225}}.

\bibitem{rullgard}
H.~Rullg{\aa}rd, {\it Topics in geometry, analysis and inverse problems}, .

\bibitem{Harnack}
A.~Harnack, {\it {\"U}ber vieltheiligkeit der ebenen algebraischen curven},
  {\em Math. Ann.} {\bf 10} (1876).

\bibitem{Gross:2000}
M.~Gross, {\it Examples of special {L}agrangian fibrations},
  \href{http://xxx.lanl.gov/abs/math.AG/0012002}{{\tt math.AG/0012002}}.

\bibitem{Martelli:2005tp}
D.~Martelli, J.~Sparks, and S.~T. Yau, {\it The geometric dual of
  $a$-maximisation for toric {S}asaki- {E}instein manifolds},
  \href{http://xxx.lanl.gov/abs/hep-th/0503183}{{\tt hep-th/0503183}}.

\bibitem{KK-KH}
K.~Hori and K.~D. Kennaway. Work in progress.

\bibitem{Stienstra:2005a}
J.~Stienstra, {\it Mahler measure variations, {E}isenstein series and instanton
  expansions},  \href{http://xxx.lanl.gov/abs/math.NT/0502193}{{\tt
  math.NT/0502193}}.

\bibitem{Stienstra:2005b}
J.~Stienstra, {\it Mahler measure, {E}isenstein series and dimers},
  \href{http://xxx.lanl.gov/abs/math.NT/0502197}{{\tt math.NT/0502197}}.

\bibitem{Koushnirenko}
A.~G. Koushnirenko, {\it Polyh\'edres de {N}ewton et nombres de {M}ilnor},
  {\em Inv. Math.} {\bf 32} (1976) 1--31.

\bibitem{Bernstein}
D.~Bernstein, {\it The number of roots of a system of equations},  {\em Func.
  Analysis and App.} {\bf 9(2)} (1975) 183--5.

\bibitem{shiff-zeld}
B.~Shiffman and S.~Zelditch, {\it Random polynomials with prescribed {N}ewton
  polytope, {I}},  \href{http://xxx.lanl.gov/abs/math/0203074}{{\tt
  math/0203074}}.

\end{thebibliography}\endgroup

\end{document}